\documentclass[prd,aps,eqsecnum,floatfix,nofootinbib,preprint,tightenlines]{revtex4}

\usepackage{latexsym}
\usepackage{graphicx}
\usepackage{multirow}
\usepackage{color}

\def\bibi{\bibitem}

\def\a{\alpha}
\def\b{\beta}

\def\d{\delta}
\def\g{\gamma}

\def\k{\kappa}

\def\m{\mu}
\def\n{\nu}

\def\p{\pi}                     
\def\r{\rho}                    
\def\s{\sigma}                  

\def\D{\Delta}

\def\G{\Gamma}

\def\P{\Pi}

\def\cq{{\cal Q}}

\def\cbo{{\,\raise-.15ex\Sc [\,}}                       
\def\gtap{\raisebox{-.4ex}{\rlap{$\sim$}} \raisebox{.4ex}{$>$}}   

\def\ddt#1{{\buildrel {\hbox{\LARGE .\kern-2pt.}} \over {#1}}}

\def\ie{\mbox{\it i.e.}}
\def\eg{\mbox{\it e.g.}}

\def\floatcaption#1#2{ \caption{ #2 \ [#1] \label{#1}} }
\def\floatcaption#1#2{ \caption{#2 \label{#1}} }

\def\bibi{\bibitem}    

\def\ttl#1{{\it #1}}
\def\ttl#1{}

\long\def\symbolfootnote[#1]#2{\begingroup%
\def\thefootnote{\fnsymbol{footnote}}\footnote[#1]{#2}\endgroup}

\long \def \blockcomment #1\endcomment{}

\def\eg{{\it e.g.}}

\def\seef{{\it cf.}}

\def\ansatz{{\it ansatz}}

\def\hw{{\hat{w}}}

\begin{document}

\rightline{TUM-HEP-961/14}
\begin{center}
\vspace{10mm}
\begin{boldmath}
{\large\bf The strong coupling from the revised ALEPH data for hadronic
$\tau$ decays
}
\end{boldmath}

\vspace{3ex}
{Diogo~Boito,$^a$ Maarten~Golterman,$^b$
 Kim~Maltman,$^{c,d}$\\ James~Osborne,$^e$ Santiago~Peris$^f$%
\\[0.1cm]
{\it
\null$^a$Physik Department T31, Technische Universit\"at M\"unchen\\
James-Franck-Stra\ss e 1, D-85748 Garching, Germany\\
\null$^b$Department of Physics and Astronomy\\
San Francisco State University, San Francisco, CA 94132, USA\\
\null$^c$Department of Mathematics and Statistics\\
York University,  Toronto, ON Canada M3J~1P3\\
\null$^d$CSSM, University of Adelaide, Adelaide, SA~5005 Australia\\
\null$^e$Physics Department, University of Wisconsin-Madison\\
1150 University Avenue, Madison, WI  53706, USA\\
\null$^f$Department of Physics, Universitat Aut\`onoma de Barcelona\\
E-08193 Bellaterra, Barcelona, Spain}}
\\[6mm]
{ABSTRACT}
\\[2mm]
\end{center}
\begin{quotation}
We apply an analysis method previously developed
for the extraction of the strong coupling
from the OPAL data to the recently revised ALEPH data for non-strange hadronic
$\tau$ decays.   Our analysis yields the values $\a_s(m_\tau^2)=0.296\pm 0.010$
using fixed-order perturbation theory, and $\a_s(m_\tau^2)=0.310\pm 0.014$
using contour-improved perturbation theory.   Averaging these values with
our previously obtained values from the OPAL data, we find 
$\a_s(m_\tau^2)=0.303\pm 0.009$, respectively, $\a_s(m_\tau^2)=0.319\pm 0.012$.
We present a critique of the analysis method employed previously,
for example in analyses by the ALEPH and OPAL collaborations,
and compare it with our own approach.
Our conclusion is that non-perturbative effects limit the accuracy
with which the strong coupling, an inherently perturbative
quantity, can be extracted at energies as low as the $\tau$ mass.
Our results further indicate that systematic errors on the
determination of the strong coupling from analyses of hadronic 
$\tau$-decay data have been underestimated in much
of the existing literature.

\end{quotation}

\vfill
\eject
\setcounter{footnote}{0}

\section{\label{intro} Introduction}
Recently, Ref.~\cite{ALEPH13}, for the ALEPH collaboration, updated and revised
previous ALEPH results for the non-strange vector ($V$) and
axial vector ($A$) spectral distributions obtained from measurements
of hadronic $\tau$ decays. In particular, Ref.~\cite{ALEPH13}
corrects a problem in the publicly posted 2005 and 2008
versions of the correlations between different energy bins uncovered
in Ref.~\cite{Tau10}.\footnote{The updated and corrected data
can be found at http://aleph.web.lal.in2p3.fr/tau/specfun13.html.}
The corrected data supersede those originally
published by the ALEPH collaboration \cite{ALEPH,ALEPH08}.

One of the hadronic quantities of interest that can be extracted from these
data is the strong coupling $\a_s(m_\tau^2)$ at the $\tau$ mass, through the use
of Finite-Energy Sum Rules (FESRs) \cite{shankar}, as advocated long
ago~\cite{Braaten88,BNP}. Both the ALEPH and OPAL \cite{OPAL} collaborations
have done so by applying an analysis strategy, developed in
Refs.~\cite{BNP,DP1992}, in which small, but non-negligible
non-perturbative effects were estimated using a truncated form of the
operator product expansion (OPE). A feature of the particular
truncation scheme employed is that it assumes that, in addition to contributions which
violate quark-hadron duality, also OPE contributions of dimension
$D>8$ unsuppressed by non-leading powers of $\alpha_s$ can be safely
neglected. Given the goal of extracting $\a_s(m_\tau^2)$ with the best
possible accuracy, these features of what we will refer to
as the ``standard analysis'' have been questioned, starting
with the work of Refs.~\cite{MY08,CGP}. In these works, it
was argued that both the OPE truncation to terms with $D\le 8$
and the neglect of
violations of quark-hadron duality lead to additional numerically
non-negligible systematic uncertainties not included in the errors
obtained on $\a_s(m_\tau^2)$ and the OPE condensates from the standard-analysis
approach. In order to remedy this
situation, in Refs.~\cite{alphas1,alphas2},
we developed a new analysis strategy designed to take both OPE and
duality-violating (DV) non-perturbative effects consistently into
account. This strategy
was then successfully applied to the OPAL data~\cite{alphas1,alphas2}.
In the present article, we apply this analysis strategy to the corrected
ALEPH data, and compare our results to those obtained from the OPAL data
in Ref.~\cite{alphas2} as well as to those of
the recent re-analysis presented in Ref.~\cite{ALEPH13}.

The calculation of the order-$\a_s^4$ term \cite{PT} in the perturbative
expansion of the Adler function in 2008 led to a renewed interest in the
determination of the strong coupling from hadronic $\tau$ decays, with many
attempts to use this new information on the theory side of the
relevant FESRs in order to sharpen the extraction of
$\a_s(m_\tau^2)$ from the
data~\cite{ALEPH13,ALEPH08,MY08,alphas1,alphas2,PT,BJ,Menke,CF,DM,Cetal}.
Since the perturbative series converges rather slowly, different partial
resummation schemes have been considered, leading to variations in the
obtained results. The majority of these post-2007
updates (Refs.~\cite{ALEPH13,ALEPH08,PT,BJ,Menke,CF,DM,Cetal}), however,
were carried out assuming that the standard-analysis treatment of
non-perturbative effects was essentially correct, with none of
the references in this subset, with the exception of Refs.~\cite{ALEPH13,ALEPH08},
redoing the analysis starting from the underlying experimental
data (the emphasis, instead, being on the merits of different resummation
schemes for the perturbative expansion). Reference~\cite{MY08}, which did
revisit the determination of the higher-$D$ OPE contributions, and
performed a more careful treatment of these contributions,
did not, however, include DV contributions in its analysis framework.
While its results were tested for self-consistency,
the absence of a representation of DV effects meant
no estimate of the residual systematic error associated with their neglect
was possible. The only articles to incorporate both the
improved treatment of higher-$D$ OPE contributions and an implementation
of a physically motivated representation of DV effects were those of
Refs.~\cite{alphas1,alphas2}, which, due to the problem with the then-existing
ALEPH covariance matrices, were restricted to analyzing OPAL data.
Our goal in this article is to reconsider the
treatment of non-perturbative effects employing the newly released
ALEPH data, which have
significantly smaller errors than the OPAL data.  We will present
results for the two most popular resummation schemes for the
perturbative (\ie, $D=0$ OPE) series: fixed-order
perturbation theory (FOPT) and contour-improved perturbation
theory (CIPT)~\cite{CIPT}, without trying to resolve the discrepancies
that arise between them (for an overview of the two methods,
see Ref.~\cite{MJ}).

This article is organized as follows. In Sec.~\ref{theory} we give a
brief overview of the necessary theory, referring to Ref.~\cite{alphas1} for
more details. In Sec.~\ref{data} we discuss the new ALEPH data set, and
check explicitly that the current publicly posted
version of the correlation matrices pass the test that led
to the identification of the problem with the previous
version~\cite{Tau10}.
We also show the comparison of the experimental ALEPH and OPAL
non-strange spectral functions. Section~\ref{strategy}
summarizes our fitting strategy, developed in Refs.~\cite{alphas1,alphas2}. Sections~\ref{fits}
and~\ref{results} present the details of the fits, and the results we
obtain from them for $\a_s(m_\tau^2)$ and dimension 6 and 8 OPE coefficients
in the $V$ and $A$ channels. 
We explore the $\chi^2$ landscape using
the Markov-chain Monte Carlo code
{\tt hrothgar}~\cite{hrothgar}, which in the case of the OPAL data
proved useful in uncovering potential ambiguities.
Also included is an estimate
for the total non-perturbative contribution to the
ratio of non-strange hadronic and electronic $\tau$
branching fractions. In Sec.~\ref{results} we check how well the two
Weinberg sum rules \cite{SW} and the sum rule for the
electro-magnetic pion mass difference \cite{EMpion} are satisfied
by our results. Finally, in Sec.~\ref{ALEPH}, we present
a critical discussion of the standard analysis employed in
Refs.~\cite{ALEPH13,ALEPH,ALEPH08,OPAL}, focusing on the most recent of these,
described in Ref.~\cite{ALEPH13}. We demonstrate explicitly the
inconsistency of this analysis with regard to the treatment of
non-perturbative effects, and
conclude that, while the standard analysis approach was
a reasonable one to attempt in the past, it must be abandoned
in current or future determinations of $\a_s(m_\tau^2)$ from hadronic $\tau$ decay data.
In our concluding section, Sec.~\ref{conclusion}, we compare our approach
with the standard-analysis method, highlighting and juxtaposing the
assumptions underlying each, and summarize our
results.

\section{\label{theory} Theory overview}
The sum-rule analysis starts from the correlation functions
\begin{eqnarray}
\label{correl}
\P_{\m\n}(q)&=&i\int d^4x\,e^{iqx}\langle 0|T\left\{J_\m(x)J^\dagger_\n(0)\right\}|0\rangle\\
&=&\left(q_\m q_\n-q^2 g_{\m\n}\right)\P^{(1)}(q^2)+q_\m q_\n\P^{(0)}(q^2)\nonumber\\
&=&\left(q_\m q_\n-q^2 g_{\m\n}\right)\P^{(1+0)}(q^2)+q^2 g_{\m\n}\P^{(0)}(q^2)\ ,\nonumber
\end{eqnarray}
where $J_\m$ stands for the non-strange $V$ or $A$ current,
$\overline{u}\g_\m d$ or $\overline{u}\g_\m\g_5 d$, while
the superscripts $(0)$ and $(1)$ label spin. The decomposition in the
third line employs the combinations $\P^{(1+0)}(q^2)$ and $q^2\P^{(0)}(q^2)$,
which are free of kinematic singularities.
Defining $s=q^2=\, -Q^2$ and the spectral function
\begin{equation}
\label{spectral}
\r^{(1+0)}(s)=\frac{1}{\p}\;\mbox{Im}\,\P^{(1+0)}(s)\ ,
\end{equation}
Cauchy's theorem and the analytical properties of $\P^{(1+0)}(s)$,
applied to the contour in Fig.~\ref{cauchy-fig}, imply the FESR
\begin{eqnarray}
\label{cauchy}
I^{(w)}_{V/A}(s_0)\equiv\frac{1}{s_0}\int_0^{s_0}ds\,w(s)\,\r^{(1+0)}_{V/A}(s)
&=&-\frac{1}{2\p i\, s_0}\oint_{|s|=s_0}
ds\,w(s)\,\P^{(1+0)}_{V/A}(s)\ ,
\end{eqnarray}
valid for any $s_0>0$ and any weight $w(s)$ analytic inside and on the
contour \cite{shankar}.

\begin{figure}
\vspace*{4ex}
\begin{center}
\includegraphics*[width=6cm]{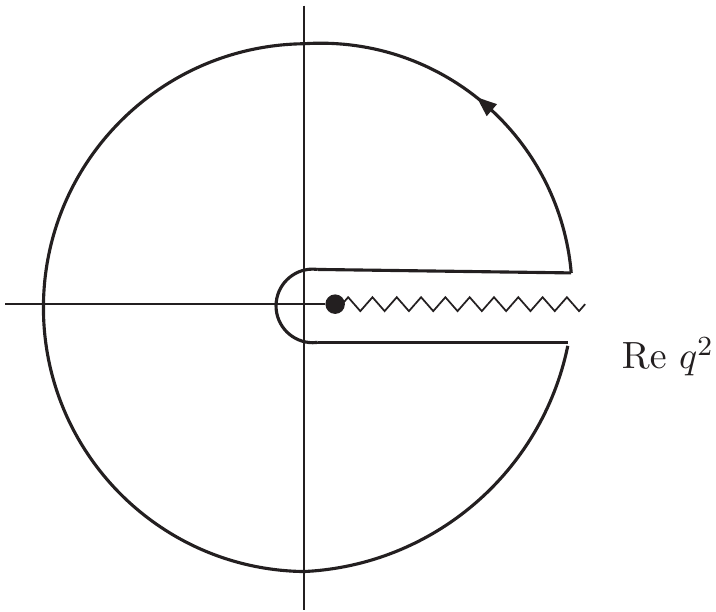}
\end{center}
\begin{quotation}
\floatcaption{cauchy-fig}%
{{\it Analytic structure of $\P^{(1+0)}(q^2)$ in the complex $s=q^2$ plane.
There is a cut
on the positive real axis starting at $s=q^2=4m_\p^2$ (a pole at $s=q^2=m_\p^2$
and a cut starting at $s=9m_\p^2$) for the $V$ ($A$)
case.
The solid curve shows the contour used in Eq.~(\ref{cauchy}).}}
\end{quotation}
\vspace*{-4ex}
\end{figure}

The flavor $ud$ $V$ and $A$ spectral functions can be
experimentally determined from the differential versions of the ratios,
\begin{equation}
\label{R}
R_{V/A;ud}=
{\frac{\G [\tau\rightarrow ({\rm hadrons})_{V/A;ud}\n_\tau (\g ) ]}
{\G [\tau\rightarrow e\bar{\n}_e \nu_\tau (\g ) ]}}\ ,
\end{equation}
of the width for hadronic decays induced by the relevant current to that
for the electron mode. Explicitly~\cite{tsai71},
\begin{equation}
\label{taukinspectral}
{\frac{dR_{V/A;ud}(s)}{ds}}= 12\pi^2\vert V_{ud}\vert^2 S_{EW}\,
{\frac{1}{m_\tau^2}} \left[ w_T(s;m_\tau^2) \rho_{V/A;ud}^{(1+0)}(s)
- w_L(s;m_\tau^2) \rho_{V/A;ud}^{(0)}(s) \right]\ ,
\end{equation}
where $S_{EW}$ is a short-distance electroweak correction and
$w_T(s;s_0)=(1-s/s_0)^2(1+2s/s_0)$, $w_L(s;s_0)=2(s/s_0)(1-s/s_0)^2$.
Apart from the pion-pole contribution, which is not chirally
suppressed, 
$\rho_{V/A;ud}^{(0)}(s) = O[(m_d\mp m_u)^2]$, and
the continuum part of $\rho_{V/A}^{(0)}(s)$ is thus numerically negligible.
As a result, the spectral functions
$\rho^{(1+0)}_{V/A;ud}(s)$ can be determined directly
from $dR_{V/A;ud}(s)/ds$. The FESR~(\ref{cauchy}) can thus be studied
for arbitrary $s_0$ and arbitrary analytic weight $w(s)$.
From now on, we will denote the experimental version of the
spectral integral on the left-hand side of Eq.~(\ref{cauchy}) by
$I_{V/A;\rm ex}^{(w)}(s_0)$
(generically, $I_{\rm ex}^{(w)}(s_0)$) and the theoretical
representation of the contour integral on the right-hand side by
$I_{V/A;\rm th}^{(w)}(s_0)$ (generically, $I_{\rm th}^{(w)}(s_0)$).

For large enough $|s|=s_0$, away from the positive real axis,
$\P^{(1+0)}(s)$ can be approximated by the OPE
\begin{equation}
\label{OPE}
\P^{(1+0)}_{\rm OPE}(s)=\sum_{k=0}^\infty \frac{C_{2k}(s)}{(-s)^{k}}\ ,
\end{equation}
with the OPE coefficients $C_{2k}$ logarithmically dependent on $s$ through
perturbative corrections. The term with $k=0$ corresponds
to the purely perturbative, mass-independent contributions, which have been
calculated to order $\a_s^4$ in Ref.~\cite{PT}, and are the
same for the $V$ and $A$ channels. The $C_{2k}$ with
$k\ge 1$ are different for the $V$ and $A$ channels, and, for $k>1$,
contain non-perturbative $D=2k$ condensate contributions. As in
Refs.~\cite{alphas1,alphas2}, we will neglect purely perturbative quark-mass
contributions to $C_2$ and $C_4$, as they are numerically very small for
the non-strange FERSs we consider in this article. For the same reason, we
will neglect the $s$-dependence of the coefficients $C_{2k}$ for $k>1$. For
the perturbative contribution, $C_0$,
we will use the result of Ref.~\cite{PT} and extract
$\a_s(m_\tau^2)$ in the $\overline{\rm MS}$ scheme. Since the coefficient $c_{51}$ of 
the order-$\a_s^5$ term has not been calculated we will use the estimate $c_{51}=283$
of Ref.~\cite{BJ} with a 100\% uncertainty. We will
also employ both FOPT and CIPT resummation schemes
in evaluating the truncated perturbative series. For more
details on the treatment of the $D>0$ OPE
contributions, we refer the reader to Ref.~\cite{alphas1}.

Perturbation theory, and in general the OPE, breaks down near the positive real
$s=q^2$ axis \cite{PQW}. We account for this by replacing the right-hand
side of Eq.~(\ref{cauchy}) by
\begin{equation}
\label{split}
-\frac{1}{2\p is_0}\oint_{|s|=s_0}ds\,w(s)\,
\left(\P^{(1+0)}_{\rm OPE}(s)+\D(s)\right)\ ,
\end{equation}
with
\begin{equation}
\label{DVdef}
\D(s)\equiv\P^{(1+0)}(s)-\P^{(1+0)}_{\rm OPE}(s)\ ,
\end{equation}
where the difference $\D(s)$ accounts, by definition, for the
quark-hadron duality violating contribution to $\Pi^{(1+0)}(s)$.
As shown in Ref.~\cite{CGP}, Eq.~(\ref{split}) can be rewritten as
\begin{equation}
\label{sumrule}
I_{\rm th}^{(w)}(s_0) = -\frac{1}{2\p is_0}\oint_{|s|=s_0}
ds\,w(s)\,\P^{(1+0)}_{\rm OPE}(s)-\frac{1}{s_0}\,
\int_{s_0}^\infty ds\,w(s)\,\frac{1}{\p}\,\mbox{Im}\,
\D(s)\ ,
\end{equation}
if $\D(s)$ is assumed to decay fast enough as $s\to\infty$. The imaginary
parts $\frac{1}{\p}\,\mbox{Im}\,\D_{V/A}(s)$ can be interpreted as
the DV parts, $\rho_{V/A}^{\rm DV}(s)$, of the $V/A$ spectral functions.

The functional form of $\D(s)$ is not known, even for large
$s$, and we thus need to resort to a model in order to account
for DVs. Following Refs.~\cite{CGP,CGPmodel,CGP05},\footnote{See
also Refs.~\cite{russians,MJ11}.} we use a model based on large-$N_c$ and Regge
considerations, choosing to parametrize $\rho_{V/A}^{\rm DV}(s)$
as\footnote{In Ref.~\cite{alphas1} we used $\k_{V/A}\equiv e^{-\d_{V/A}}$; in
Ref.~\cite{alphas2} we switched to $\d_{V/A}$.}
\begin{equation}
\label{ansatz}
\rho_{V/A}^{\rm DV}(s)=
e^{-\d_{V/A}-\g_{V/A}s}\sin{(\a_{V/A}+\b_{V/A}s)}\ .
\end{equation}
This introduces, in addition to $\a_s$ and the $D\ge 4$ OPE condensates, four
new parameters in each channel. As in Refs.~\cite{alphas1,alphas2}, we will
assume that Eq.~(\ref{ansatz}) holds for $s\ge s_{\rm min}$, with $s_{\rm min}$ to be
determined from fits to the data. This, in turn, assumes that we can take
$s_{\rm min}$ significantly smaller than $m_\tau^2$, \ie, that both the OPE and
the {\it ansatz}~(\ref{ansatz}) can be used in some interval below $m_\tau^2$.

Let us pause at this point to revisit the basic ideas 
underlying the DV \ansatz~(\ref{ansatz}).
Since there exists, as yet, no theory of DVs starting from first principles in
QCD, the \ansatz~(\ref{ansatz}) represents simply our best,
physically motivated, guess as to an appropriate form of DV contributions
to the $V$ and $A$ spectral functions. The 
damped oscillatory form employed
is, however, far from arbitrary. First, it reflects the fact that 
DVs are expected to produce almost harmonic oscillations around the 
perturbative continuum, in line with expectations from
Regge theory, in which resonances occur with equal squared-mass
spacings on the relevant daughter trajectories. 
Second, the exponential damping factor in the \ansatz\ 
reflects the understanding that the OPE is (at best) an asymptotic, and 
not a convergent, expansion. It is certainly the case that the OPE 
representation is more successful for euclidean $Q^2$ $\sim 2$ GeV$^2$ 
than for comparable Minkowski scales, $q^2\sim 2$ GeV$^2$, where DV 
contributions are clearly visible in the spectral 
functions. Once DVs are identified as representing the irreducible error present 
in this asymptotic expansion, it is natural to assume that their 
contribution should exhibit an exponentially suppressed dependence 
on $s=q^2$, as in our \ansatz~(\ref{ansatz}). These qualitative expectations are also reflected
in the explicit Regge- and large-$N_c$-motivated model
discussed in much more detail in Refs.~\cite{CGP,CGPmodel,CGP05,russians,MJ11}. 
These plausibility arguments aside, we will use the precise ALEPH data to 
subject the parametrization~(\ref{ansatz}) to non-trivial tests described in 
detail in Sec.~\ref{ALEPH}.

Several considerations underlie our choice of weight functions $w(s)$.
First, we will choose
weight functions which are likely to be well-behaved in perturbation theory,
based on the findings of Ref.~\cite{BBJ12}. In particular, we will exclude weight
functions with a term linear in $s$, and require
the ones we use to include a constant term (which we will normalize to one).
Second, because it is not known at which order the OPE might start to
diverge (for the values of $s_0$ of interest), we wish to avoid terms in
Eq.~(\ref{OPE}) with $D>8$, about which essentially nothing is known. That means
that if we do not want to arbitrarily set the coefficients
$C_D$ with $D>8$ equal to zero, our weight functions
are restricted to polynomials with degree not larger than three. Combining
these constraints, we are left with the form
\begin{equation}
\label{weightform}
w(s;s_0)=1+a(s/s_0)^2+b(s/s_0)^3\ .
\end{equation}

This allows us at most three independent weight functions, and limits the
extent to which we can use sufficiently pinched weights, \ie, weights with
a (multiple) zero at $s=s_0$, which help to suppress DVs \cite{KM98,DS99}.
The upshot is that, if we want to exploit the $s_0$ dependence of the data
(instead of fitting only at $s_0=m_\tau^2$, as was done in
Refs.~\cite{ALEPH13,ALEPH,ALEPH08,OPAL}) and treat the OPE consistently,
modeling DVs is unavoidable \cite{alphas1}. We emphasize that the $s_0$ dependence of
fit results provides a crucial test of the validity of FESR fits to the
data, as we will see below.
As in Refs.~\cite{alphas1,alphas2}, we choose to consider the weight functions
\begin{eqnarray}
\hw_0(x)&=&1\ ,\label{weights}\\
\hw_2(x)&=&1-x^2\ ,\nonumber\\
\hw_3(x)&=&(1-x)^2(1+2x)=1-3x^2+2x^3=w_T(s;s_0)\ ,\nonumber\\
x&\equiv&s/s_0\ .\nonumber
\end{eqnarray}
The first choice, $\hat{w}_0$, is predicated on the fact that pinching is
known to suppress DV contributions and we need at least one weight which
is sufficiently sensitive to DV contributions to fix the DV parameters.
The remaining two weights $\hat{w}_2$ and $\hat{w}_3$ are singly and
doubly pinched, respectively.  For a more detailed
discussion of our choices, we refer to Ref.~\cite{alphas1}.
An important observation is that these choices for what goes
into the parametrization of $I_{\rm th}^{(w)}(s_0)$ did remarkably
well in the analysis of the OPAL data.   It therefore makes sense
to see what happens if
we apply the same strategy to the ALEPH data.

\section{\label{data} The ALEPH data}

In this section, we discuss the revised ALEPH data, which are available
from Ref.~\cite{datahtml}. First, we perform a minor rescaling, in order to account for
more precise values
of some ``external'' quantities (\ie, quantities not directly
measured by ALEPH, but used in their
analysis of the data); this is discussed in
Sec.~\ref{renormalization}, where we also specify our other inputs.
Then, in Sec.~\ref{correlations} we apply to the corrected covariance
matrices the test of Ref.~\cite{Tau10} that led us to uncover the problem with the
previously published versions, and verify
that the revised covariances pass this test. Finally, we compare the
$V$ and $A$ spectral functions obtained from the ALEPH data with
those from the OPAL data.

\subsection{\label{renormalization} Data and normalization}
We will use the following input values in our analysis:
\begin{eqnarray}
\label{input}
m_\tau&=&1.77682(16)~\mbox{GeV}\ ,\\
B_e&=&0.17827(40)\ ,\nonumber\\
V_{ud}&=&0.97425(22)\ ,\nonumber\\
S_{EW}&=&1.0201(3)\ ,\nonumber\\
m_\p&=&139.57018(35)~\mbox{MeV}\ ,\nonumber\\
f_\p&=&92.21(14)~\mbox{MeV}\ .\nonumber
\end{eqnarray}
Here $B_e$ is the branching fraction for the decay
$\tau\to e\overline{\n}_e\n_\tau$ and we have used the result of an HFAG
fit of the $\tau$ branching fractions which incorporates $\p_{\m 2}$ and
$K_{\m 2}$ data and Standard Model expectations based on these data for
the $\p$ and $K$ branching fractions
\cite{hfagpimu2kmu2fit11}; $f_\pi$ is the $\pi$ decay constant.
The value for $V_{ud}$ is from Ref.~\cite{htrpp10}, that for $S_{EW}$
from Ref.~\cite{SEW}, and the values for $m_\tau$, $m_\pi$ and $f_\pi$ are from the
Particle Data Group \cite{PDG}. Only the error on $B_e$ has a significant
effect in our analysis; errors on the other input quantities are too small
to affect the final analysis errors in any significant way.

To the best of our knowledge, Ref.~\cite{ALEPH13} uses the values
$B_e=0.17818(32)$ and $S_{EW}=1.0198$. This value for $B_e$ we infer from
the ALEPH values for $R_V=1.782(9)$ \cite{ALEPH13} and the corresponding
branching fraction $B_V=0.31747$ \cite{datahtml} specified in the publicly
posted $V$ data file (no error quoted). The continuum
(pion-less) axial branching fraction
$B_{A,{\rm cont}}=0.19369$
with $B_e=0.17818$ translates into $R_{A,{\rm cont}}=1.08705$.
From these values, and the quoted value $R_{ud}=3.475(11)$
\cite{ALEPH13}, it follows that the ALEPH value for $R_\p$,
the pion pole contribution to $R_{ud}$,
is $R_\p=0.606$. However, if one employs the very
precisely known value of $f_\p$ quoted above, obtained from $\p_{\m 2}$
decays, together with the quoted values for $S_{EW}$ and $V_{ud}$, one
finds instead the more precisely determined expectation $R_\p=0.6101$.
Using this latter value as well as the ALEPH value $R_{ud}=3.475(11)$
leads to $R_V+R_{A,{\rm cont}}=2.865$, instead of the ALEPH value
$(B_V+B_{A,{\rm cont}})/B_e=(0.31747+0.19369)/0.17818=2.8688$.
We employ the more precise $\p_{\m 2}$ expectation for the important
$A$ channel pion-pole contribution, and take this difference into
account by rescaling the $V$ and continuum
$A$ non-strange spectral functions by the common factor
$2.865/2.8688=0.9987$, since we have no information on whether this
rescaling should affect the $V$ and $A$ channels asymmetrically.
Our rescaling is thus imperfect, but it is to be noted that the effect
of this rescaling lowers our value for $\a_s(m_\tau^2)$ by less than one
percent, a much smaller shift than that allowed
by the total error, see Sec.~\ref{results}.

The new ALEPH data use a variable bin width, with the highest bin,
number 80, centered at ${\tt sbin}(80)=3.3375$~GeV$^2$, which is above
$m_\tau^2=3.1571$~GeV$^2$. The next-highest bin, number 79, is centered
at ${\tt sbin}(79)=3.0875$~GeV$^2$, with a width
${\tt dsbin}(79)=0.1750$~GeV$^2$, so that also
${\tt sbin}(79)+{\tt dsbin}(79)/2>m_\tau^2$.
In order to avoid using values of
$s$ larger than $m_\tau^2$, we will modify these values to
\begin{eqnarray}
\label{binadjust}
{\tt sbin}(79)&=&3.07854~\mbox{GeV}^2\ ,\\
{\tt dsbin}(79)&=&0.157089~\mbox{GeV}^2\ ,\nonumber
\end{eqnarray}
so that ${\tt sbin}(79)+{\tt dsbin}(79)/2=m_\tau^2$.

Finally, ALEPH provides binned spectral data for ${\tt sfm2}({\tt sbin})$,
which are related to the spectral functions by
\begin{equation}
\label{ALEPHform}
{\tt sfm2}({\tt sbin})=
100\times \frac{12\p^2|V_{ud}|^2 S_{EW}B_e}{m_\tau^2}\,\D
w^T({\tt sbin};m_\tau^2)\rho^{(1+0)}({\tt sbin})\ ,
\end{equation}
in which
\begin{equation}
\label{binwidthav}
\D w^T({\tt sbin};m_\tau^2)=\int_{{\tt sbin}-{\tt dsbin}/2}^{{\tt sbin}
+{\tt dsbin}/2}ds\, w^T(s;m_\tau^2)\ .
\end{equation}
For infinitesimal ${\tt dsbin}=ds$ one has
$\D w^T(s;m_\tau^2)=w^T(s;m_\tau^2)ds$, but for finite
bin width we have to make a choice in how we construct moments with
other weights from the spectral functions obtained from
Eq.~(\ref{ALEPHform}). We choose to use
the definition
\begin{equation}
\label{defIex}
I^{(w)}_{\rm ex}(s_0)=\sum_{{\tt sbin}\le s_0}\left(\int_{{\tt sbin}-
{\tt dsbin}/2}^{{\tt sbin}+{\tt dsbin}/2}ds\,
w(s;s_0)\right)\rho^{(1+0)}({\tt sbin})\
\end{equation}
for all moments considered in this article.

\subsection{\label{correlations} Correlations}
\begin{figure}[t]
\begin{center}
\includegraphics*[width=11cm]{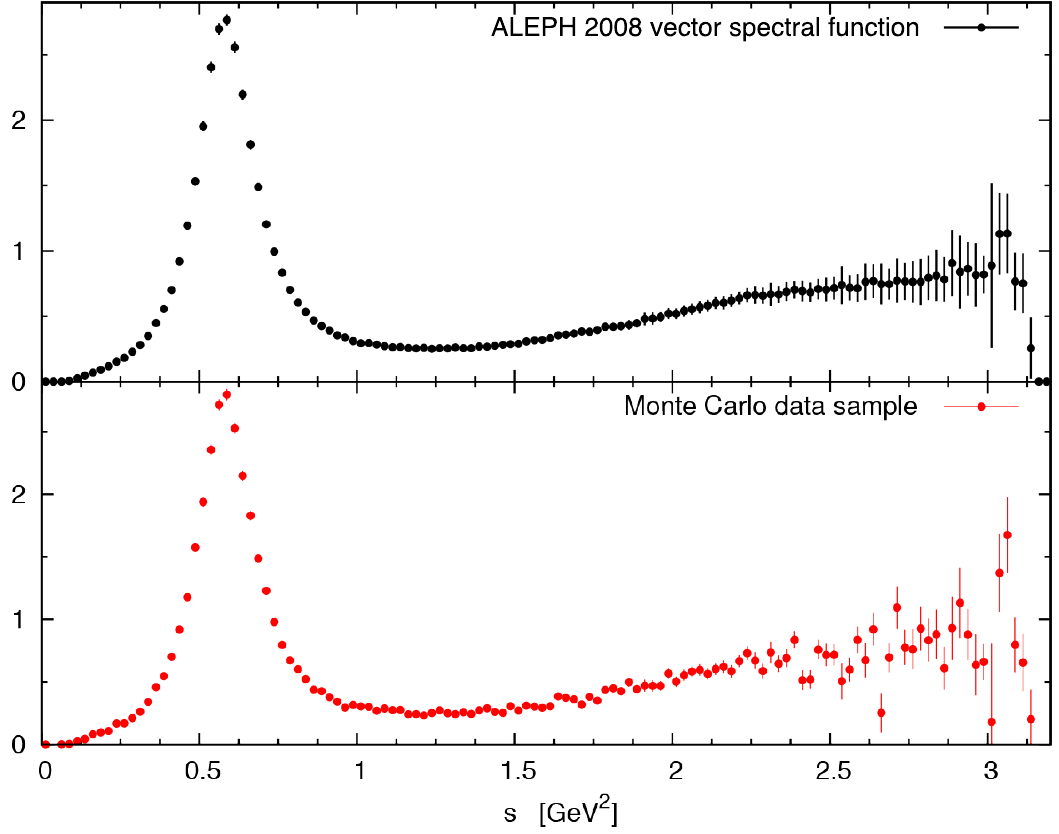}
\end{center}
\begin{quotation}
\floatcaption{test1}%
{{\it Vector spectral function times $2\p^2$. Top panel: ALEPH data from 2008
{\rm\cite{ALEPH08}}; bottom panel: Monte Carlo sample with 2008 covariance
matrix. }}
\end{quotation}
\vspace*{-4ex}
\end{figure}
\begin{figure}[t]
\begin{center}
\includegraphics*[width=11cm]{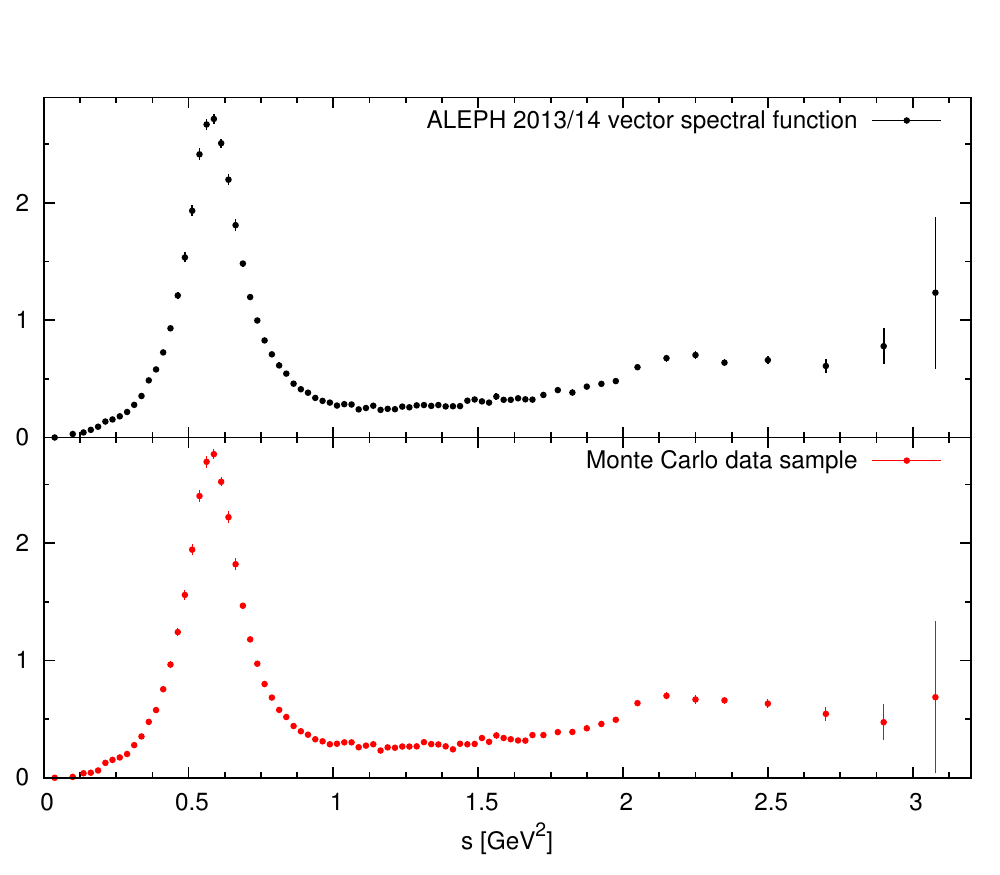}
\end{center}
\begin{quotation}
\floatcaption{test2}{{\it Vector spectral function times $2\p^2$. Top panel:  ALEPH data from 2013
{\rm\cite{ALEPH13}}; bottom panel: Monte Carlo sample with 2013 covariance
matrix.}}
\end{quotation}
\vspace*{-4ex}
\end{figure}
As shown in Ref.~\cite{Tau10}, there was a problem with the
publicly posted 2005 and 2008 versions of the ALEPH covariance matrices.
This problem, since corrected in Ref.~\cite{ALEPH13}, turns out to have
resulted from an inadvertent omission of contributions to the correlations
between different bins induced by the unfolding procedure. The problem was
discovered by producing fake data sets from a multivariate gaussian
distribution based on the posted ALEPH data and covariance matrices,
and then comparing the resulting fake data to the actual ALEPH data. The
result of this test is shown in Fig.~\ref{test1}, which is the same
as Fig.~3 of Ref.~\cite{Tau10}. The top panel shows the experimental data
taken from  Ref.~\cite{ALEPH08}, the bottom panel a typical fake data set
produced using the corresponding covariance matrix. The absence of the
strong correlations seen in the actual data from the corresponding fake
data is what signals the existence of the problem with the previous
version of the ALEPH covariance matrix. Figure~\ref{test2} shows the
result of performing the same test on the updated and corrected results
reported in Ref.~\cite{ALEPH13}, the top panel again showing the actual ALEPH
data and the bottom panel a typical fake data set. The fake data (red
points) obviously behave much more like the corresponding real data
than was the case previously.%
\footnote{Even though the new wider binning near the kinematic endpoint
makes it somewhat harder to see such differences in this region.}
We have examined many such fake data sets with the same conclusion.
A similar exercise was, of course, carried
out for the $A$-channel case.

\subsection{\label{OPAL} Comparison with OPAL data}
In Fig.~\ref{ALEPH-OPAL} we show the vector and axial spectral functions as
measured by ALEPH \cite{ALEPH13,datahtml} and OPAL \cite{OPAL}. The
normalizations of the spectral functions for both experiments have
been updated to take into account modern values for relevant branching
fractions; for the normalization of ALEPH
data, see Sec.~\ref{renormalization} above, and for the normalization of
OPAL data, see Sec.~III of Ref.~\cite{alphas2}.

While there is in general good agreement between the ALEPH and OPAL spectral
functions, a detailed inspection reveals some tension between the
two, given the size of the errors, for instance in the regions below
$0.5$~GeV$^2$ and around $2$~GeV$^2$ in the vector channel, with possibly
anti-correlated tensions in the same regions in the axial channel.
The presence of a large $D=0$, 1-loop $\alpha_s$-independent 
contribution in the weighted OPE integrals enhances the impact of
such small discrepancies on the output $\alpha_s(m_\tau^2)$. 
We quantify the impact of these differences below, showing that they lead
to some tension between the values for $\a_s(m_\tau^2)$ obtained from the 
two data sets, though the results turn out to agree within total 
estimated errors. 

\begin{figure}
\begin{center}
\includegraphics*[width=13cm]{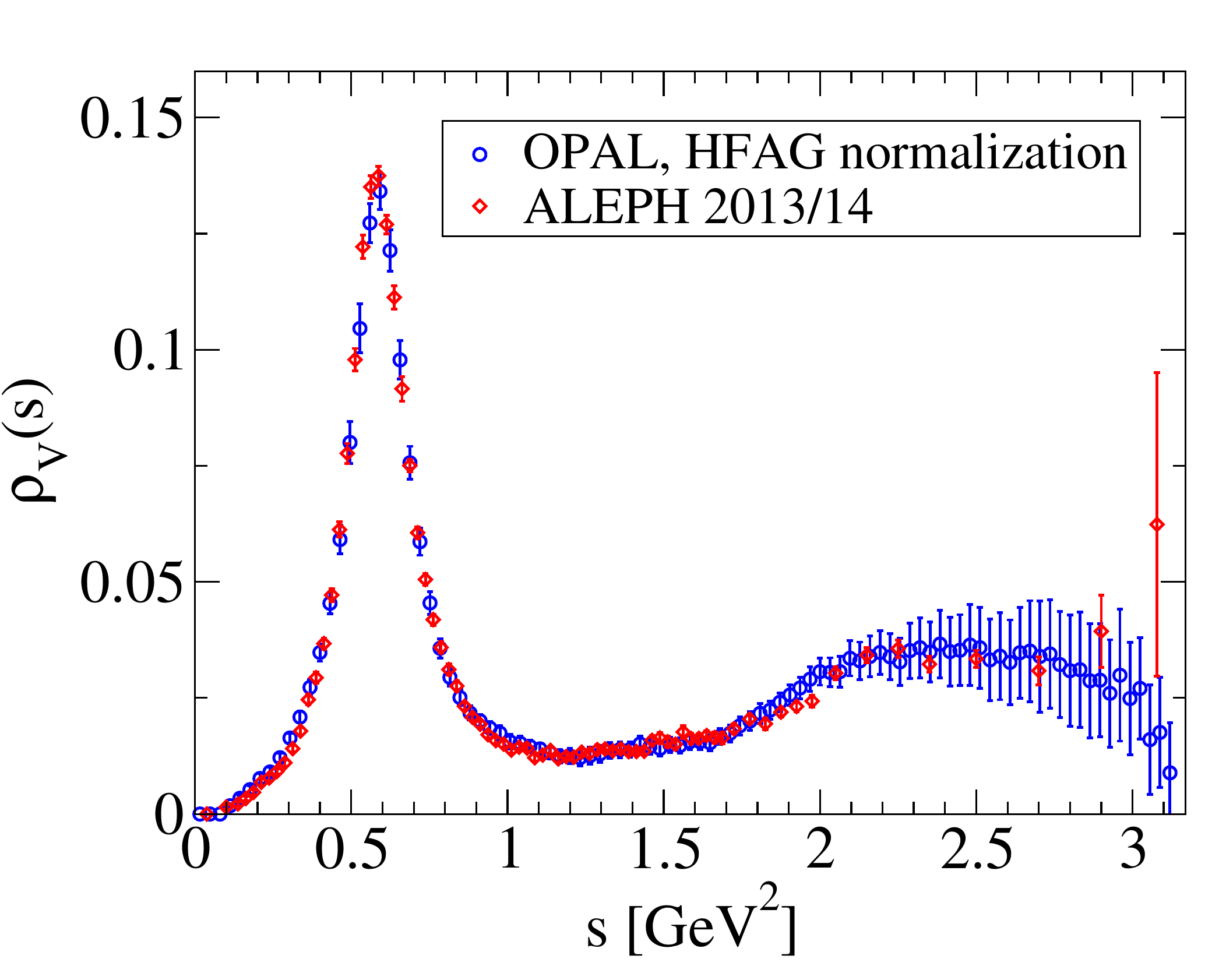}
\vspace{0.0cm}
\includegraphics*[width=13cm]{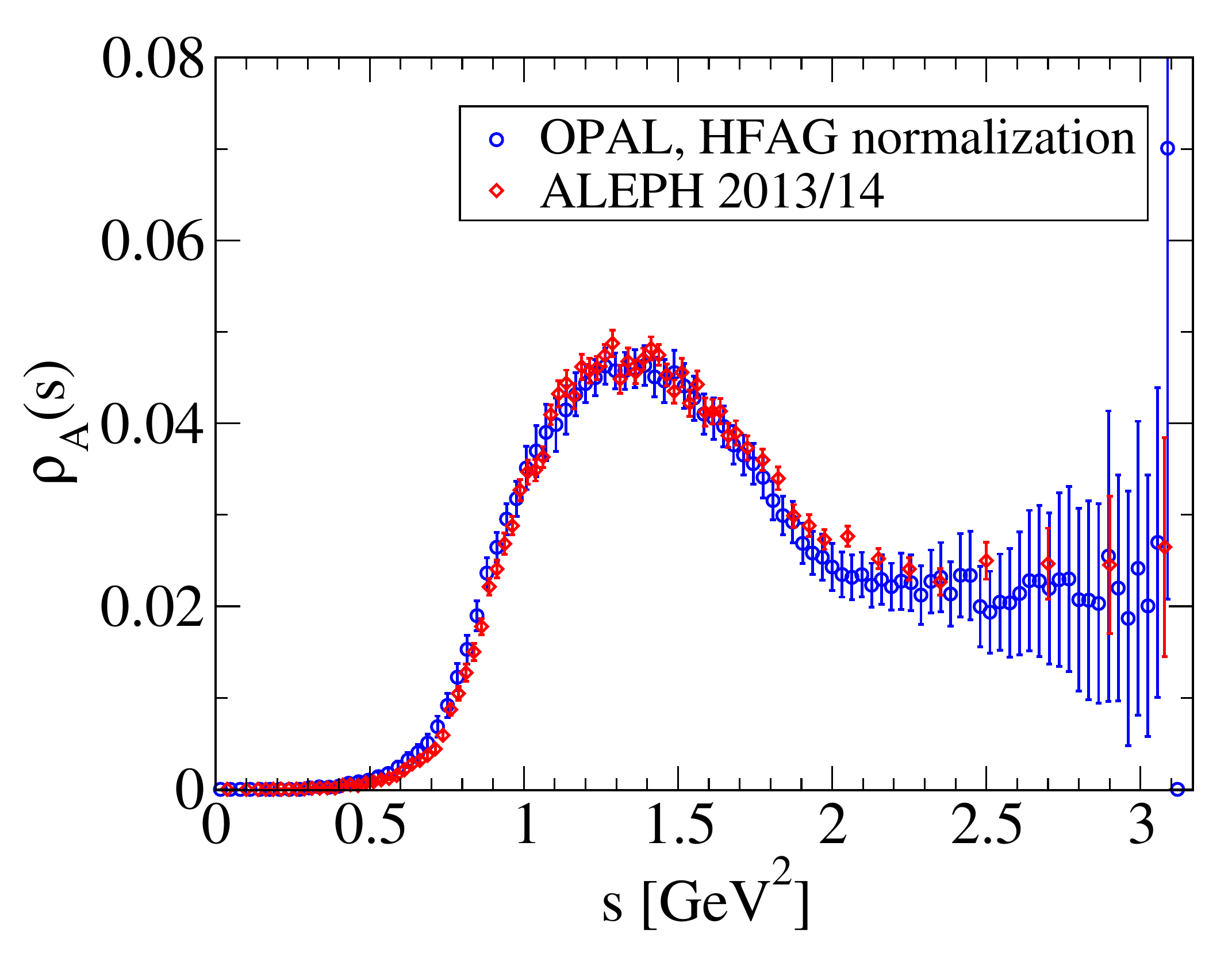}
\end{center}
\begin{quotation}
\floatcaption{ALEPH-OPAL}%
{{\it Comparison of ALEPH and OPAL data for the spectral functions. Top panel:
$I=1$ vector channel; bottom panel: $I=1$ continuum (pion-pole subtracted)
axial channel.}}
\end{quotation}
\vspace*{-4ex}
\end{figure}

\section{\label{strategy} Fitting strategy}
As already explained in Sec.~\ref{theory}, and in more detail in
Refs.~\cite{alphas1,alphas2}, non-pinched weights are needed in order to get
a handle on the DV parameters of Eq.~(\ref{ansatz}). The simplest and most robust
choice of weight allowing us to extract these parameters is the weight
$\hw_0(x)=1$. In order to check the stability of these simple fits, we also
perform simultaneous fits of the weights $\hw_0$ and $\hw_2$, and of
$\hw_0$, $\hw_2$ and $\hw_3$, as in Ref.~\cite{alphas1,alphas2}. This gives us
access to the $D=6$ and $D=8$ terms in the OPE, but also allows us to test
for the consistency of the values
of $\a_s(m_\tau^2)$ and the DV parameters between our different fits.

The values we obtain for $I^{(w)}_{\rm ex}(s_0)$ from the ALEPH data are
highly correlated, both between different values of $s_0$, and between
different weight functions. If we consider only fits using
$I^{(\hw_0)}_{\rm ex}(s_0)$ for a range of $s_0$ values, it turns out that
fully correlated $\chi^2$ fits are possible, but if we also include
$I^{(\hw_2)}_{\rm ex}(s_0)$ and
$I^{(\hw_3)}_{\rm ex}(s_0)$ in the fits, the complete correlation matrices
become too singular. For fits with multiple weights, we will follow
Refs.~\cite{alphas1,alphas2}, using instead the block-diagonal ``fit quality''
\begin{equation}
\label{blockcorr}
\cq^2=\sum_w\sum_{s_0^i,\, s_0^j}
\left(I_{\rm ex}^{(w)}(s_0^i)-I_{\rm th}^{(w)}(s_0^i;{\vec p})\right)
\left(C^{(w)}\right)^{-1}_{ij}
\left(I_{\rm ex}^{(w)}(s_0^j)-I_{\rm th}^{(w)}(s_0^j;{\vec p})\right)\ ,
\end{equation}
where we have made the dependence of $I_{\rm th}^{(w)}$ on the fit parameters
${\vec p}$ explicit. The matrix $C^{(w)}$ is the (block-diagonal) covariance
matrix of the set of moments with fixed weight $w$ and $s_0$ running over
the chosen fit window range. The sums over $s_0^i$ and
$s_0^j$ are over bins $i$ and $j$, and the sum over $w$ is over $\hw_0$
and $\hw_2$, or over $\hw_0$, $\hw_2$ and $\hw_3$.\footnote{If only one
weight is included in the sum, $\cq^2$ reverts to the standard $\chi^2$.}
The motivation for this choice
is that the cross-correlations between two moments arise mainly
because the weight functions used in multiple-moment fits appear to be
 close to being linearly dependent in practice (even though, as
a set of polynomials, of course they are not).
This near-linear dependence is possibly caused by the relatively large
errors on the data for values of $s$ toward $m_\tau^2$, because it is
primarily in this region
that the weights $\hw_0$, $\hw_2$ and $\hw_3$ differ from each other.

An important observation is that we can freely choose our fit quality
$\cq^2$, as long as errors are propagated taking the full data
correlation matrix into account. In our case, we choose to estimate
fit errors for fits using Eq.~(\ref{blockcorr}) by propagating the
data covariance matrix through a small fluctuation analysis; for details
on how this is done, we refer to the appendix of Ref.~\cite{alphas1}.
We note that the fit quality $\cq^2$ does not follow a standard $\chi^2$
distribution, so that no absolute meaning can be attached to the minimum
value obtained in a fit of this type.

The theoretical moments $I_{\rm th}^{(w)}(s_0;{\vec p})$ are non-linear
functions of (some of) the fit parameters ${\vec p}$, and it is thus
not obvious what the probability distribution of the model parameters
looks like. As in Ref.~\cite{alphas2}, we will therefore also explore the
posterior probability distribution of the model parameters, assuming
that the input data follow a multivariate gaussian distribution.
In order to map out this probability distribution, we use the same
Markov-chain Monte Carlo code {\tt hrothgar} \cite{hrothgar} as was used
in Ref.~\cite{alphas2}, to which we refer for more details.
The distribution generated by {\tt hrothgar} is proportional to
$\mbox{exp}[-\cq^2({\vec p})/2]$ on the space of parameters, given the data.

\section{\label{fits} Fits}
In this section we present our fits, leaving the discussion of $\a_s$ and
other parameters obtained from these fits to Sec.~\ref{results}. We first present fits
to moments constructed from the $V$ spectral function only, followed by
fits using both the $V$ and $A$ spectral moments.
We have considered $\chi^2$ fits to $I^{(\hw_0)}_{\rm ex}$ and combined fits
using fit qualities of the form~(\ref{blockcorr}) to $I^{(\hw_0)}_{\rm ex}$,
$I^{(\hw_2)}_{\rm ex}$, and $I^{(\hw_3)}_{\rm exp}$. Below we will show only
the $\chi^2$ fits to $I^{(\hw_0)}_{\rm ex}$ and the $\cq^2$ fits to all three
moments. The results from $\cq^2$ fits to the two moments
$I^{(\hw_0)}_{\rm ex}$ and $I^{(\hw_2)}_{\rm ex}$ are completely consistent with these,
and we therefore omit them below in the interest of brevity.

As reviewed above, and discussed in much more detail in
Refs.~\cite{alphas1,alphas2}, the necessity to fit not only OPE parameters,
but also DV parameters,
makes it impossible to fit spectral moments for the sum of the $V$ and $A$
spectral functions. Already for $I^{(\hw_0)}_{\rm ex}$ this would entail a
9-parameter fit, and  with the existing data such fits turn out
to be unstable. Reference~\cite{ALEPH13} did perform fits to moments of
the $V+A$ spectral function at the price of neglecting duality violations
and contributions from $D>8$ terms in the OPE;
we will compare our fits with those of
Ref.~\cite{ALEPH13} in detail in Sec.~\ref{ALEPH} below.

{}From Fig.~\ref{ALEPH-OPAL}, we see that the only ``feature'' in the
$A$ channel is the peak corresponding to the $a_1$ meson. In contrast,
the $V$ channel data indicate the existence of more resonance-like
features than just the $\r$ meson peak around $s=0.6$~GeV$^2$, even though
the resolution is not good enough to resolve multiple resonances beyond
the $\r$. If we wish to avoid making the assumption that already the
lowest peak in each channel is in the asymptotic regime in which the
\ansatz~(\ref{ansatz}) is valid, we should limit ourselves to fits to the
$V$ channel only. However, we will present also fits to the
combined $V$ and $A$ channels below, and see that the results are consistent
with those from fits to only the $V$ channel.

In all cases, we find it necessary to include the moment
$\hw_0$ in our fits in order to determine both $\a_s(m_\tau^2)$ and
the DV parameters. While one might consider fits to the
spectral function itself, such fits are found to be insufficiently
sensitive to the parameter $\a_s(m_\tau^2)$, and hence have not been
pursued.\footnote{It is important to distinguish fits
to the spectral function itself from fits to {\it the moments of}
the spectral function; they are quite different.  Even in the case of the
$\hat{w}_0$ moment, the integral $I^{(\hw_0)}_{V;\rm ex}(s_0)$ contains all the data from threshold to $s_0$ and always includes, in particular, the $\rho$ peak. On the other hand, a fit of the DV \ansatz~(\ref{ansatz}) to the vector spectrum would probably only include data for $s_0$  between $s_{\rm min}$ and $m_{\tau}^2$, and, since one needs to choose $s_{\rm min}\gg m_\r^2$,  the $\rho$ peak  is clearly excluded. The change in $I^{(\hw_0)}_{V/A;\rm ex}(s_0)$
as $s_0$ is increased from the upper edge of bin $k$ to the upper
edge of bin $k+1$ is, of course, equal to the average value of the
relevant spectral function, $\rho_{V/A}$, in bin $k+1$. As such,
in fits which employ all possible $s_0\ge s_{\rm min}$,
the fact that the $s_0$-dependence of $I^{(\hw_0)}_{V/A;\rm ex}(s_0)$
is one of the key elements entering the fit means that
spectral function values in the interval $s_{\rm min}\le s\le m_\tau^2$
are part of the input, but clearly 
not the {\it only} input. \\
Let us be even more specific.
First, as already noted, even for single-weight $\hat{w}_0$ fits,
the integral of the experimental spectral function over the region from
threshold to $s_{\rm min}$  enters the $\hat{w}_0$ moment for all $s_0$.   
While this is a region in which the OPE and the DV \ansatz\ are not
valid, this additional input turns out to be crucial; fits for both
$\alpha_s(m_\tau^2)$ and the DV parameters are not possible without
including it.  Second, as seen in our previous analysis employing
the OPAL data, fit results are not changed if, rather than using integrated data for
all available $s_0>s_{\rm min}$, one instead employs a winnowed set
thereof in the analysis. For such a winnowed set, it is only the sums
of the experimental spectral function values over the bins lying between
adjacent winnowed $s_0$, and not the full set of spectral function
values in all bins in those intervals, that determine the $s_0$ variation
entering the fit. Finally, all of the multi-weight fits we employ
involve weights, $w(x=s/s_0)$, which are themselves $s_0$ dependent. 
This means that the $s_0$ dependence
of the DV part of the corresponding theory moments results not just
from the values of $\rho^{\rm DV}(s)$ in the inteval $s_0\le s\le m_\tau^2$
(where experimental constraints exist), but also involve
$s_0$- and $w(x)$-dependent weighted integrals of the DV \ansatz\ form
in the interval from $m_\tau^2$ to $\infty$. It would thus be
incorrect to characterize the moment-based fit analysis we
employ as in any way representing simply
a fit to the experimental
spectral functions.} 
Fits involving only pinched moments such as $\hw_2$ and $\hw_3$, on the
other hand, are insufficiently sensitive to the DV parameters.
All our fits will thus include the spectral moments
$I^{(\hw_0)}_{\rm ex}(s_0)$, either in the $V$ channel alone, or in the
combined $V$ and $A$ channels. In the latter case, there is a separate
set of DV parameters for each of these channels, but the fit parameter
$\a_s(m_\tau^2)$ is, of course, common to both.\footnote{The
$D>2$ OPE coefficients are also generally different between the $V$ and
$A$ channels \cite{BNP}. In the case of $C_4$ (which, due to the
absence of terms linear in $x$, does not enter for the weights we
employ, in the approximation of dropping contributions higher-than-leading order
in $\alpha_s$) the full gluon condensate and leading-order quark
condensate contributions are the same for the $V$ and $A$ channels.
For polynomial weights with a term linear in $x$,
$D=4$ contributions would be present, and one could
impose the resulting near-equality of $C_4$ in the $V$ and $A$
channels. This was done in the version of the analysis
performed by OPAL but not in the analyses of the
ALEPH collaboration, including Ref.~\cite{ALEPH13}. The fact that
the fitted value of the gluon condensate obtained from independent
$V$ and $A$ channel fits in Ref.~\cite{ALEPH13} are not close
to agreeing within errors is, in fact, a clear
sign of the unphysical nature of these fits, see Sec.~\ref{ALEPH} below.}

\subsection{\label{V} Fits to vector channel data}
We begin with fits to the single moment
$I^{(\hw_0)}_{\rm ex}(s_0)$, as a function of $s_{\rm min}$, with
$s_{\rm min}$ defined to be the minimum value of $s_0$ included in the fit.
\begin{table}[t]
\begin{center}
\begin{tabular}{|c|c|c|c|c|c|c|c|c|c|}
\hline
$s_{\rm min}$ (GeV$^2$) & $\chi^2$/dof & $p$-value (\%) & $\alpha_s$ & $\delta_V$ & $\gamma_V$ & $\alpha_V$ & $\beta_V$ \\
\hline
1.425 & 33.0/21 & 5 & 0.312(11) & 3.36(36) & 0.66(22) & -0.33(61) & 3.27(33)  \\
1.475 & 29.5/19 & 6 & 0.304(11) & 3.32(41) & 0.70(25) & -1.21(73) & 3.72(39)  \\
1.500 & 29.5/18 & 4 & 0.304(11) & 3.32(41) & 0.70(25) & -1.19(87) & 3.71(45)  \\
1.525 & 29.0/17 & 3 & 0.302(11) & 3.37(43) & 0.68(26) & -1.49(94) & 3.86(48)  \\
1.550 & 24.5/16 & 8 & 0.295(10) & 3.50(50) & 0.62(29) & -2.43(94) & 4.32(48)  \\
1.575 & 23.5/15 & 8 & 0.298(11) & 3.50(47) & 0.62(28) & -2.1(1.0) & 4.15(53)  \\
1.600 & 23.4/14 & 5 & 0.297(12) & 3.50(48) & 0.62(28) & -2.1(1.1) & 4.16(56)  \\
1.625 & 23.4/13 & 4 & 0.298(13) & 3.47(50) & 0.63(28) & -2.0(1.2) & 4.12(62)  \\
1.675 & 23.1/11 & 2 & 0.301(15) & 3.35(60) & 0.68(31) & -1.7(1.4) & 3.96(70)  \\
\hline
\hline
1.425 & 33.2/21 & 4 & 0.331(15) & 3.20(34) & 0.74(21) & -0.30(61) & 3.24(33)  \\
1.475 & 29.5/19 & 6 & 0.320(14) & 3.16(40) & 0.78(24) & -1.20(73) & 3.70(39)  \\
1.500 & 29.5/18 & 4 & 0.320(15) & 3.16(40) & 0.78(24) & -1.19(87) & 3.69(45)  \\
1.525 & 28.9/17 & 4 & 0.317(14) & 3.22(42) & 0.75(25) & -1.51(93) & 3.85(48)  \\
1.550 & 24.3/16 & 8 & 0.308(13) & 3.36(49) & 0.69(28) & -2.48(93) & 4.33(48)  \\
1.575 & 23.3/15 & 8 & 0.311(14) & 3.35(46) & 0.69(27) & -2.2(1.0) & 4.17(52)  \\
1.600 & 23.3/14 & 6 & 0.311(15) & 3.36(47) & 0.69(27) & -2.2(1.1) & 4.19(56)  \\
1.625 & 23.2/13 & 4 & 0.312(16) & 3.33(49) & 0.70(28) & -2.1(1.2) & 4.15(62)  \\
1.675 & 23.0/11 & 2 & 0.314(19) & 3.23(58) & 0.74(30) & -1.8(1.5) & 4.02(74)  \\
\hline
\end{tabular}
\end{center}
\floatcaption{VVw1paper}{\it $V$ channel fits to $I^{(\hw_0)}_{\rm ex}(s_0)$
from $s_0=s_{\rm min}$ to $s_0=m_\tau^2$. FOPT results
are shown above the double line, CIPT below;
no $D>0$ OPE terms included in the fit. 
{$\gamma_V$ and $\beta_V$ in units of {\rm GeV}$^{-2}$.}}
\end{table}%
Since these are $\chi^2$ fits,
one may estimate the $p$-values for these fits; they are shown in the
third column of Tab. \ref{VVw1paper}.
We note that the $p$-values are not large, but they are not small enough
to exclude the validity of our fit function based on the ALEPH data.
Judged by $p$-value, the fits with $s_{\rm min}=1.55$ and $1.575$~GeV$^2$ are the
best fits, and we thus take the average value of the central values for the
fit parameters from these two fits as our best value, with a statistical
error that is
the larger of the two (noting that these are essentially equal in size).
For the strong coupling, we find
\begin{eqnarray}
\label{ashw0}
\a_s(m_\tau^2)&=&0.296(11)\ ,\qquad\mbox{(FOPT)}\ ,\\
&=&0.310(14)\ ,\qquad\mbox{(CIPT)}\ .\nonumber
\end{eqnarray}
The difference between the FOPT and CIPT
results reflects the well-known fact that the two
prescriptions show no sign of converging to one another as
the truncation order is increased~\cite{PT,BJ}.
We observe that the $p$-value starts to decrease again from
$s_{\rm min}=1.6$~GeV$^2$, indicating that the data become too sparse
for an optimal fit. We investigated the sensitivity of these fits to
omitting the data in up to four bins with the largest values of $s$,
and found no significant difference. This is no surprise, given the
errors shown in Fig.~\ref{CIFOw0fit}.
For illustration, we show the parameter correlation
matrix for the FOPT fit with $s_{\rm min}=1.55$~GeV$^2$ in Tab.~\ref{corr}.
\begin{table}[t]
\begin{center}
\begin{tabular}{|c|ccccc|}
\hline
 & $\a_s$ & $\d_V$ & $\g_V$ & $\a_V$ & $\b_V$ \\
\hline
$\a_s$ & 1 & 0.600 & -0.606 & 0.689 & -0.653 \\
$\d_V$ & 0.600 & 1 & -0.994 & 0.310 & -0.297 \\
$\g_V$ & -0.606 & -0.994 & 1 & -0.330 & 0.315 \\
$\a_V$ & 0.689 & 0.310 & -0.330 & 1 & -0.996 \\
$\b_V$ & -0.653 & -0.297 & 0.315 & -0.996 & 1 \\
\hline
\end{tabular}
\end{center}
\begin{quotation}
\floatcaption{corr}{{\it Parameter correlation matrix for the $V$
channel $\hat{w}_0$ FOPT fit
with $s_{\rm min}=1.55$~{\rm GeV}$^2$ shown in
Tab.~\ref{VVw1paper}.}}
\end{quotation}
\vspace*{-4ex}
\end{table}%

\begin{figure}[t]
\begin{center}
\includegraphics*[width=7cm]{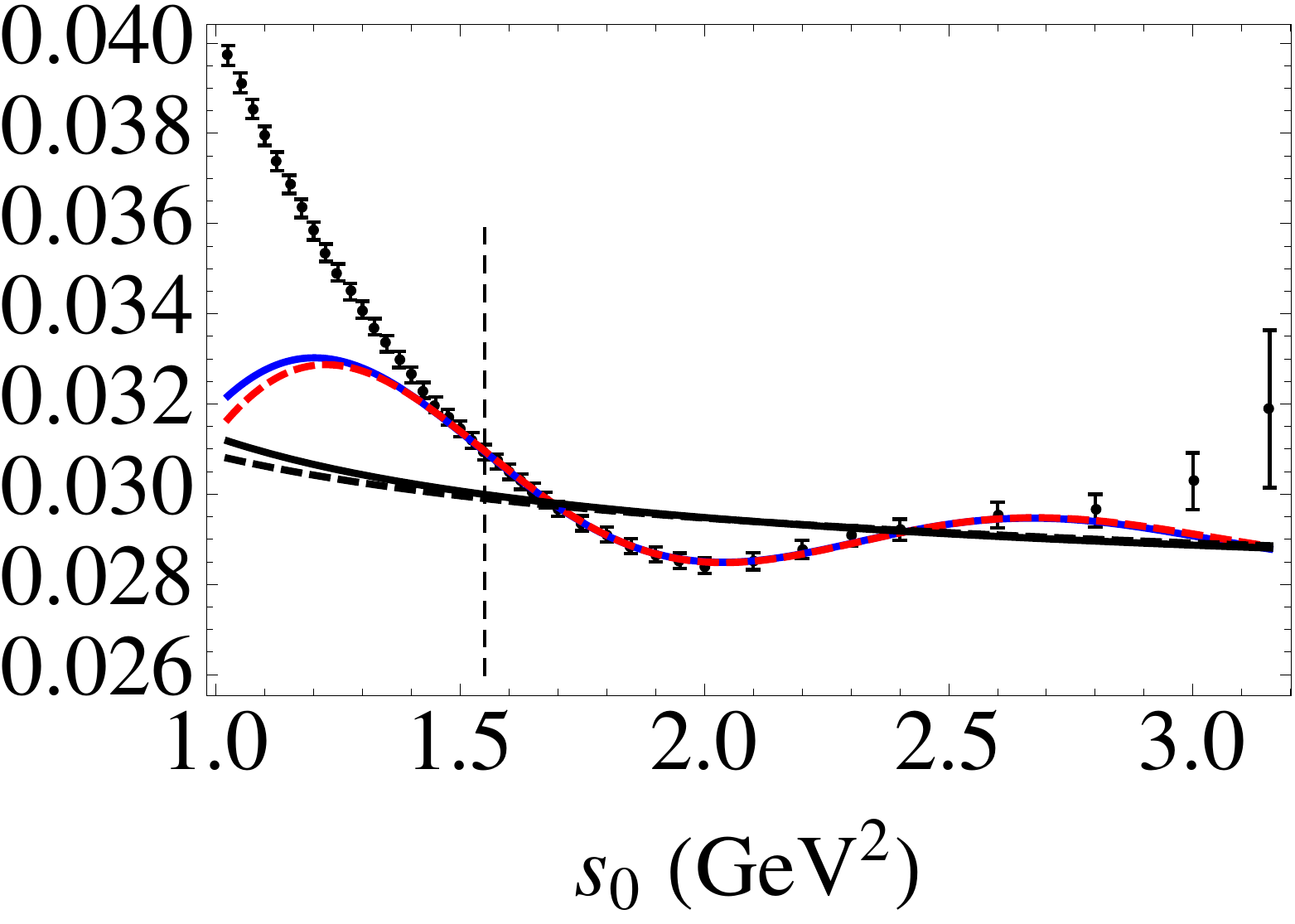}
\hspace{0.5cm}
\includegraphics*[width=7cm]{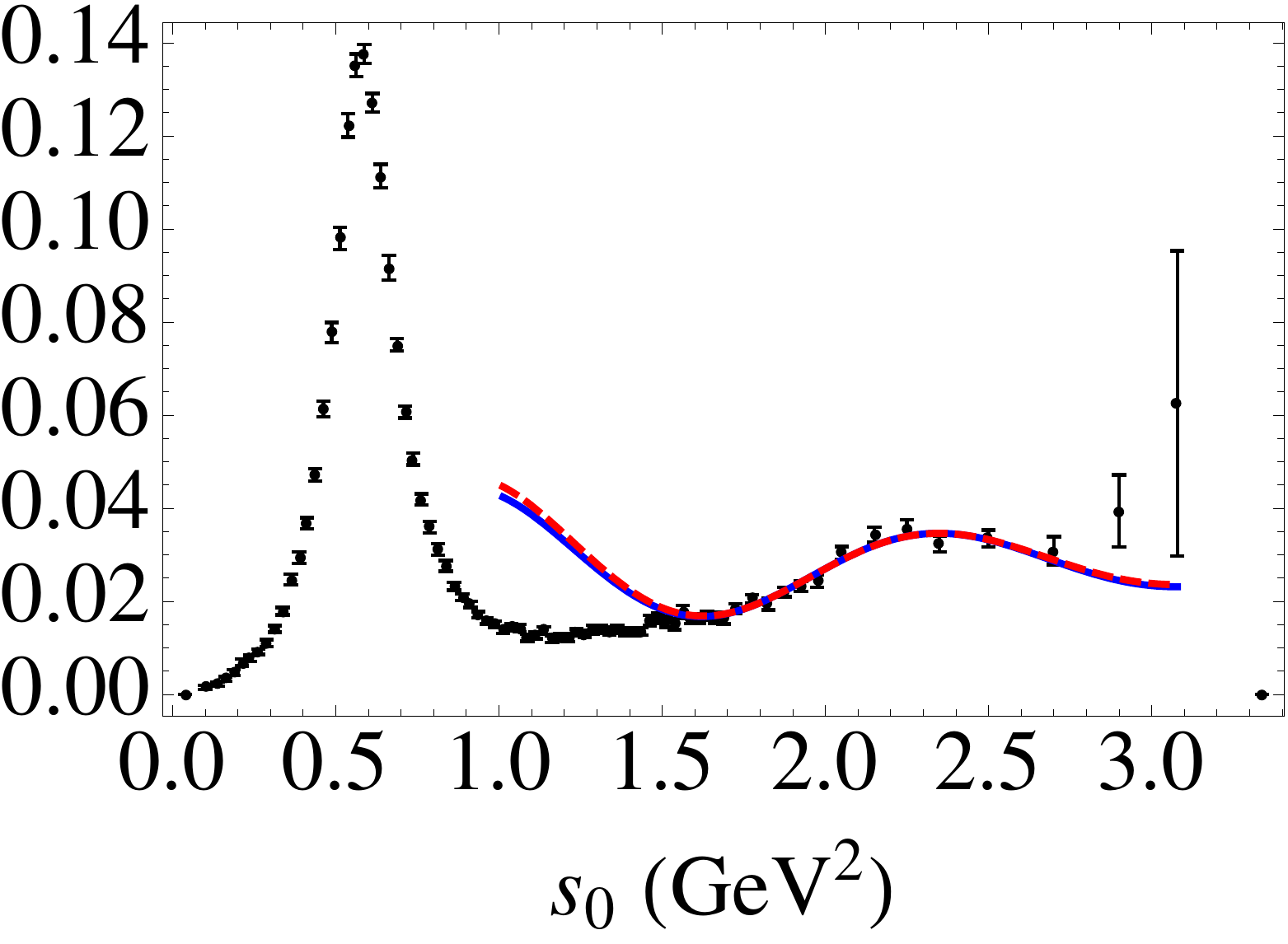}
\end{center}
\begin{quotation}
\floatcaption{CIFOw0fit}{{\it Left panel:
comparison of $I^{(\hw_0)}_{\rm ex}(s_0)$ and $I^{(\hw_0)}_{\rm th}(s_0)$
for the $s_{\rm min}=1.55\ {\rm GeV}^2$ $V$ channel fits of
Tab.~\ref{VVw1paper}. Right panel: comparison of the theoretical
spectral function resulting from this fit with the experimental results.
CIPT fits are shown in red (dashed) and FOPT in blue (solid).
The (much flatter) black curves in the left panel represent the OPE parts of the
fits, \ie, the fit results with the DV parts
removed.
The vertical dashed line indicates the location of $s_{\rm min}$.
}}
\end{quotation}
\vspace*{-4ex}
\end{figure}
In Fig.~\ref{CIFOw0fit} we show the results of CIPT and FOPT fits to
$I^{(\hw_0)}_{\rm ex}(s_0)$ for $s_{\rm min}=1.55$~GeV$^2$. The left panel shows
the results of the fits for the moment,
the right-hand panel the OPE+DV versions of the spectral functions
resulting from these fits. 

\begin{figure}[t]
\begin{center}
\includegraphics*[width=10cm]{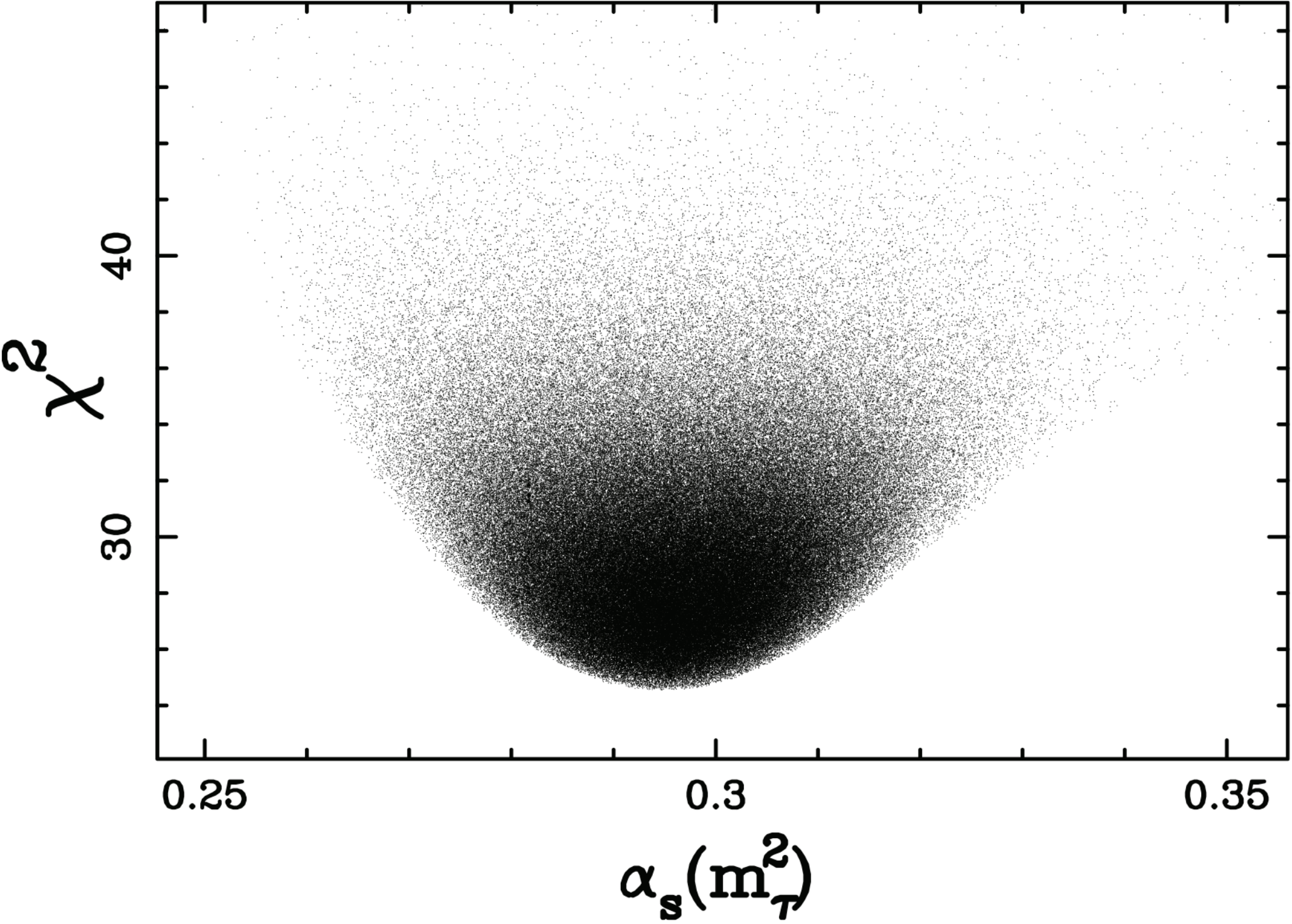}
\end{center}
\begin{quotation}
\floatcaption{chi2}{{\it $\chi^2$ versus $\a_s(m_\tau^2)$, FOPT,
$s_{\rm min}=1.55$~{\rm GeV}$^2$, 1250000 points.}}
\end{quotation}
\vspace*{-4ex}
\end{figure}
\begin{figure}[h!]
\vspace*{4ex}
\begin{center}
\includegraphics*[width=7cm]{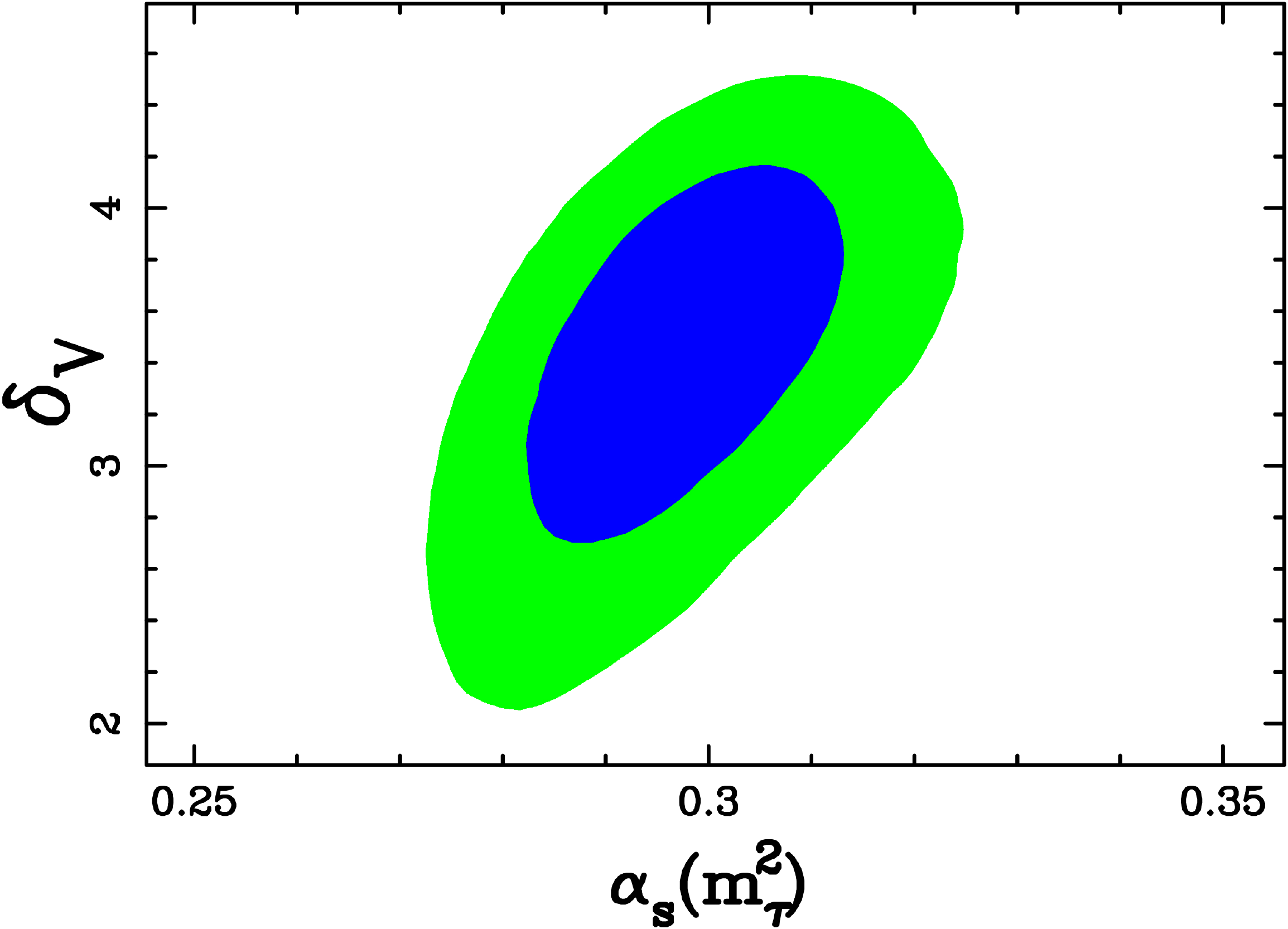}
\hspace{0.5cm}
\includegraphics*[width=7cm]{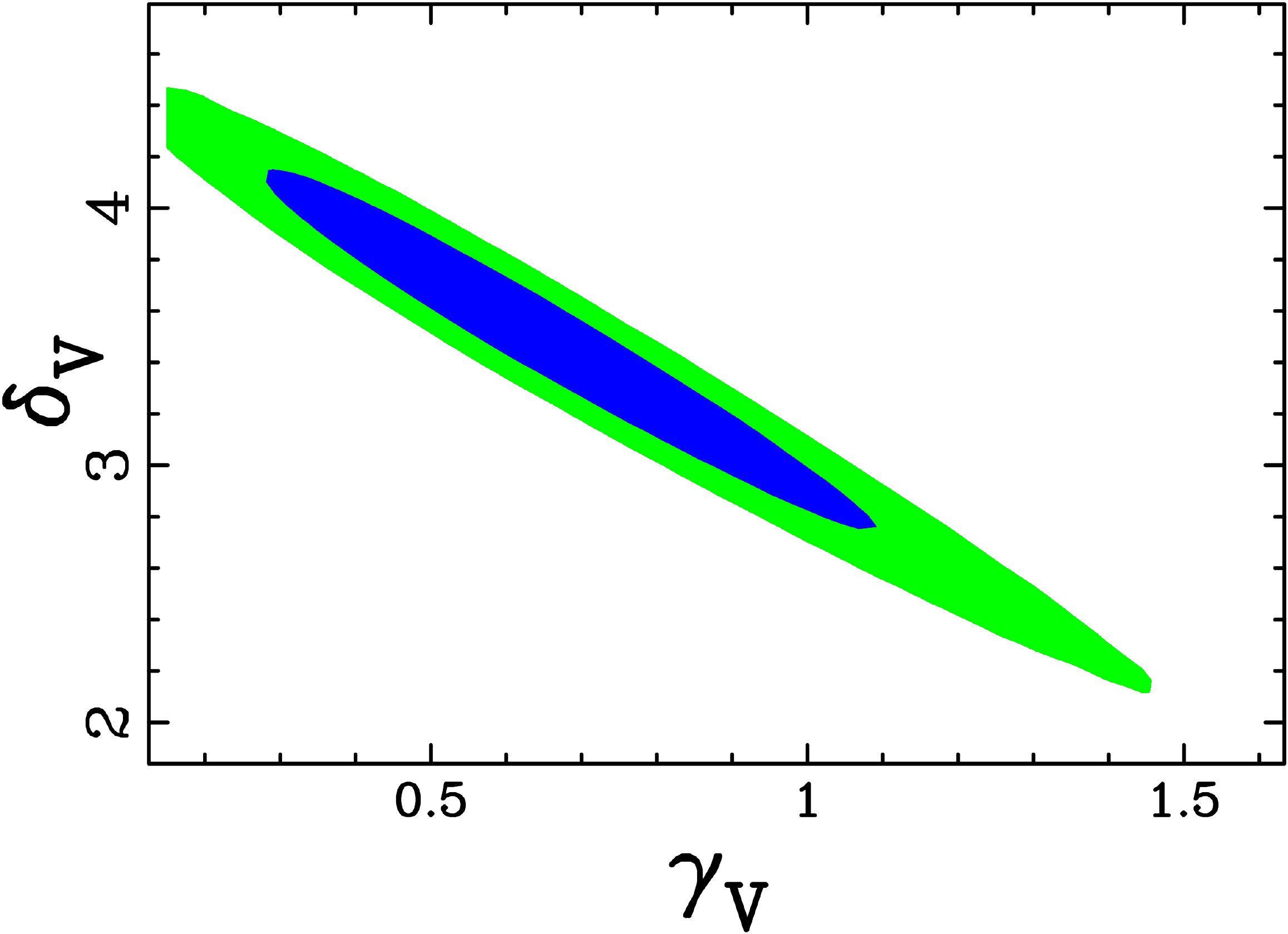}
\end{center}
\begin{quotation}
\floatcaption{contour}{{\it Two-dimensional contour plots showing $\d_V$ versus
$\a_s(m_\tau^2)$ (left) and $\d_V$ versus $\g_V$ (right)
for the $V$ channel $\hat{w}_0$ FOPT, $s_{\rm min}=1.55$~{\rm GeV}$^2$
fit. Blue (darker) areas and green (lighter) areas contain 68\%,
respectively, 95\% of the distribution.
$\gamma_V$ in units of {\rm GeV}$^{-2}$}}
\end{quotation}
\vspace*{-4ex}
\end{figure}
As in Ref.~\cite{alphas2}, we studied the posterior probability distribution,
using the same Markov-chain Monte~Carlo code, {\tt hrothgar} \cite{hrothgar}.
We remind the reader that it is not obvious what this distribution looks
like, even if we assume that the data errors follow a multivariate
gaussian distribution. For the fits of Tab.~\ref{VVw1paper},
this code generates points in the 5-dimensional parameter space, and
computes the $\chi^2$ value associated with each of these points.
These points are distributed as ${\rm exp}[-\chi^2({\vec p})]$,
with $\vec p$ the parameter vector, and $\chi^2$ evaluated on the
ALEPH data (including the full covariance matrix) and the values of
the parameters at these points.

In Fig.~\ref{chi2} we show $\chi^2$ as a function of $\a_s(m_\tau^2)$,
choosing the FOPT fit with $s_{\rm min}=1.55$~GeV$^2$. Since for each
$\a_s(m_\tau^2)$ points with many different values for the other four
parameters are generated stochastically, the distribution appears as
the cloud shown in the figure. This distribution shows a unique minimum
for the value of $\chi^2$, at approximately $\a_s(m_\tau^2)=0.295$, consistent
with Tab.~\ref{VVw1paper}. The width of the distribution is also roughly
consistent with the error of $\pm 0.010$, but we see that
the distribution of points is not entirely symmetric around the
minimum. There is no alternative (local) minimum, as was the case
with the OPAL data \cite{alphas2}.

We also find the parameters
$\d_V$ and $\gamma_V$ to be much better constrained than was
the case for the corresponding fits to the OPAL data in
Ref.~\cite{alphas2}. The distributions in the $\delta_V$--$\alpha_s(m_\tau^2)$
and $\delta_V$--$\gamma_V$ planes are shown in the left and right
panels of Fig.~\ref{contour}.\footnote{Note that the vertical
axis covers the interval $\d_V\in[2,5]$, to be compared with the
significantly larger interval $\d_V\in[-2,5]$ in Fig.~2 of
Ref.~\cite{alphas2}.}
Since for all other fits presented in the rest of this article the
conclusions about the posterior probability distribution found with
{\tt hrothgar} are similar, we will refrain from showing the analogues of
Figs.~\ref{chi2} and~\ref{contour}
for those fits. 

\begin{table}[t!]
\hspace{0cm}\begin{tabular}{|c|c|c|c|c|c|c|c|c|c|}
\hline
$s_{\rm min}$ (GeV$^2$) & $\cq^2$/dof & $\alpha_s$ & $\delta_{V}$ & $\gamma_{V}$ & $\alpha_{V}$ & $\beta_{V}$ & $10^2C_{6V}$ & $10^2C_{8V}$\\
\hline
1.425 & 106.0/71=1.49  & 0.305(10) & 3.02(38) & 0.87(24) & -0.68(56) & 3.43(31) & -0.59(17) & 0.94(29) \\
1.475 & 93.3/65=1.43  & 0.302(10) & 3.07(44) & 0.85(27) & -1.41(68) & 3.81(36) & -0.71(16) & 1.19(28) \\
1.500 & 93.2/62=1.50  & 0.302(10) & 3.08(45) & 0.85(27) & -1.40(77) & 3.80(40) & -0.71(18) & 1.19(30)\\
1.525 & 85.6/59=1.45  & 0.298(10) & 3.21(49) & 0.78(29) & -1.96(78) & 4.08(41) & -0.79(16) & 1.36(27) \\
1.550 & 76.3/56=1.36  & 0.295(10) & 3.30(52) & 0.74(30) & -2.48(81) & 4.33(41) & -0.86(14) & 1.50(24) \\
1.575 & 74.5/53=1.41  & 0.297(10) & 3.29(51) & 0.74(29) & -2.25(87) & 4.22(44) & -0.83(16) & 1.43(27) \\
1.600 & 74.2/50=1.48  & 0.297(11) & 3.31(51) & 0.73(30) & -2.27(92) & 4.23(47) & -0.83(16) & 1.44(29) \\
1.625 & 73.8/47=1.57  & 0.298(11) & 3.28(54) & 0.74(31) & -2.16(99) & 4.18(50) & -0.81(18) & 1.40(32) \\
1.675 & 72.0/41=1.76  & 0.299(12) & 3.28(63) & 0.74(34) & -2.1(1.1) & 4.13(57) & -0.80(21) & 1.37(39) \\
\hline
\hline
1.425 & 98.6/71=1.39 & 0.328(16) & 3.17(39) & 0.77(25) & -0.43(61) & 3.30(32) & -0.60(19) & 0.83(35)\\
1.475 & 89.5/65=1.38 & 0.319(14) & 3.11(44) & 0.81(27) & -1.24(71) & 3.72(37) & -0.76(16) & 1.18(31)\\
1.500 & 89.4/62=1.44 & 0.319(15) & 3.11(44) & 0.81(27) & -1.20(81) & 3.70(42) & -0.76(18) & 1.16(34)\\
1.525 & 82.1/59=1.39 & 0.314(14) & 3.22(48) & 0.77(28) & -1.81(80) & 4.00(42) & -0.85(15) & 1.37(28)\\
1.550 & 73.7/56=1.32 & 0.309(13) & 3.28(51) & 0.74(30) & -2.39(82) & 4.28(42) & -0.93(13) & 1.53(25)\\
1.575 & 71.8/53=1.35 & 0.311(14) & 3.28(50) & 0.74(29) & -2.12(89) & 4.15(45) & -0.89(15) & 1.45(28)\\
1.600 & 71.7/50=1.43 & 0.311(14) & 3.28(51) & 0.74(29) & -2.16(94) & 4.17(48) & -0.90(15) & 1.46(29)\\
1.625 & 71.5/47=1.52 & 0.312(15) & 3.24(53) & 0.75(30) & -2.0(1.0) & 4.11(51) & -0.88(17) & 1.42(34)\\
1.675 & 69.8/41=1.70 & 0.313(16) & 3.22(63) & 0.76(33) & -1.9(1.2) & 4.04(59) & -0.86(20) & 1.38(42)\\
\hline
\end{tabular}
\vspace*{4ex}
\floatcaption{VVwtaupaper}{\it $V$ channel fits to $I^{(\hw_0)}_{\rm ex}(s_0)$,
$I^{(\hw_2)}_{\rm ex}(s_0)$ and $I^{(\hw_0)}_{\rm ex}(s_3)$ from $s_0=s_{\rm min}$ to
$s_0=m_\tau^2$, FOPT results are shown above the double line, CIPT below;
$D=6,\ 8$ OPE terms included in the fit.
$\gamma_V$ and $\beta_V$
in units of {\rm GeV}$^{-2}$, $C_{6V}$ in units of {\rm GeV}$^6$ and $C_{8V}$ in
units of {\rm GeV}$^8$.}
\end{table}%
Next, we consider simultaneous fits to the moments $I^{(\hw_0)}_{\rm ex}(s_0)$,
$I^{(\hw_2)}_{\rm ex}(s_0)$ and $I^{(\hw_0)}_{\rm ex}(s_3)$; results for the same
values of $s_{\rm min}$ as before are given in Tab.~\ref{VVwtaupaper}.
These fits are performed by minimizing $\cq^2$ as defined in
Eq.~(\ref{blockcorr}), with correlations between different moments omitted.
However, the full correlation matrix, including correlations between
different moments, has been taken into account
in the parameter fit error estimates shown in the table. These errors
were determined by linear propagation of the full data covariance matrix;
for a detailed explanation of the method, we refer to the appendix of
Ref.~\cite{alphas1}.

Judging by the values of $\cq^2/$dof,\footnote{Which, given the fact that
$\cq^2$ is not equal to $\chi^2$ for these fits, cannot easily be
translated into $p$-values.} again the two fits for $s_{\rm min}=1.55$
and $1.575$~GeV$^2$ are the optimal ones.
Averaging parameter values between these two fits, we find
\begin{eqnarray}
\label{ashw023}
\a_s(m_\tau^2)&=&0.296(10)\ ,\qquad\mbox{(FOPT)}\ ,\\
&=&0.310(14)\ ,\qquad\mbox{(CIPT)}\ ,\nonumber
\end{eqnarray}
in excellent agreement with Eq.~(\ref{ashw0}). We have also considered fits
involving only the two
moments $I^{(\hw_0)}_{\rm ex}(s_0)$ and $I^{(\hw_2)}_{\rm ex}(s_0)$, and
find results very similar those contained in Tabs.~\ref{VVw1paper}
and \ref{VVwtaupaper}. In Fig.~\ref{CIFOw023fit} we show the quality of the
fits of Tab.~\ref{VVwtaupaper} for $s_{\rm min}=1.55$~GeV$^2$.

\begin{figure}[t!]
\begin{center}
\includegraphics*[width=7cm]{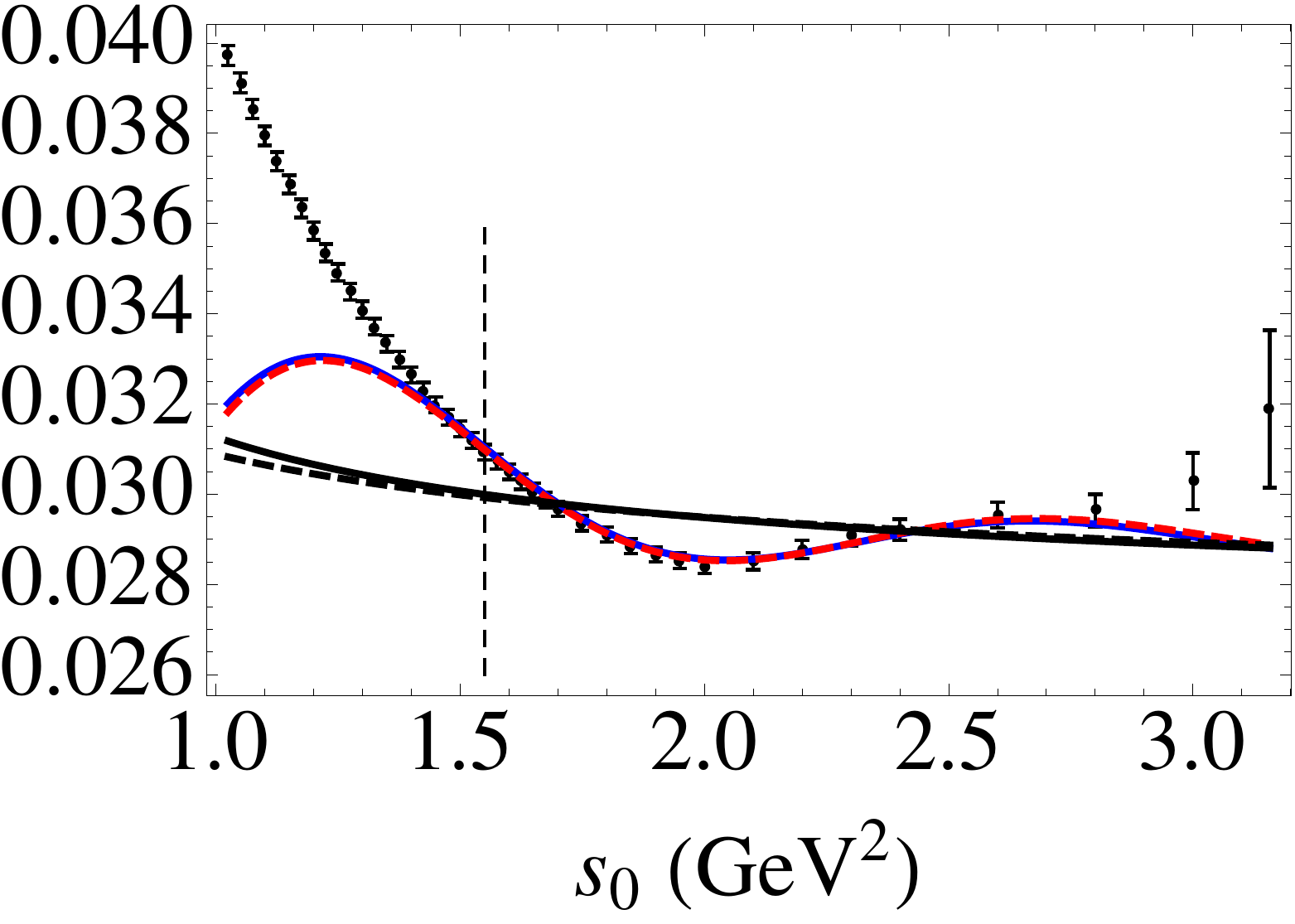}
\hspace{0.5cm}
\includegraphics*[width=7cm]{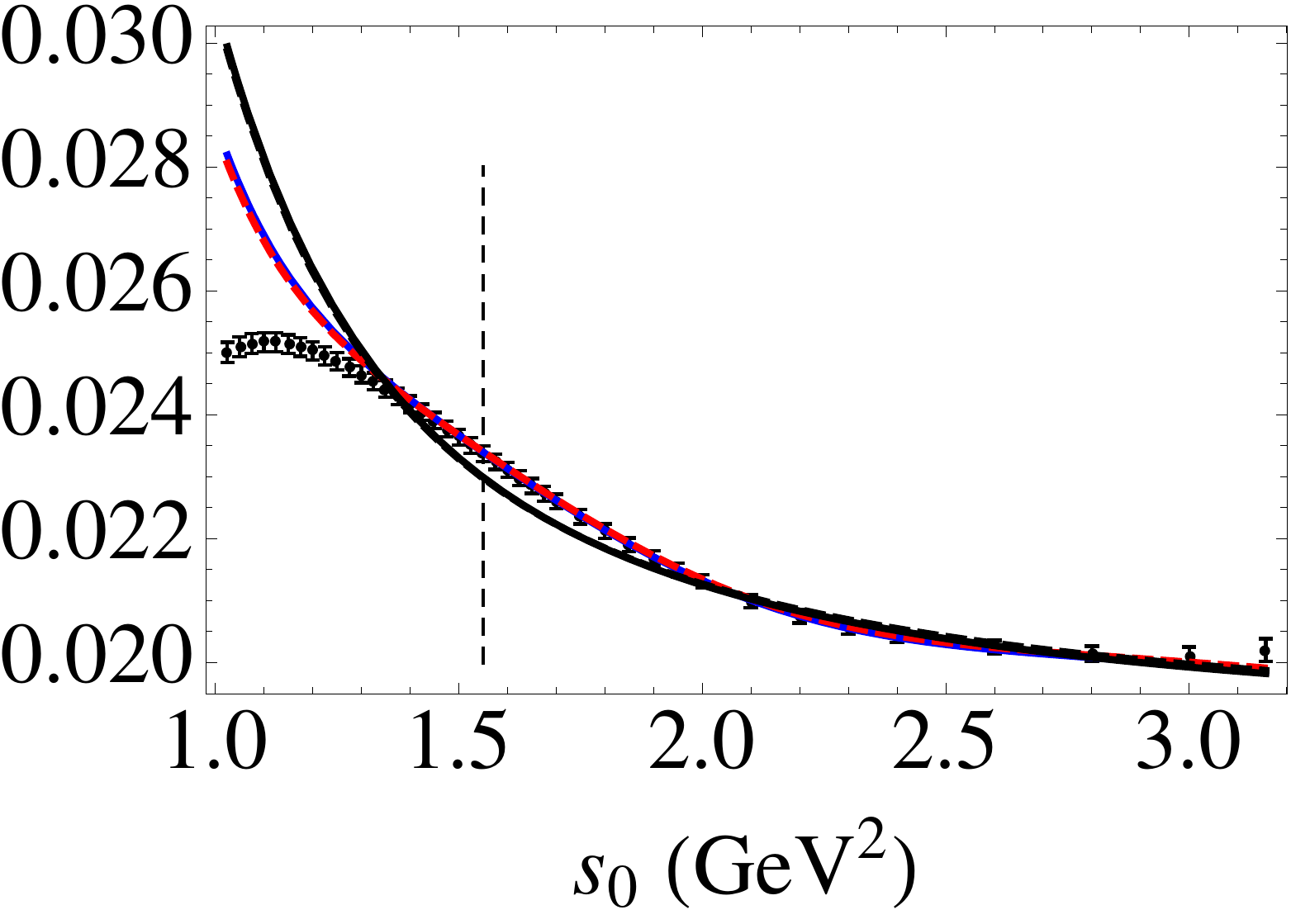}
\vspace{1cm}
\includegraphics*[width=7cm]{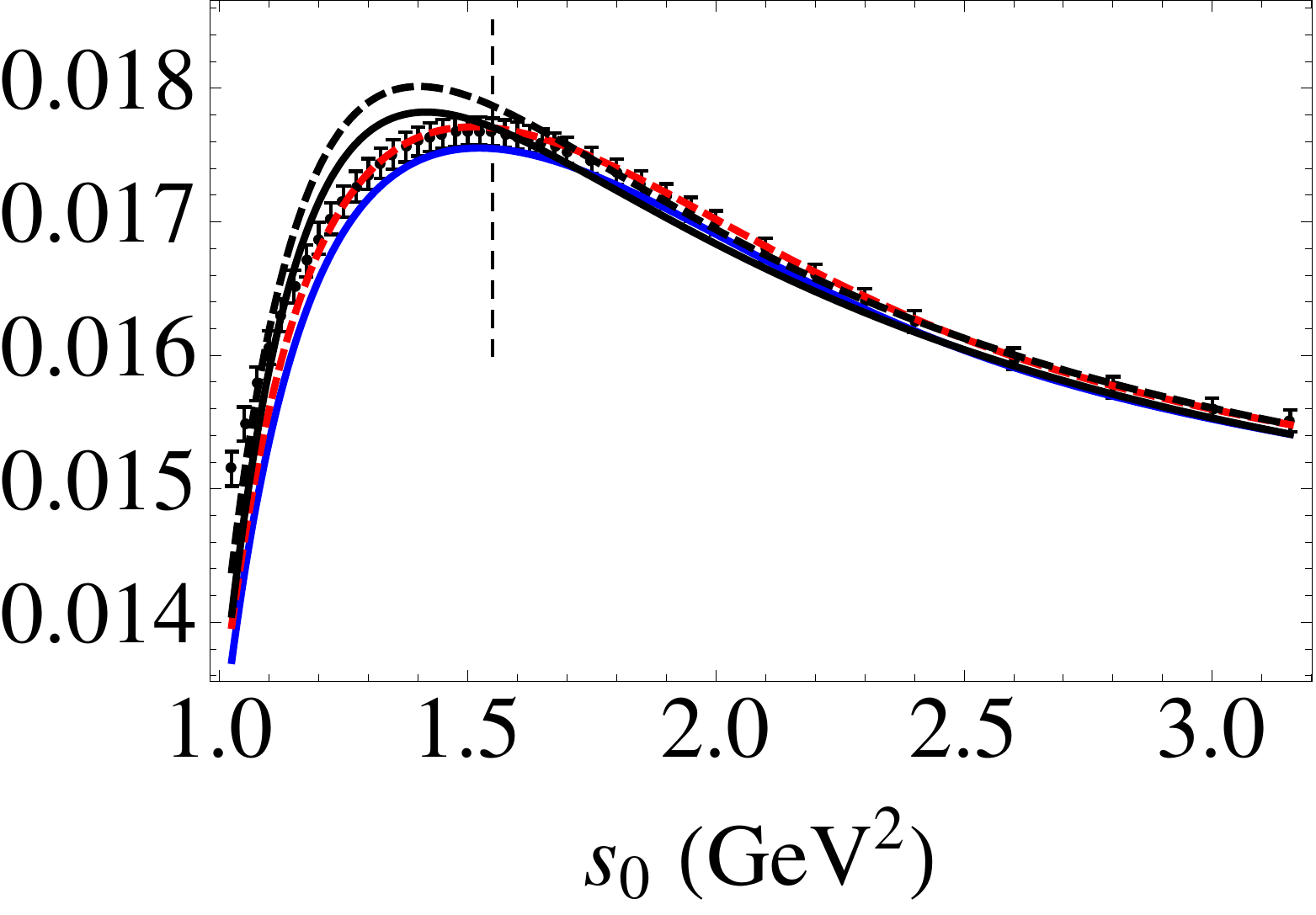}
\hspace{0.5cm}
\includegraphics*[width=7cm]{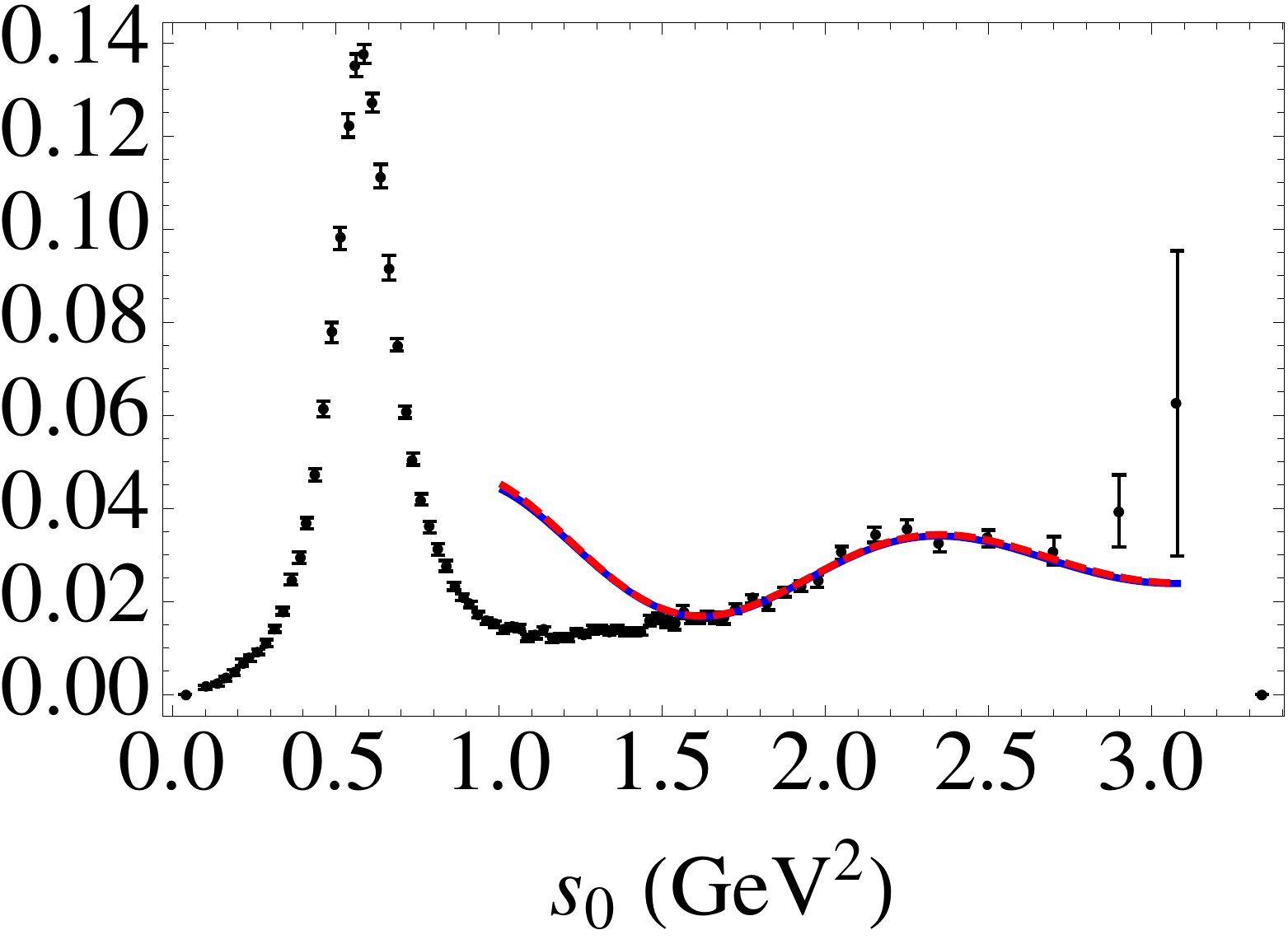}
\end{center}
\vspace*{-6ex}
\begin{quotation}
\floatcaption{CIFOw023fit}{{\it Upper left panel:
comparison of $I^{(\hw_0)}_{\rm ex}(s_0)$ and $I^{(\hw_0)}_{\rm th}(s_0)$
for the $s_{\rm min}=1.55\ {\rm GeV}^2$ $V$ channel fits of
Tab.~\ref{VVwtaupaper}. Lower left panel and upper right panel:
analogous comparisons for $I^{(\hw_2)}(s_0)$ (upper right panel) and
$I^{(\hw_3)}(s_0)$ (lower left panel). CIPT fits are shown in red (dashed) and
FOPT in blue (solid).
Lower right panel: comparison
of the theoretical spectral function resulting
from this fit with the experimental results.
The black curves (which are much
flatter for the $\hat{w}_0$ case)
represent the OPE parts of the fits. The vertical dashed line
indicates the location of $s_{\rm min}$.}}
\end{quotation}
\vspace*{-4ex}
\end{figure}
We end this subsection with several comments. First, we see that pinching
indeed serves to suppress the role of DV
contributions. The upper right panel in Fig.~\ref{CIFOw023fit} shows
the singly pinched $\hat{w}_2$ case and the lower left panel shows the
doubly pinched $\hat{w}_3$ case.
There is also a significant difference between the colored
and black curves in all panels, though with the
onset of this difference shifting to lower $s_0$ as the
degree of pinching is increased. The existence of these differences implies
that, with the errors on the ALEPH data, the presence of duality
violations is evident for all three moments. This, in turn, implies
that omitting duality violations from the theory side of
the corresponding FESRs has the potential to produce a significant
additional systematic error on $\a_s(m_\tau^2)$ (and the higher $D$
OPE coefficients) that cannot be estimated if only fits without
DV parameters are attempted.
We will return to this point in Sec.~\ref{ALEPH} below.
Second, we note that the spectral function itself below $s=s_{\rm min}$ is not very well
described by the curves obtained from the fits.
While the form of Eq.~(\ref{ansatz}) constitutes a reasonable assumption
for asymptotically large $s$, we do not know {\it a priori} what a
reasonable value of $s_{\rm min}$ should be. It is clear, however, that
our \ansatz\ works
reasonably well for $s\,\gtap\, 1.5$~GeV$^2$, but that
the asymptotic regime definitely does not include the region around
the $\r$ peak.

\begin{table}[!t]
\begin{center}
\begin{tabular}{|c|c|c|c|c|c|c|c|c|}
\hline
$s_{\rm min}$ (GeV$^2$) & $\chi^2$/dof & $p$-value (\%) & $\alpha_s$ & $\d_V$ & $\g_V$ & $\a_V$ & $\b_V$ \\
& & & &  $\d_A$ & $\g_A$ & $\a_A$ & $\b_A$ \\
\hline
1.500 & 49.8/37 & 8 & 0.310(14) & 3.45(40) & 0.62(24) & -1.0(1.0) & 3.60(53)  \\
& & & & 1.85(38) & 1.38(20) & 4.5(1.2) & 2.46(59)  \\
1.525 & 48.6/35 & 6 & 0.309(15) & 3.53(42) & 0.59(25) & -1.2(1.2) & 3.71(60)  \\
& & & & 1.99(40) & 1.31(20) & 4.4(1.2) & 2.49(62)  \\
1.550 & 40.0/33 & 19 & 0.297(11) & 3.57(48) & 0.58(28) & -2.33(97) & 4.27(50)  \\
& & & & 1.56(49) & 1.44(22) & 5.43(89) & 1.99(46)  \\
1.575 & 38.7/31 & 16 & 0.300(12) & 3.57(45) & 0.58(26) & -1.9(1.1) & 4.08(55)  \\
& & & & 1.67(51) & 1.41(23) & 5.22(94) & 2.10(48)  \\
1.600 & 37.2/298 & 14 & 0.300(12) & 3.56(46) & 0.59(27) & -2.0(1.2) & 4.10(59)  \\
& & & & 1.41(57) & 1.52(25) & 5.4(1.0) & 2.01(52)  \\
1.625 & 35.4/27 & 13 & 0.300(13) & 3.50(48) & 0.62(27) & -1.9(1.3) & 4.07(64)  \\
& & & & 0.90(72) & 1.73(29) & 5.8(1.2) & 1.82(60)  \\
\hline
\hline
1.500 & 49.7/37  & 8 & 0.327(18) & 3.29(39) & 0.70(24) & -1.0(1.0) & 3.59(53)  \\
& & & & 1.92(39) & 1.35(20) & 4.5(1.1) & 2.50(60)  \\
1.525 & 48.5/35  & 6 & 0.326(19) & 3.37(40) & 0.66(24) & -1.2(1.2) & 3.70(60)  \\
& & & & 2.06(41) & 1.28(21) & 4.4(1.2) & 2.54(62)  \\
1.550 & 39.7/33  & 20 & 0.311(13) & 3.43(47) & 0.65(27) & -2.38(96) & 4.28(49)  \\
& & & & 1.61(49) & 1.43(22) & 5.36(87) & 2.04(45)  \\
1.575 & 38.4/31  & 17 & 0.315(15) & 3.42(44) & 0.65(26) & -2.0(1.1) & 4.10(56)  \\
& & & & 1.72(52) & 1.39(24) & 5.15(92) & 2.14(48)  \\
1.600 & 36.9/29  & 15 & 0.314(15) & 3.41(45) & 0.66(26) & -2.1(1.2) & 4.13(59)  \\
& & & & 1.46(58) & 1.50(25) & 5.33(98) & 2.06(51)  \\
1.625 & 35.1/27  & 14 & 0.314(16) & 3.36(48) & 0.68(27) & -2.0(1.3) & 4.11(64)  \\
& & & & 0.96(72) & 1.71(29) & 5.7(1.1) & 1.87(58)  \\
\hline
\end{tabular}
\end{center}
\vspace*{4ex}
\caption{\it Combined $V$ and $A$ channel fits to $I^{(\hw_0)}_{\rm ex}(s_0)$
from $s_0=s_{\rm min}$ to $s_0=m_\tau^2$. FOPT results
are shown above the double line, CIPT below; no $D>0$ OPE terms
are included in the fit. $\gamma_{V,A}$ and $\beta_{V,A}$
in units of {\rm GeV}$^{-2}$.}
\label{VAw1paper}
\end{table}%

\subsection{\label{VandA} Combined fits to vector and axial channel data}
We now consider fits analogous
to those of the preceding subsection,
involving simultaneous fitting of
the $V$ and $A$ spectral moments as a function of $s_{\rm min}$.
The fit parameter $\a_s(m_\tau^2)$ is common to the two channels,
while the $D>0$
OPE and DV parameters are distinct for each.
Fits to $I^{(\hw_0)}_{ex,V}(s_0)$ and $I^{(\hw_0)}_{ex,A}(s_0)$
are shown in Tab.~\ref{VAw1paper}; we displayed fewer values of
$s_{\rm min}$ for the sake of brevity.

\begin{figure}[t]
\begin{center}
\includegraphics*[width=7cm]{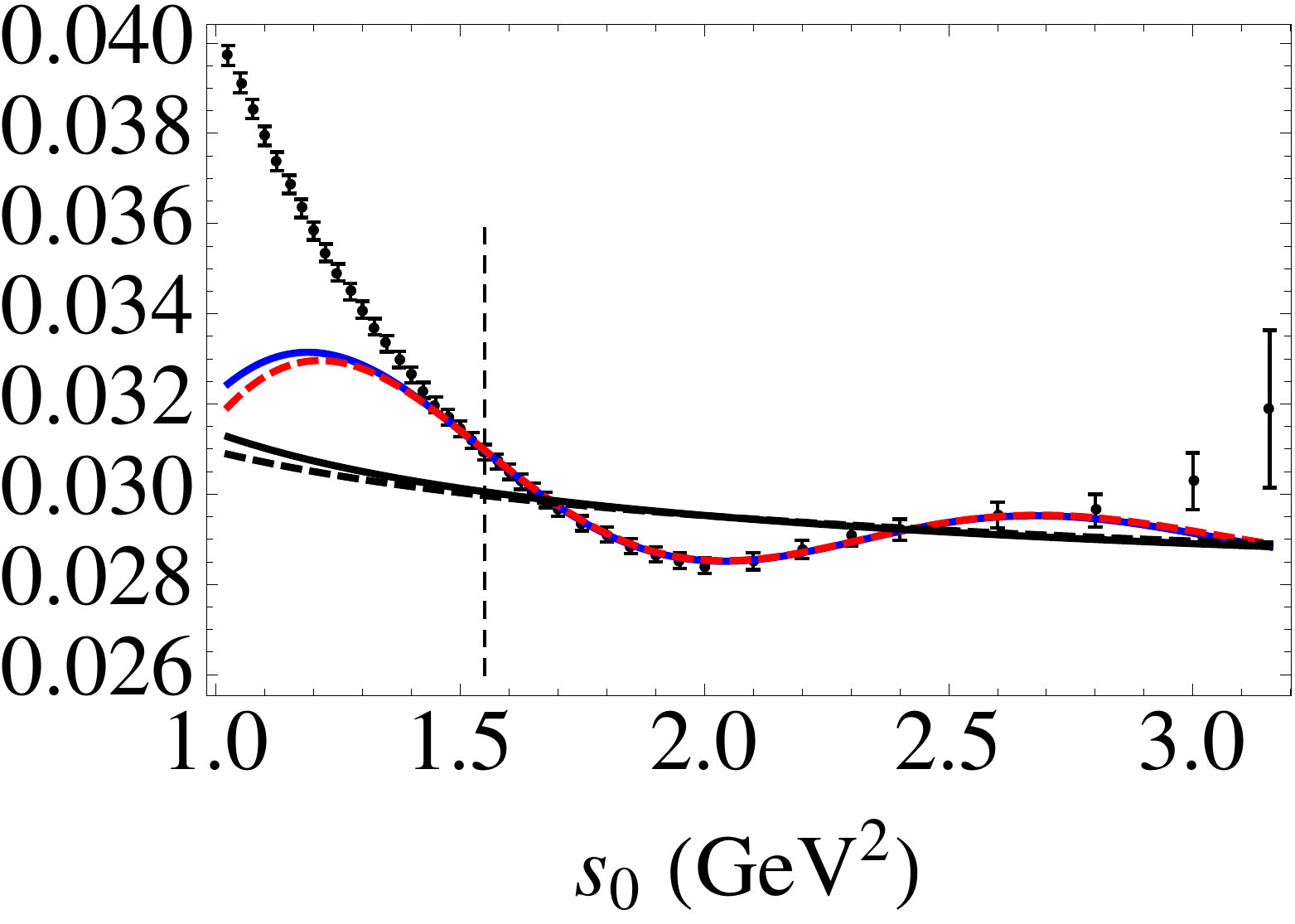}
\hspace{0.5cm}
\includegraphics*[width=7cm]{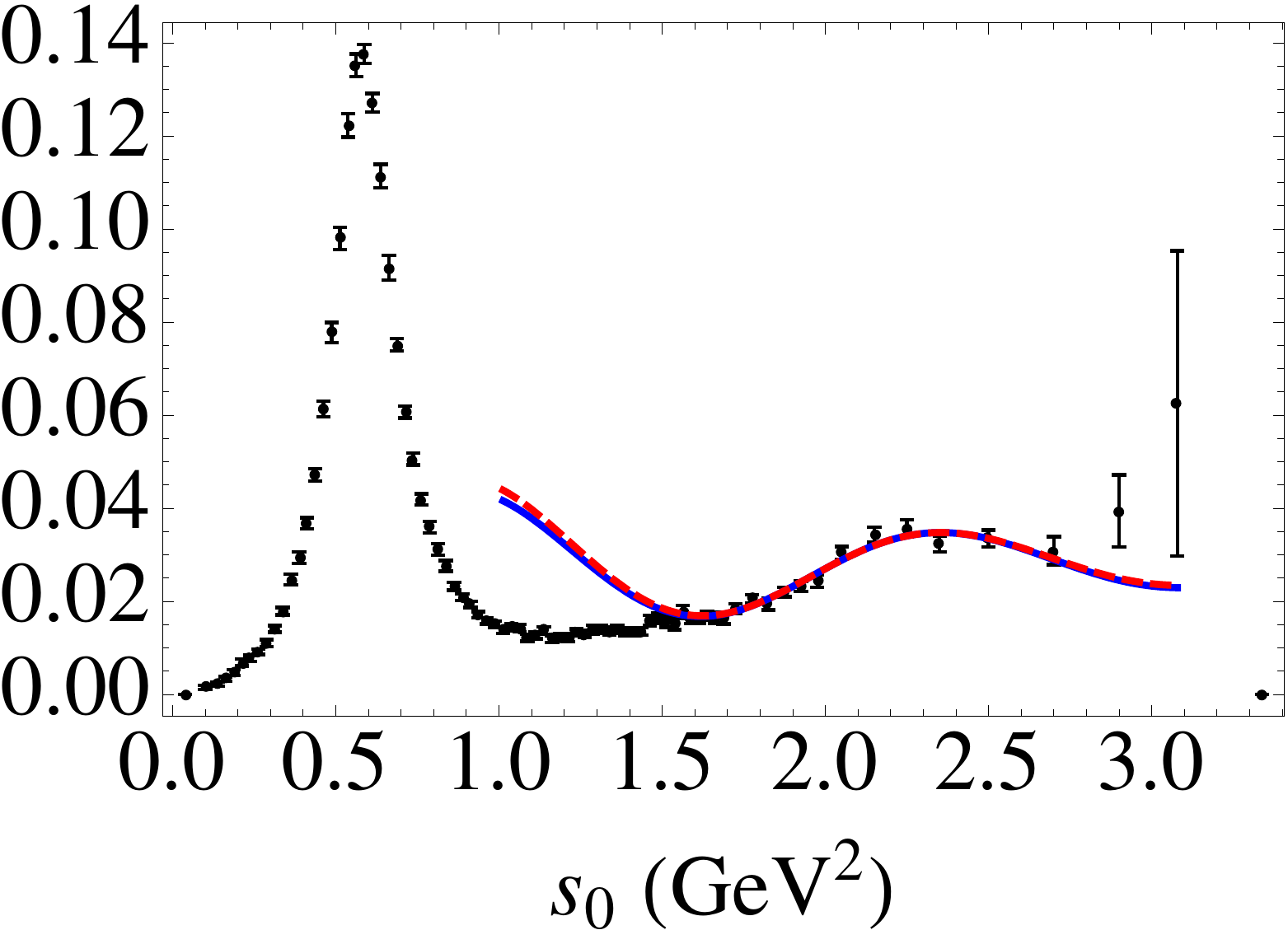}
\vspace{0.5cm}
\includegraphics*[width=7cm]{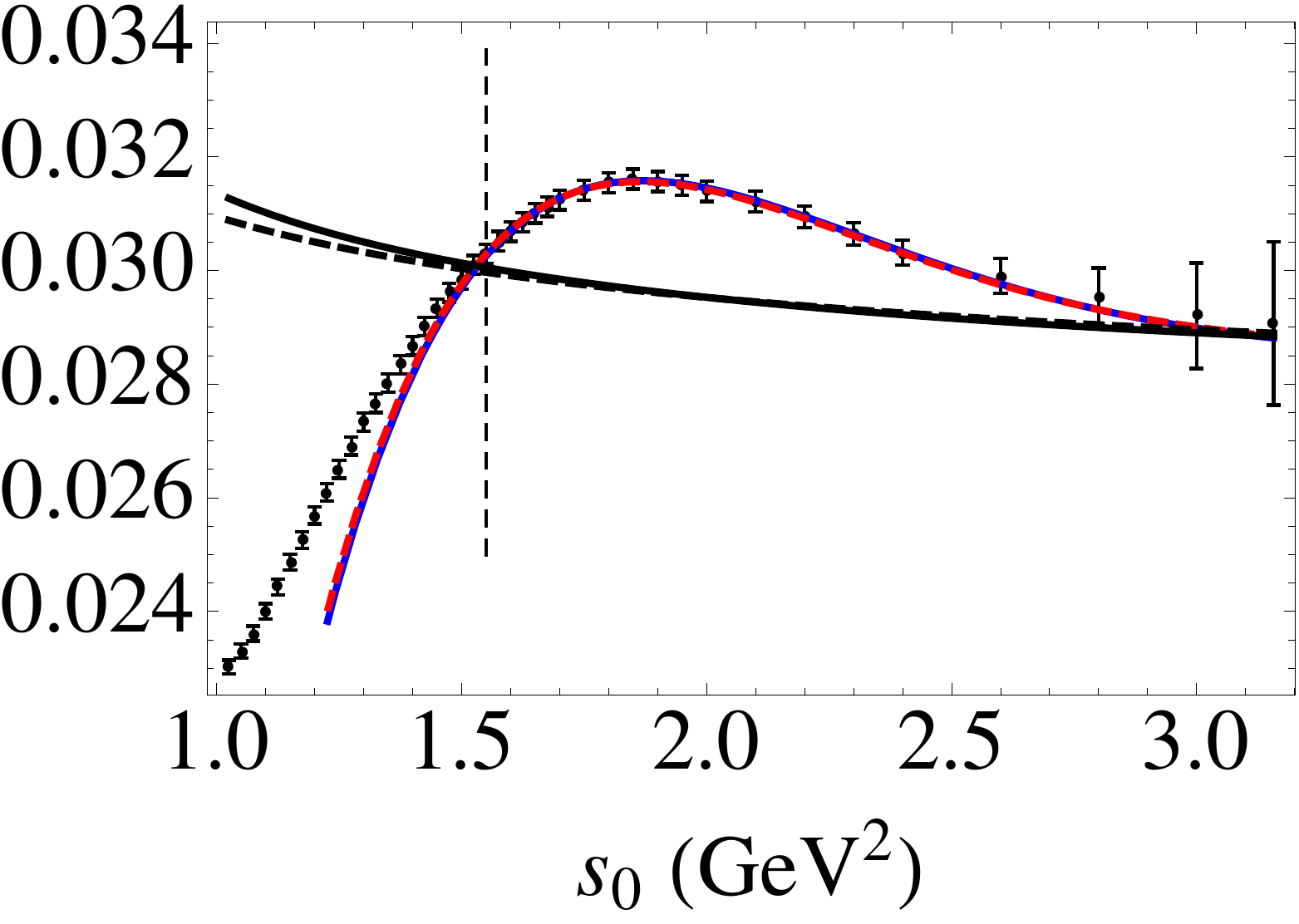}
\hspace{0.5cm}
\includegraphics*[width=7cm]{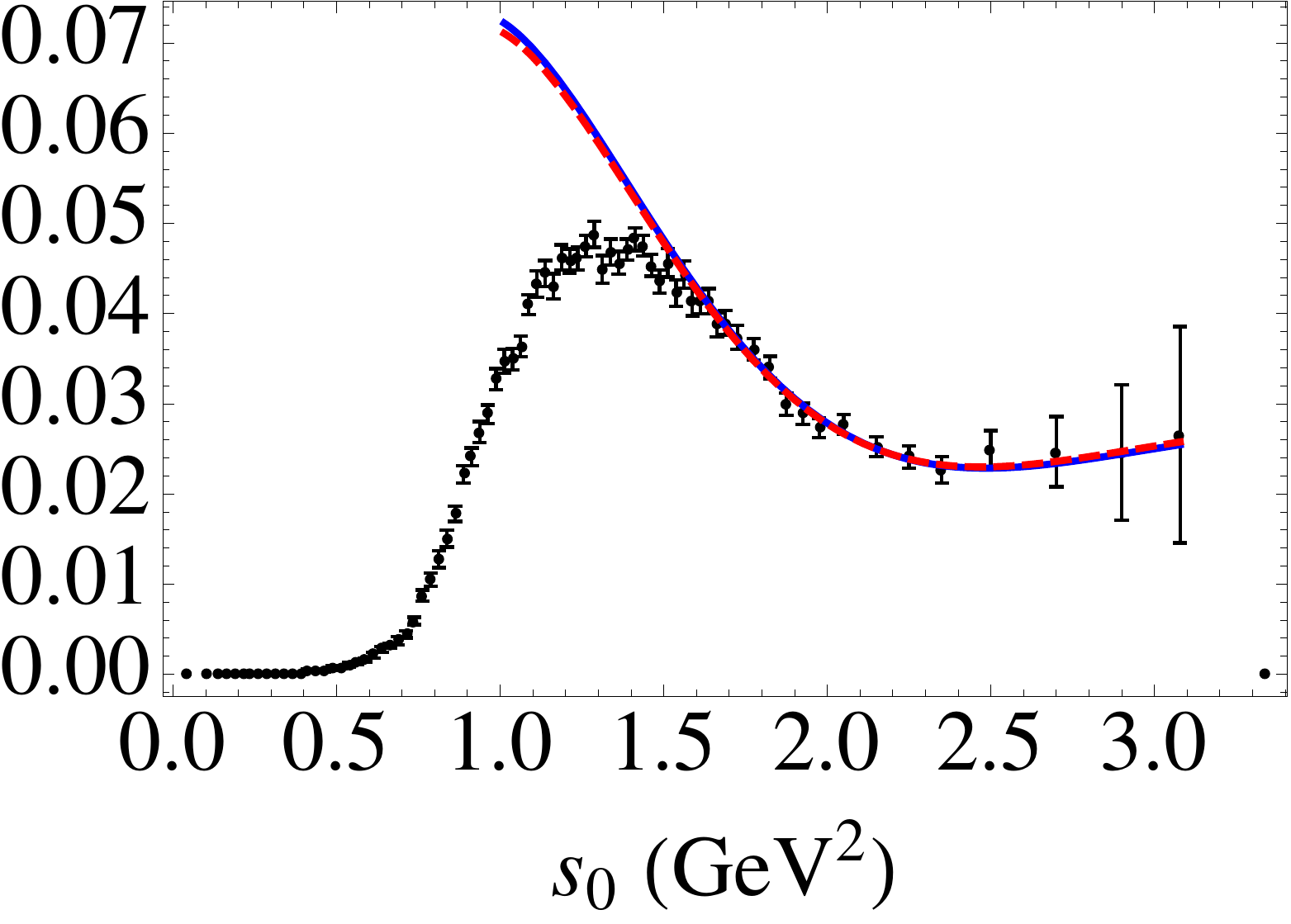}
\end{center}
\begin{quotation}
\floatcaption{CIFOw0VAfit}{{\it Left panels: comparison of
$I^{(\hw_0)}_{\rm ex}(s_0)$ and $I^{(\hw_0)}_{\rm th}(s_0)$ for the
$s_{\rm min}=1.55\ {\rm GeV}^2$ combined $V$ and $A$
channel fits of Tab.~\ref{VAw1paper} ($V$ top, $A$ bottom).
Right panels: comparison of the theoretical spectral function resulting
from this fit with the experimental results ($V$ top, $A$ bottom).
CIPT fits are shown in red (dashed) and FOPT in blue (solid).
The (much flatter) black curves on the left represent the OPE parts of the fits.
The vertical dashed line indicates the location of $s_{\rm min}$.}}
\end{quotation}
\vspace*{-4ex}
\end{figure}
Fits with $s_{\rm min}=1.55$ and $1.575$~GeV$^2$ have the highest $p$-values,
as before. Averaging the parameter values for these fits, we find
\begin{eqnarray}
\label{ashw0VA}
\a_s(m_\tau^2)&=&0.299(12)\ ,\qquad\mbox{(FOPT)}\ ,\\
&=&0.313(15)\ ,\qquad\mbox{(CIPT)}\ ,\nonumber
\end{eqnarray}
slightly higher values than those of Eqs.~(\ref{ashw0}) and ~(\ref{ashw023}),
but consistent within errors. The errors are $\chi^2$ errors, since all
correlations were taken into account in the fit; they are slightly larger
than those found in the $V$-channel fits.

\begin{table}[!h]
\hspace{0cm}\begin{tabular}{|c|c|c|c|c|c|c|c|c|}
\hline
$s_{\rm min}$ (GeV$^2$) & $\cq^2$/dof & $\alpha_s$ & $\delta_{V,A}$ & $\gamma_{V,A}$ & $\alpha_{V,A}$ & $\beta_{V,A}$ & $10^2C_{6V,A}$ & $10^2C_{8V,A}$ \\
\hline
1.475 & 182/131=1.39  & 0.297(7) & 2.90(42) & 0.95(26) & -1.61(65) & 3.91(35) & -0.78(13) & 1.31(23) \\
&&&2.26(35) & 1.13(18) & 4.92(58) & 2.25(30) & -0.08(35) & 1.12(96) \\
1.500 & 160/125=1.28  & 0.297(8) & 2.92(43) & 0.94(26) & -1.62(73) & 3.91(39) & -0.78(14) & 1.31(25) \\
&&&1.90(44) & 1.29(21) & 5.26(69) & 2.08(36) & -0.26(44) & 1.8(1.4) \\
1.525 & 149/119=1.25  & 0.294(8) & 3.08(48) & 0.86(28) & -2.16(75) & 4.18(40) & -0.85(13) & 1.46(23) \\
&&&1.86(48) & 1.30(22) & 5.38(72) & 2.02(37) & -0.38(49) & 2.1(1.6) \\
1.550 & 126/113=1.11  & 0.292(9) & 3.19(51) & 0.80(30) & -2.65(79) & 4.42(41) & -0.90(13) & 1.57(22) \\
&&&1.53(56) & 1.42(24) & 5.73(84) & 1.84(43) & -0.63(61) & 3.0(2.2) \\
1.575 & 124/107=1.16  & 0.293(9) & 3.18(51) & 0.81(29) & -2.47(84) & 4.33(43) & -0.88(14) & 1.52(24) \\
&&&1.57(61) & 1.41(26) & 5.67(86) & 1.87(44) & -0.57(61) & 2.8(2.2) \\
1.600 & 116/101=1.15  & 0.293(9) & 3.20(52) & 0.80(30) & -2.51(89) & 4.35(46) & -0.89(14) & 1.53(25) \\
&&&1.14(74) & 1.59(29) & 6.0(1.0) & 1.72(53) & -0.73(72) & 3.6(2.7) \\
1.625 & 112/95=1.18  & 0.294(10) & 3.20(55) & 0.79(31) & -2.43(95) & 4.31(48) & -0.87(15) & 1.50(28) \\
&&&0.85(92) & 1.71(34) & 6.2(1.2) & 1.61(63) & -0.80(80) & 4.0(3.2) \\
\hline
\hline
1.475 & 159/131=1.21 & 0.338(13) & 3.45(32) & 0.61(20) & -0.63(67) & 3.42(35) & -0.58(16) & 0.83(31)\\
&&&2.23(33) & 1.25(21) & 3.45(81) & 3.02(42) & 0.59(25) & -0.64(58)\\
1.500 & 146/125=1.17 & 0.328(15) & 3.26(39) & 0.72(24) & -0.92(79) & 3.56(41) & -0.67(18) & 1.00(35)\\
&&& 1.96(41) & 1.34(22) & 4.41(89) & 2.53(46) & 0.25(40) & 0.3(1.0)\\
1.525 & 136/119=1.14 & 0.320(13) & 3.35(44) & 0.69(26) & -1.59(79) & 3.90(41) & -0.80(15) & 1.26(29)\\
&&& 1.93(46) & 1.32(23) & 4.76(83) & 2.35(43) & 0.05(43) & 0.78(12)\\
1.550 & 118/113=1.04 & 0.312(13) & 3.35(49) & 0.70(29) & -2.28(81) & 4.23(42) & -0.90(13) & 1.48(25)\\
&&& 1.59(55) & 1.44(25) & 5.37(89) & 2.03(46) & -0.33(56) & 2.0(1.8)\\
1.575 & 115/107=1.07 & 0.315(13) & 3.35(48) & 0.70(28) & -1.98(88) & 4.09(45) & -0.86(15) & 1.39(29) \\
&&& 1.65(59) & 1.42(27) & 5.23(92) & 2.11(47) & -0.22(55) & 1.6(1.7)\\
1.600 & 108/101=1.07 & 0.314(14) & 3.33(49) & 0.71(29) & -2.04(93) & 4.12(47) & -0.87(15) & 1.41(30)\\
&&& 1.23(70) & 1.60(30) & 5.6(1.1) & 1.95(55) & -0.37(64) & 2.2(2.2)\\
1.625 & 105/95=1.10 & 0.315(15) & 3.28(53) & 0.73(30) & -1.9(1.0) & 4.06(51) & -0.85(17) & 1.37(34)\\
&&& 0.96(85) & 1.71(35) & 5.7(1.2) & 1.87(63) & -0.42(71) & 2.4(2.5)\\
\hline
\end{tabular}
\vspace*{4ex}
\floatcaption{VAwtaupaper}{\it Combined $V$ and $A$ channel fits to
$I^{(\hw_0)}_{\rm ex}(s_0)$, $I^{(\hw_2)}_{\rm ex}(s_0)$ and
$I^{(\hw_3)}_{\rm ex}(s_0)$
from $s_0=s_{\rm min}$ to $s_0=m_\tau^2$. FOPT results
are shown above the double line, CIPT below;
$D=6,\ 8$ OPE terms included in the fit.
$\gamma_{V,A}$ and $\beta_{V,A}$
in units of {\rm GeV}$^{-2}$, $C_{6V,A}$ in units of {\rm GeV}$^6$ and $C_{8V,A}$ in
units of {\rm GeV}$^8$.
}
\end{table}%

For $s_{\rm min}=1.55$~GeV$^2$ we show the quality of the fits
in the left panels of Fig.~\ref{CIFOw0VAfit} and the $V$ and $A$
spectral-function comparisons obtained using
parameter values from the fit in the corresponding
right-hand panels.
We note that the fit curves in the axial case are
essentially determined by the shoulder of the $a_1$ resonance, in
contrast to what happens in the vector case, where the $\r$ peak is
well away from the region relevant for the shape of the fit curves.

Tab.~\ref{VAwtaupaper} shows the results of
the combined $V$ and $A$ channel fits to the three moments
$I^{(\hw_0)}_{\rm ex}(s_0)$, $I^{(\hw_2)}_{\rm ex}(s_0)$ and $I^{(\hw_0)}_{\rm ex}(s_3)$.
Judging by the values of $\cq^2$/dof, the best fits are again those with
$s_{\rm min}=1.55$ and $1.575$~GeV$^2$, leading to
\begin{eqnarray}
\label{ashw023VA}
\a_s(m_\tau^2)&=&0.293(9)\ ,\qquad\mbox{(FOPT)}\ ,\\
&=&0.313(13)\ ,\qquad\mbox{(CIPT)}\ .
\nonumber \end{eqnarray}
These values are in good agreement with those of
the other fits reported above.
As before, fits to just the pair of moments $I^{(\hw_0)}_{\rm ex}(s_0)$ and
$I^{(\hw_2)}_{\rm ex}(s_0)$ do not lead to any surprises.
We show the {quality of the fits of Tab.~\ref{VAwtaupaper} for the
moments $I^{(\hw_0)}_{\rm ex}(s_0)$ and the
comparison of the resulting spectral functions
to the experimental ones
for both channels in Fig.~\ref{CIFOw023VAfit}.
The fits for the other two moments look very similar to
those in Fig.~\ref{CIFOw023fit} for the $V$ channel, and show a
similar quality in the $A$ channel.

\begin{figure}[t]
\begin{center}
\includegraphics*[width=7cm]{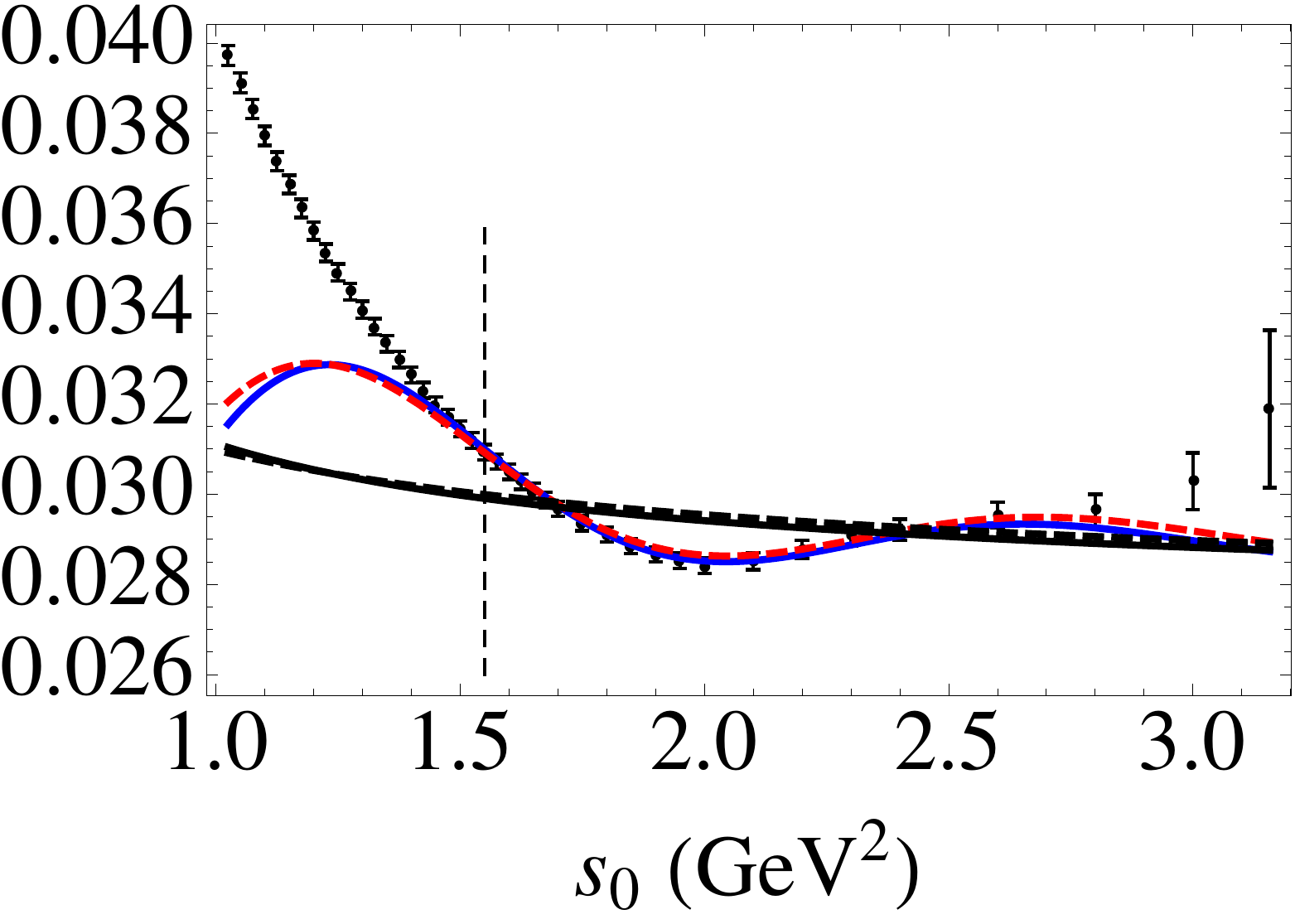}
\hspace{0.5cm}
\includegraphics*[width=7cm]{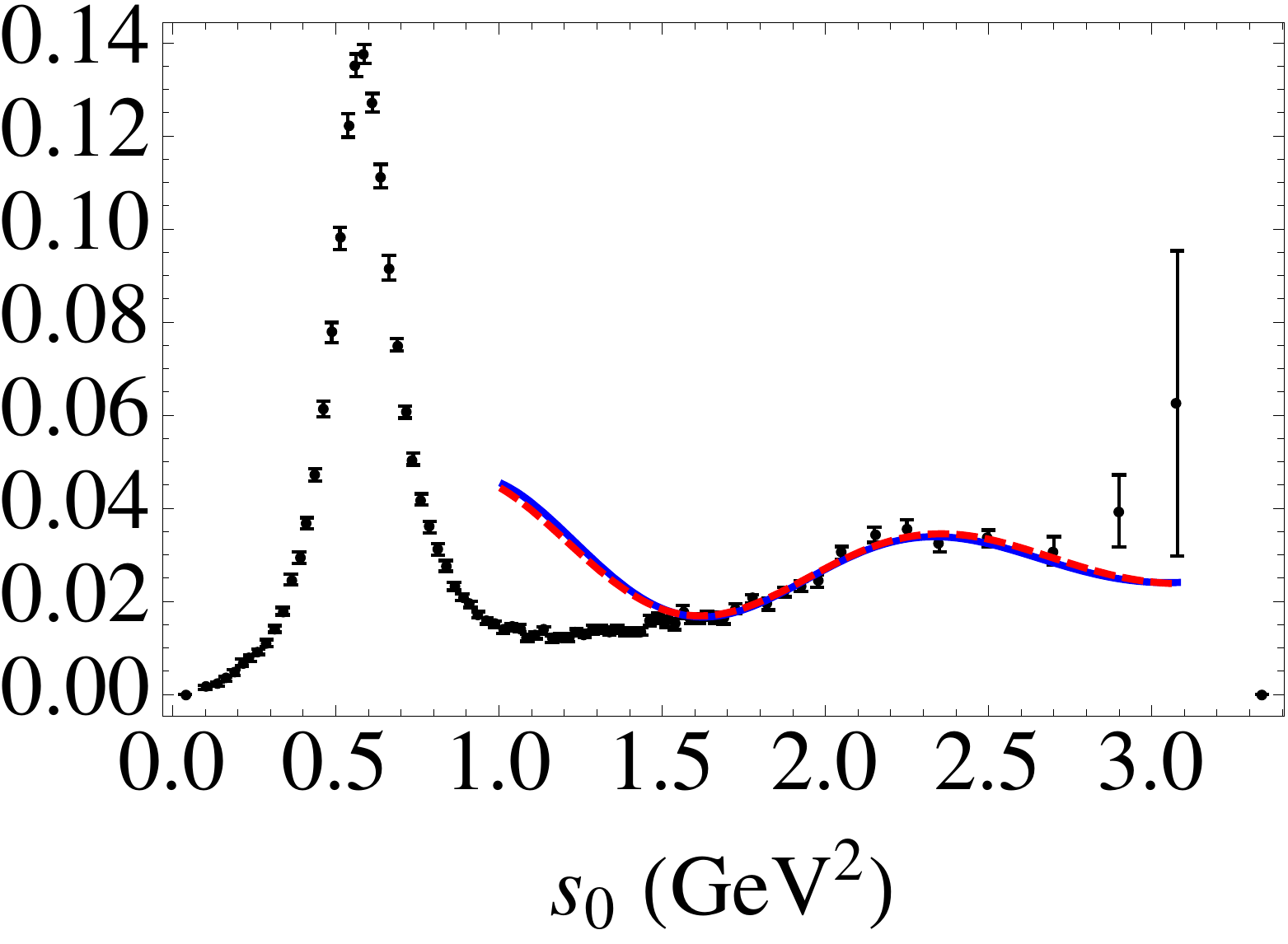}
\vspace{0.5cm}
\includegraphics*[width=7cm]{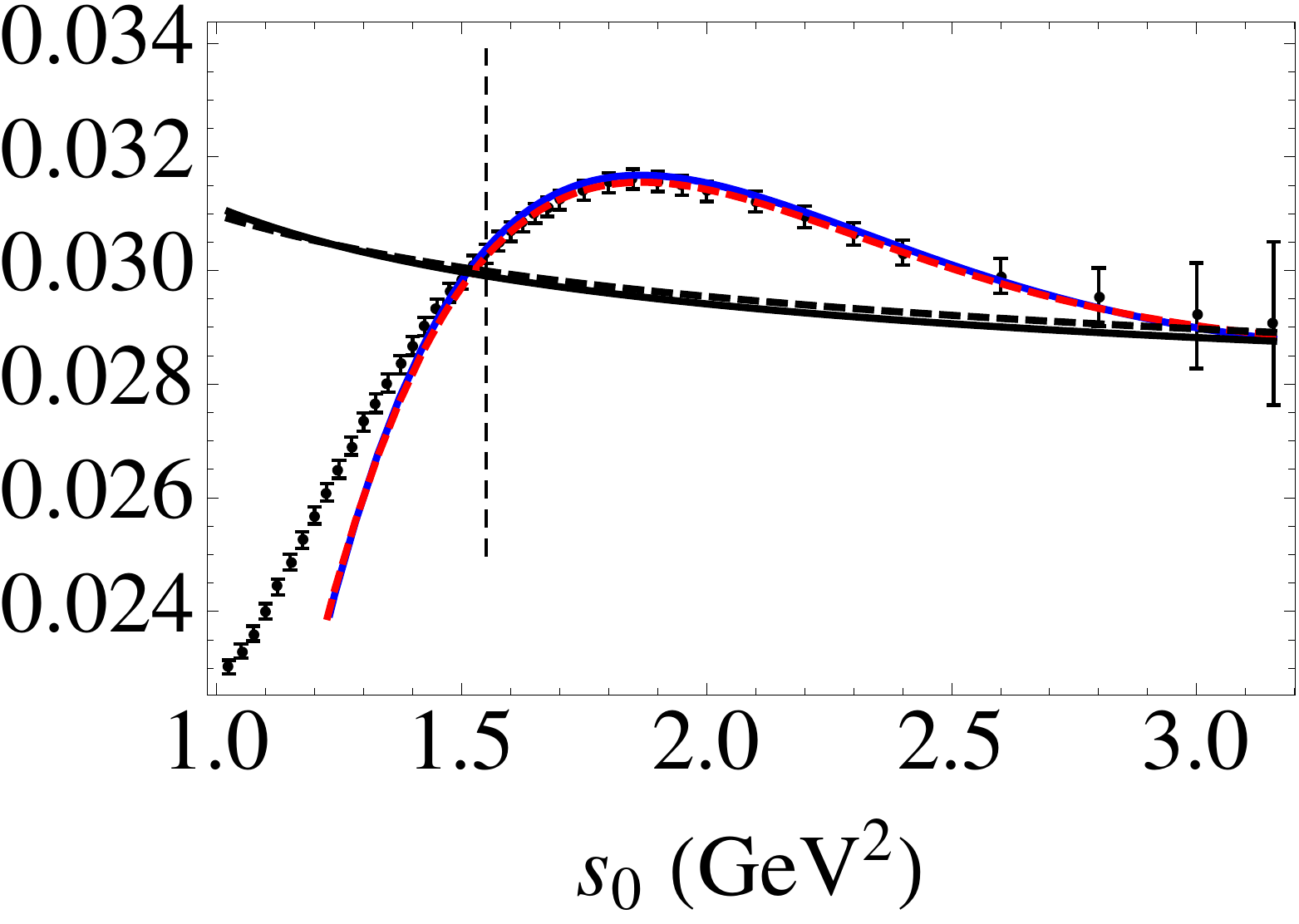}
\hspace{0.5cm}
\includegraphics*[width=7cm]{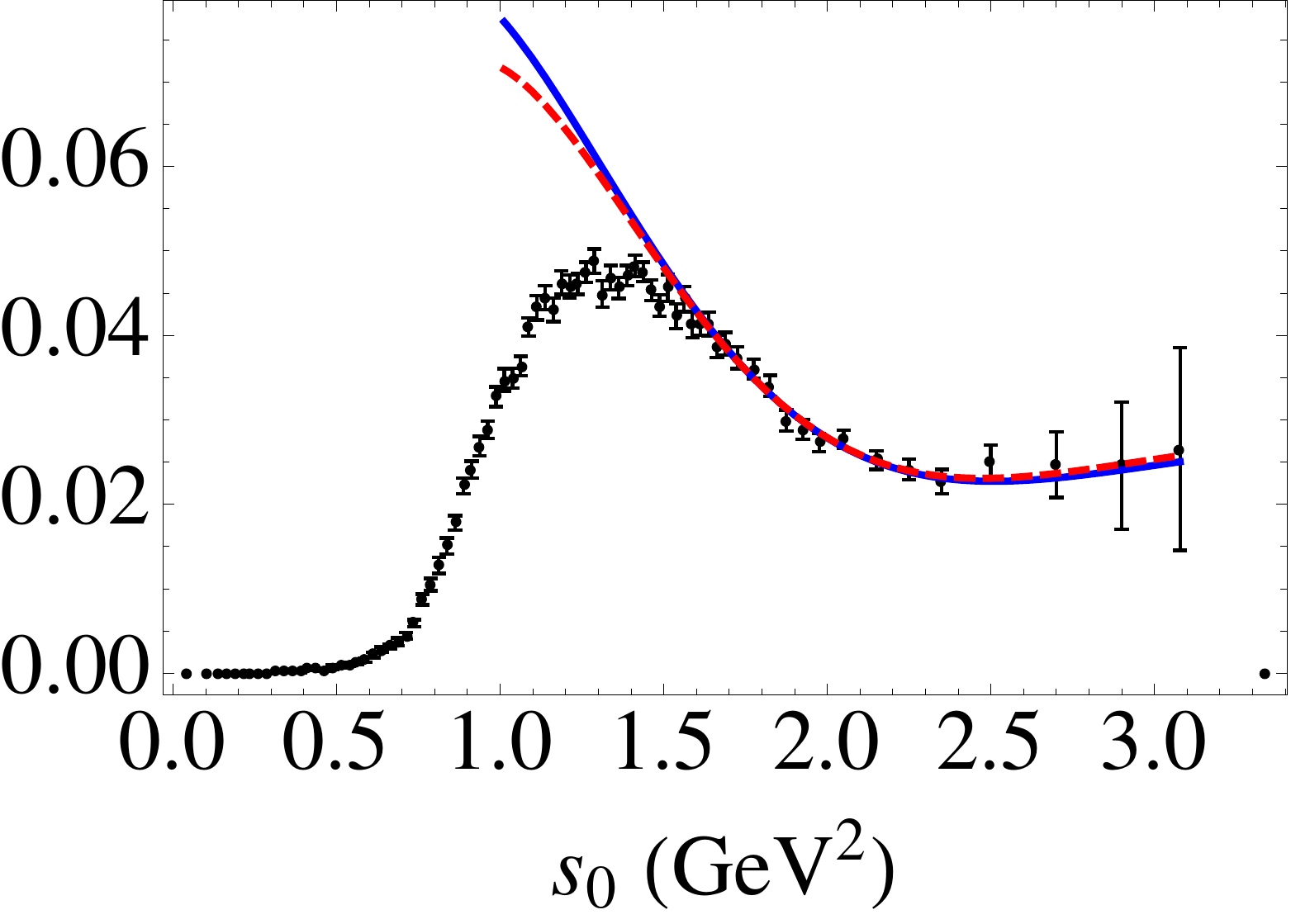}
\end{center}
\begin{quotation}
\floatcaption{CIFOw023VAfit}{\it Left panels: comparison of
$I^{(\hw_0)}_{\rm ex}(s_0)$ and $I^{(\hw_0)}_{\rm th}(s_0)$ for the
$s_{\rm min}=1.55\ {\rm GeV}^2$ combined $V$ and $A$ channel
fits of Tab.~\ref{VAwtaupaper} ($V$ top, $A$ bottom).
Right panels: comparison of the theoretical spectral function resulting
from this fit with the experimental results ($V$ top, $A$ bottom).
CIPT fits are shown in red (dashed) and FOPT in blue (solid).
The (much flatter) black curves on the left represent the OPE parts of the fits.
The vertical dashed line indicates the location of $s_{\rm min}$.}
\end{quotation}
\vspace*{-4ex}
\end{figure}
\section{\label{results} Tests and results}
There are a number of consistency checks that can be applied once values
for $\a_s(m_\tau^2)$ as well as the $D>0$
OPE and DV parameters have been obtained from a fit. We will present some
of these in Sec.~\ref{tests}. Then, in Sec.~\ref{alphas}, we will present our
final number for $\a_s(m_\tau^2)$, following this in Sec.~\ref{nonpert} by a
determination of the non-perturbative contribution to
$R_{V+A;ud}$ and a comparison of the $D=6$ OPE coefficients with the results
of estimates based on the vacuum saturation approximation (VSA).
In Sec.~\ref{OPALresults} we will compare the present results with those from
our fits to the OPAL data.

\begin{figure}[t]
\begin{center}
\includegraphics*[width=10cm]{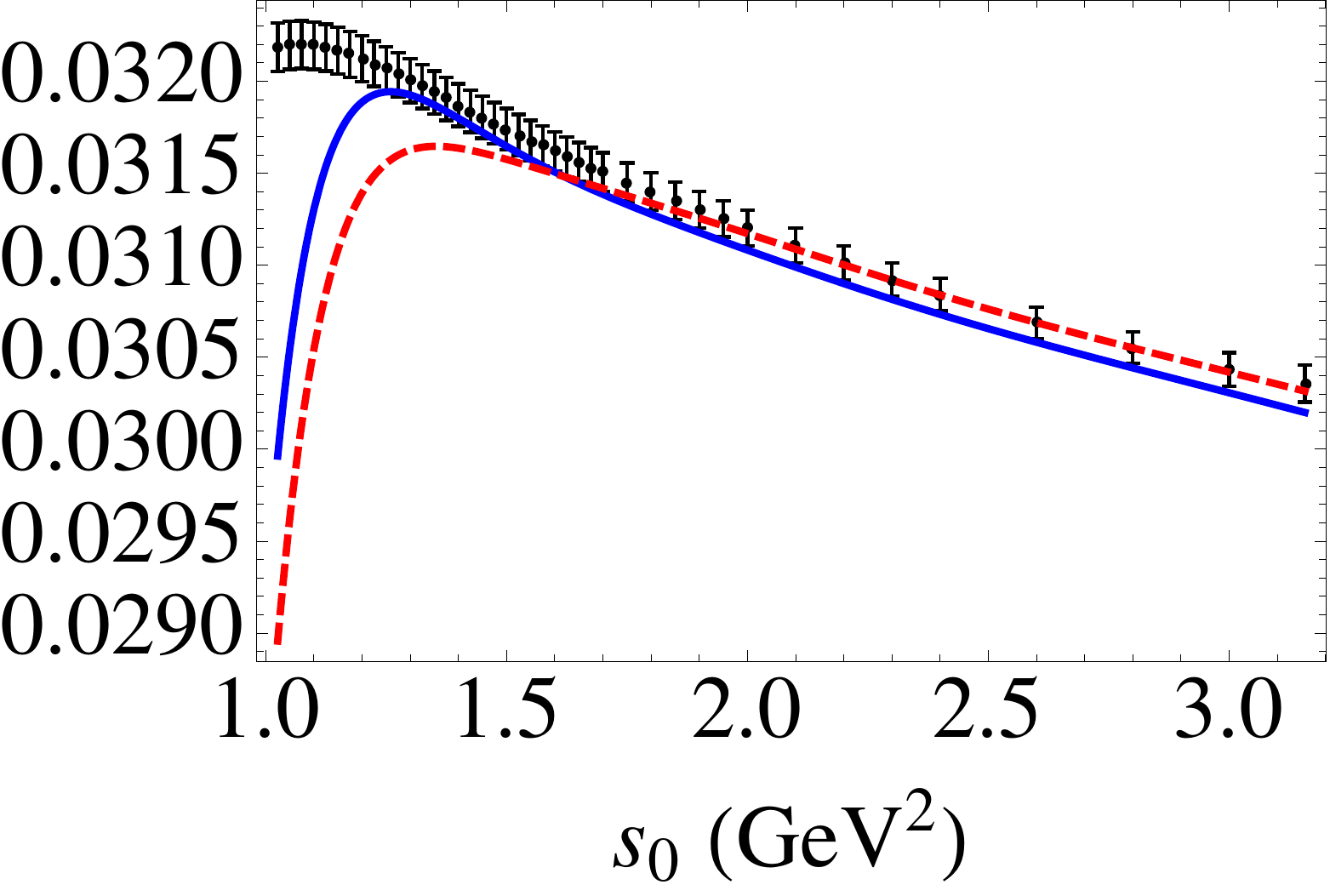}
\end{center}
\begin{quotation}
\floatcaption{CIFOw023VAfitw3VplusA}{\it The rescaled version of $R_{V+A;ud}(s_0)$
(the RHS of eq.~(\ref{Rdef}))
as a function of $s_0$. Theory curves from
$s_{\rm min}=1.55$~{\rm GeV}$^2$ entries of
Tab.~\ref{VAwtaupaper}; CIPT (red, dashed) and FOPT (blue, solid).}
\end{quotation}
\vspace*{-4ex}
\end{figure}
\subsection{\label{tests} Tests}
We consider first the comparison of the experimental value of
\begin{equation}
\label{Rdef}
I^{(\hw_3)}_{ex,V}(s_0)+I^{(\hw_3)}_{ex,A}(s_0) =
{\frac{m_\tau^2}{12\pi^2 \vert V_{ud}\vert^2 S_{EW}}}
\, R_{V+A;ud}(s_0)
\end{equation}
with the function obtained from the fit.
In Fig.~\ref{CIFOw023VAfitw3VplusA} we show this
comparison, using the parameter values for $s_{\rm min}=1.55$~GeV$^2$ from
Tab.~\ref{VAwtaupaper}. The fitted curves are in good agreement
everywhere above $s_0\approx 1.3$~GeV$^2$ ($s_0\approx 1.5$~GeV$^2$)
for the FOPT (CIPT) fits.\footnote{We recall that even though
correlations between different spectral moments are not included in the
fit quality $\cq^2$, those between bins within one spectral moment are
included, making these fits strongly correlated.}
We include this test because (in rescaled form)
it was originally advocated as an important confirmation
of the analysis of Ref.~\cite{ALEPH}. One can see that our 
fits satisfy this test at least as well (see \eg\ Fig.~73 of Ref.~\cite{ALEPH}). 
In other words, this test
is not able to discriminate between the results of our analysis and 
those Refs.~\cite{ALEPH13,ALEPH,ALEPH08}.
For more discussion on the comparison between our analysis
and that of Refs.~\cite{ALEPH13,ALEPH,ALEPH08} we refer
to Sec.~\ref{ALEPH}.

\begin{figure}[t]
\begin{center}
\includegraphics*[width=7cm]{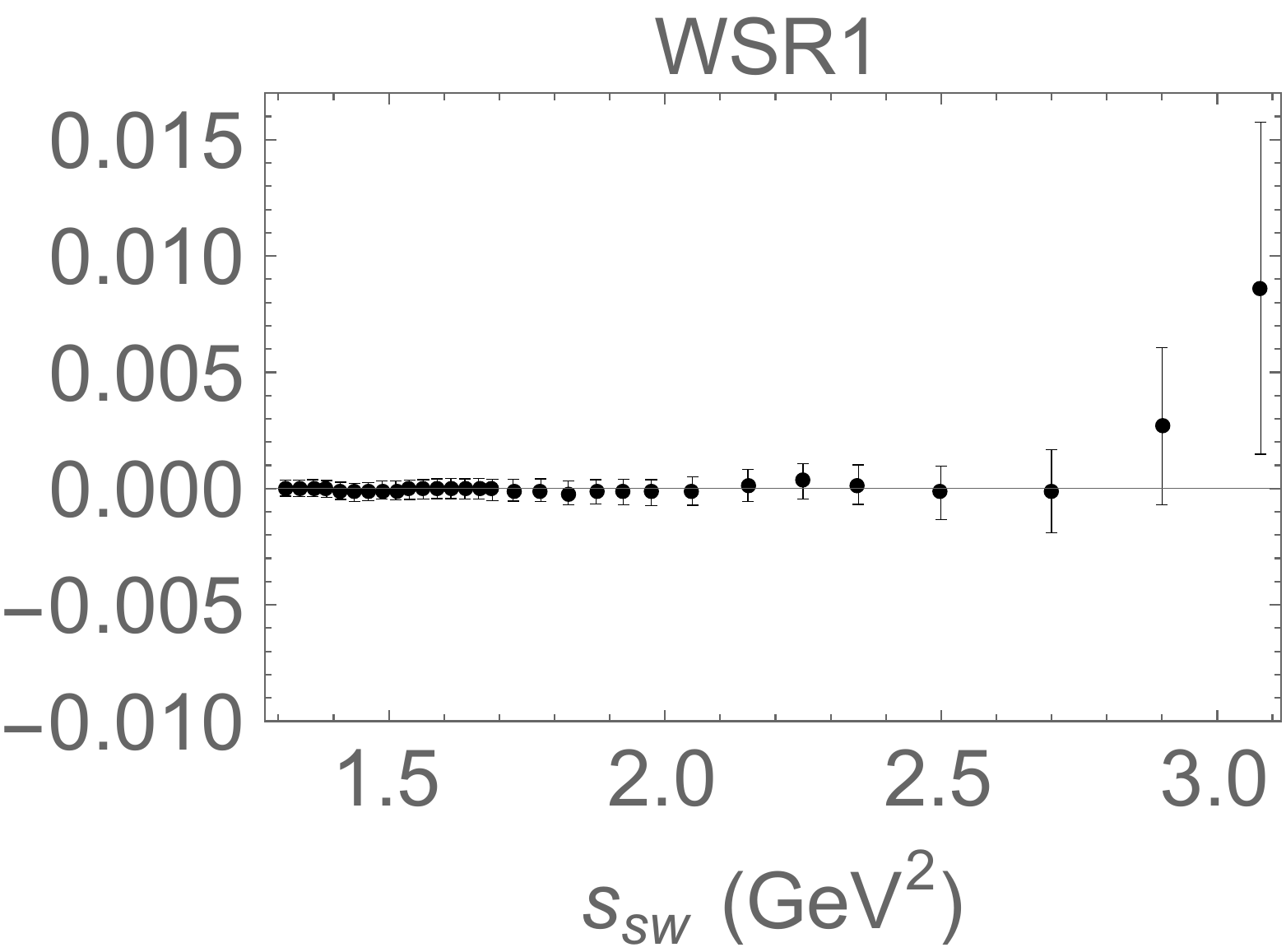}
\hspace{0.5cm}
\includegraphics*[width=7cm]{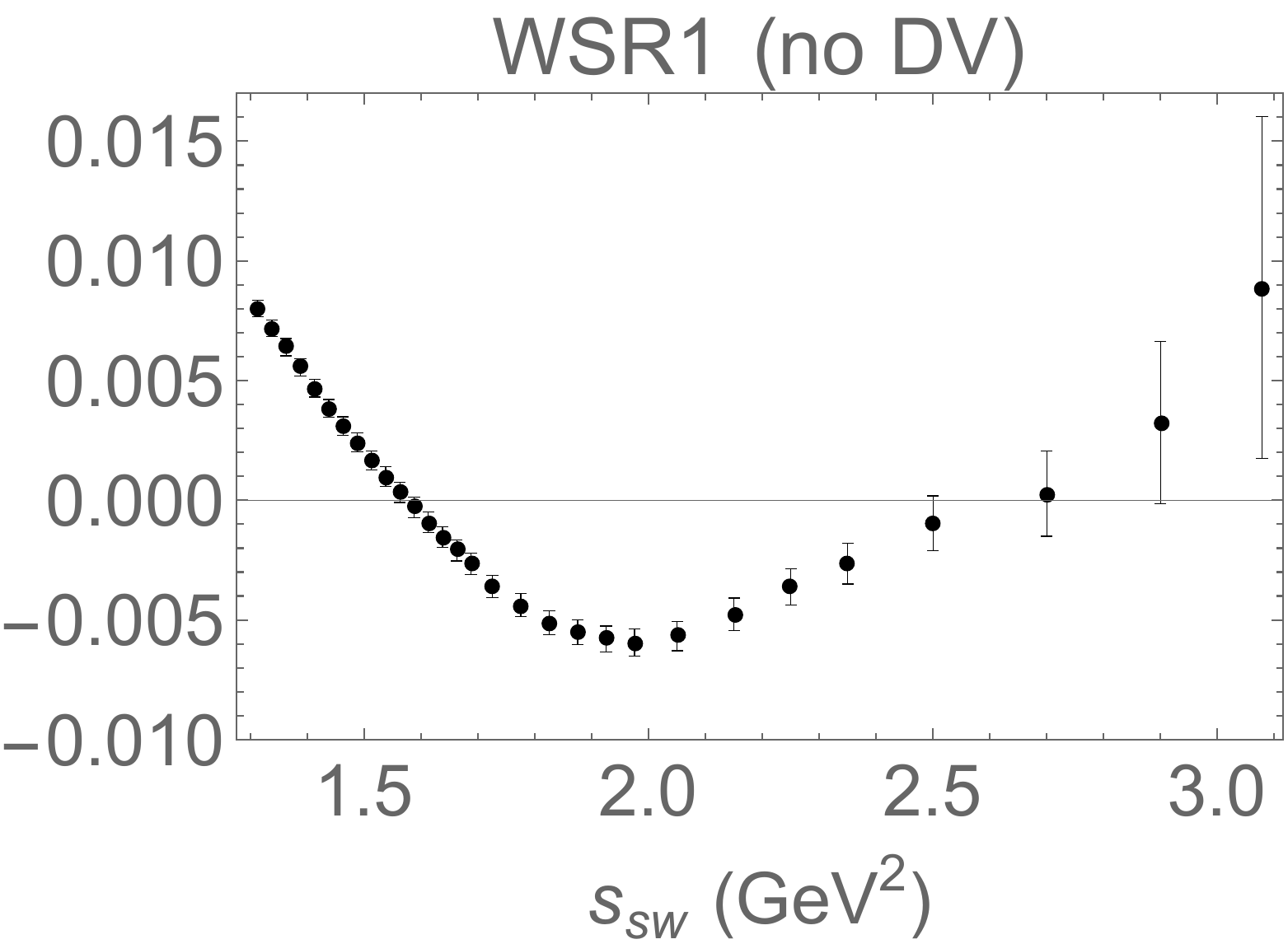}
\end{center}
\begin{quotation}
\floatcaption{WSR}{\it The first Weinberg sum rule, with DVs (left
panel) and without DVs (right panel), both in {\rm GeV}$^2$.
Data have been used for $s<s_{\rm sw}$, while the DV \ansatz~(\ref{ansatz})
with parameter values
obtained from the $s_{\rm min}=1.55$~{\rm GeV}$^2$ fit
has been used for $s>s_{\rm sw}$.
The figures shown use CIPT fits.}
\end{quotation}
\vspace*{-4ex}
\end{figure}
As in Ref.~\cite{alphas1}, we may also consider the first and second Weinberg
sum rules (WSRs) \cite{SW}, as well as the DGMLY sum rule for the pion
electro-magnetic mass splitting \cite{EMpion}.
These sum rules can be written as
\begin{eqnarray}
\label{WSRdefs}
\int_0^\infty ds\,\left(\r^{(1+0)}_V(s)-\r^{(1+0)}_A(s)\right)
&=&\int_0^\infty ds\,\left(\r^{(1)}_V(s)-\r^{(1)}_A(s)\right)-2f_\p^2=0\ ,\\
\int_0^\infty ds\,s\left(\r^{(1+0)}_V(s)-\r^{(1+0)}_A(s)\right)
&=&\int_0^\infty ds\,s\left(\r^{(1)}_V(s)-\r^{(1)}_A(s)\right)-2m_\p^2
f_\p^2=0\ ,\nonumber\\
\int_0^\infty ds\,s\log{(s/\m^2)}\left(\r^{(1)}_V(s)-\r^{(1)}_A(s)\right)
&=&\frac{8\p f_0^2}{3\a}
\left(m_{\p^\pm}^2-m_{\p^0}^2\right)\ ,\nonumber
\end{eqnarray}
where $f_0$ is the pion decay constant in the chiral limit, and $\a$ is
the fine-structure constant.
For the second WSR we assume that terms of order $m_im_j$, $i,j=u,d$ can be
neglected. Without this assumption, the integral is linearly divergent,
forcing us to cut it off. If we cut off the integral at $s_0$,
there would be an extra contribution proportional to $m_im_j\a_s^2s_0$
in this sum rule. This contribution is still very small at $s_0=m_\tau^2$
(of order a few percent of the contribution $2m_\p^2 f_\p^2$), allowing us
to assume that we are effectively in the chiral limit with regard to the
second WSR. Even the term  $2m_\p^2 f_\p^2$, while dominating the term
proportional to $m_im_j\a_s^2s_0$, vanishes in the chiral limit, and
itself turns out to be numerically negligible within errors.
Also the DGMLY sum rule holds only in the chiral limit, and in that limit the
integral on the left-hand side is independent
of $\m$ because of the second WSR.

In Fig.~\ref{WSR} we show the first integral in Eq.~(\ref{WSRdefs}) as a
function of the ``switch'' point $s_{\rm sw}$ below which we use the
experimental data, and above which
we use the DV \ansatz~(\ref{ansatz}) with parameters from the CIPT fit with
$s_{\rm min}=1.55$~GeV$^2$ of Tab.~\ref{VAwtaupaper} in order to evaluate the
integral. Using parameter values from Tab.~\ref{VAw1paper} or FOPT fits
leads to almost identical figures.\footnote{The contribution from OPE
terms to the spectral functions $\r_{V,A}$ is suppressed by an extra power of
$\a_s$, and small enough to be negligible \cite{CGP,alphas1}.}
The figure on the left includes the contribution from Eq.~(\ref{ansatz}),
while the figure on the right omits such contributions.
The latter is equivalent to the
upper right panel of Fig. 8 in the first paper in Ref.~\cite{ALEPH}.
Clearly, the first WSR is very well satisfied by our fits,
but only if duality violations are taken into account.
We do not show similar figures for the second WSR and the
DGMLY sum rule, because our conclusions for these sum rules
are very similar. Just as in Ref.~\cite{alphas1,alphas2},
these sum rules are satisfied within errors, but only if
duality violations are taken into account. In particular,
within errors, one may assume that our representation of the
spectral functions is in the chiral limit, for the purpose of these
three sum rules.

\subsection{\label{alphas} The strong coupling}
The presence of duality violations forces us to make several assumptions
in order to extract a value for $\a_s(m_\tau^2)$.  These assumptions have been
checked against the data, \seef\ Figs.~\ref{CIFOw0fit} and \ref{CIFOw023fit}--\ref{WSR}. First,
we need to assume that Eq.~(\ref{ansatz}) provides a
satisfactory description of duality violations for
asymptotically large $s$. Second, we need to assume that
$s\,\gtap\, 1.5$~GeV$^2$ is already in the asymptotic region.
And, finally, if we wish to also use the axial data, we need
to assume that this is true both in the $V$ and $A$ channels.
As already discussed above, this would amount to
the assumption that the upper shoulder of the
$a_1$ resonance is already more or less
in the asymptotic region. Using only the $V$-channel fits, we avoid
having to make this latter assumption,
and doing so we find, from the results quoted in
Eq.~(\ref{ashw023}),

\begin{eqnarray}
\label{alphasfinal}
\a_s(m_\tau^2)&=&0.296(10)(1)(2)=0.296\pm 0.010\
,\qquad(\overline{\mbox{MS}},\ n_f=3,\ \mbox{FOPT})\ ,\\
&=&0.310(14)(1)(1)=0.310\pm 0.014\
,\qquad(\overline{\mbox{MS}},\ n_f=3,\ \mbox{CIPT})\ ,\nonumber
\end{eqnarray}
where the first error is the statistical fit error
already given in Eq.~(\ref{ashw023}), while the second
represents half the difference between the $s_{\rm min}=1.55$ and
$1.575$ GeV$^2$ results of Tab.~\ref{VVwtaupaper} from which the
average is derived. The third error represents the change induced
by varying the estimated 6-loop $D=0$ coefficient $c_{51}=283$ \cite{BJ}
by the assumed $100\%$ uncertainty about its central value, as
in Ref.~\cite{alphas1,alphas2}.
The error from this latter uncertainty
would be about $\pm 0.004$ for both FOPT and CIPT if it were estimated
from fits using only the moment with weight $\hw_0$; this would raise
both final errors by $0.001$. We observe that
the final errors we find are of the same order of magnitude as
the difference between the FOPT and CIPT values of $\a_s(m_\tau^2)$.
We also note that in all tables the value
of $\a_s(m_\tau^2)$ is very stable as a function of
$s_{\rm min}$ for all values of $s_{\rm min}$ included in these tables,
except for possibly the lowest $s_{\rm min}$ shown.

Equation~(\ref{alphasfinal}) constitutes our final result for
$\a_s(m_\tau^2)$ from the revised ALEPH data. Converting these results
into values for $\a_s$ at the $Z$ mass using the
standard self-consistent combination of 4-loop running with
3-loop matching at the flavor thresholds \cite{cks97},
we find
\begin{eqnarray}
\label{alphasZ}
\a_s(m_Z^2)&=&0.1155\pm 0.0014\ ,\qquad(\overline{\mbox{MS}},
\ n_f=5,\ \mbox{FOPT})\ ,\\
&=&0.1174\pm 0.0019\ ,\qquad(\overline{\mbox{MS}},\ n_f=5,
\ \mbox{CIPT})\ .\nonumber
\end{eqnarray}

\subsection{\label{nonpert} Non-perturbative quantities}

As in Ref.~\cite{alphas2}, we would like to estimate the relative
deviation of the aggregate dimension-6 condensates $C_{6,V/A}$ from
the values given by the VSA. We express these condensates in terms
of the VSA-violating parameters
$\r_1$ and $\r_5$ by \cite{BNP}
\begin{equation}
\label{C6VA}
C_{6,V/A} \,=\, \frac{32}{81}\,\p \a_s(m_\tau^2)\,
\langle\bar qq\rangle^2 \!\left(\!\!\begin{array}{c} 2\,\rho_1 -
9\,\rho_5 \\ 11\,\rho_1 \end{array}\!\!\right)\! \ ,
\end{equation}
with VSA results for $C_{6,V/A}$
corresponding to $\r_1=\r_5=1$.
Using $\langle\bar qq(m_\tau^2)\rangle=(-272\ \mbox{MeV})^3$ \cite{jam02},
and the averages of the results for $C_{6,V}$ and $C_{6,A}$
from the $s_{\rm min}=1.55$ and $1.575$ GeV$^2$ fits of Tab.~\ref{VAwtaupaper},
we find\footnote{We
neglected the smaller errors on $\a_s$ and $\langle\bar qq\rangle$.}
\begin{eqnarray}
\label{rho1rho5}
\rho_1 &\!\!=\!\!& \,-4 \pm 4 \,, \quad
\rho_5 \,=\, 5.9 \pm 0.9 \qquad \mbox{(FOPT)} \ , \\
\rho_1 &\!\!=\!\!& \,-2 \pm 3 \,, \quad
\rho_5 \,=\, 5.9 \pm 0.8 \qquad \mbox{(CIPT)} \ .\nonumber
\end{eqnarray}
While no conclusion can be drawn about the accuracy of
the VSA for $\rho_1$, it is clear that the VSA is
a poor approximation for $\rho_5$.
The value for $\r_5$ is consistent with the one we found from OPAL
data in Ref.~\cite{alphas2}.

It is conventional to characterize the size of
non-perturbative contributions to the ratio
$R_{V+A;ud}=R_{V;ud}+R_{A;ud}$ of the total non-strange hadronic
decay width to the electron decay width, where $R_{V/A;ud}$
have been defined in Eq.~(\ref{taukinspectral}), by the parametrization
\begin{equation}
\label{Rtau}
R_{V+A;ud}
=N_c S_{\rm EW}|V_{ud}|^2\left(1+\d_P+\d_{NP}\right)\ ,
\end{equation}
where $\d_P$ stands for the perturbative, and $\d_{NP}$ for the
non-perturbative contributions beyond the parton model.
If one knows $\d_{NP}$, the quantity $\d_P$, and hence $\a_s(m_\tau^2)$
can be determined from the experimental value of $R_{V+A;ud}$.
In such an approach, the error on $\a_s(m_\tau^2)$ is thus directly
correlated with that on $\d_{NP}$. As in
Ref.~\cite{alphas2}, our fits give access to the values of $\d_{NP}$,
as well as those of $\d^{(6)}$, $\d^{(8)}$, and $\d^{\rm DV}$,
the contributions to $\d_{NP}$ from the $D=6$ and $D=8$
terms in the OPE as well as the DV term.
From the $s_{\rm min}=1.55$~GeV$^2$ fits of Tab.~\ref{VAwtaupaper}, we find
\begin{eqnarray}
\label{deltas}
\d^{(6)}&=&0.058\pm 0.026   \ ,\qquad\quad\ \d^{(8)}=-0.036\pm 0.017   \ ,\\
\d_{\rm DV}&=&-0.0016\pm 0.0011    \ \qquad\mbox{(FOPT)}\ ,\nonumber\\
\null\nonumber\\
\d^{(6)}&=&0.040\pm 0.024    \ ,\qquad\quad\ \d^{(8)}=-0.024\pm 0.015   \ ,\nonumber\\
\d_{\rm DV}&=&-0.0009\pm 0.0009    \ \qquad\mbox{(CIPT)}\ .\nonumber
\end{eqnarray}
The FOPT and CIPT estimates for these quantities are consistent with each
other.    There is a strong correlation between $\d^{(6)}$ and $\d^{(8)}$, 
about $-0.97$ in the FOPT case.

The values for $\d^{NP}$ derived from these
results are
\begin{eqnarray}
\label{dNP}
\d^{NP}&=&0.020\pm 0.009\qquad\mbox{(FOPT)}\ ,\\
\d^{NP}&=&0.016\pm 0.010\qquad\mbox{(CIPT)}\ ,\nonumber
\end{eqnarray}
which differ by $1.6\ \s$, respectively, $1.2\ \s$ from the values
found using the the OPAL data
in Ref.~\cite{alphas2}. With the value $R_{V+A;ud}=3.475(11)$
quoted in Ref.~\cite{ALEPH13},
one finds $\d_P\approx 0.18$, an order of magnitude
larger than $\d_{NP}$,
indicating that $R_{V+A;ud}$}
is a dominantly
perturbative quantity. However, as in Ref.~\cite{alphas2}, we
find an error on $\d_{NP}$ much larger
than that reported by standard analyses
in the literature,
almost an order of magnitude so, for example,
when compared to
Ref.~\cite{ALEPH13}. The result is that the error on
$\a_s(m_\tau^2)$ is underestimated
in the standard analysis;
for further discussion, we again refer to
Sec.~\ref{ALEPH} below.

\subsection{\label{OPALresults}Comparison with the fits of Ref.~\cite{alphas2} to OPAL data}
A particularly interesting check is to look for consistency of
the results from our fits to the ALEPH
data with those we obtained by fitting the OPAL data in Ref.~\cite{alphas2}.
For the strong coupling, our results from OPAL data were
\begin{eqnarray}
\label{alphasOPAL}
\a_s(m_\tau^2)&=&0.325\pm 0.018\ ,\qquad(\overline{\mbox{MS}},\ n_f=3,
\ \mbox{FOPT,\ OPAL,\ Ref.~\cite{alphas2}})\ ,\\
&=&0.347\pm 0.025\ ,\qquad(\overline{\mbox{MS}},\ n_f=3,
\ \mbox{CIPT,\ OPAL,\ Ref.~\cite{alphas2}})\ .\nonumber
\end{eqnarray}
The values~(\ref{alphasfinal}) we find from the ALEPH data are $1.4$,
respectively, $1.3$ $\s$ lower than the OPAL values, assuming that
the errors on the ALEPH and OPAL values are independent.
We also note that the fits in Ref.~\cite{alphas2} were not entirely unambiguous;
a choice about the preferred range for $\d_V$ had to be made.
The fact that the difference between our central ALEPH-
and OPAL-based values, as well as that between our
central CIPT- and FOPT-based
results, is, in each case, comparable in size to the error obtained
in any of these analyses supports the notion that any improvement
in the precision with which $\a_s(m_\tau^2)$ can be determined from
hadronic $\tau$ decays will require significant improvements to
the data.
Of course, this assumes that the fit
\ansatz\ employed is valid in the region of $s_0$ larger than
about $1.5$~GeV$^2$.
We will return to this point in Sec.~\ref{ALEPH} below, as
well as in the Conclusion.

The coupling $\a_s(m_\tau^2)$ is, of course,
not the only fit parameter. One may for instance compare
the values of the OPE and DV parameters between
Tab.~\ref{VVwtaupaper} above and Tab.~4 of Ref.~\cite{alphas2}
for $s_{\rm min}\approx 1.5$~GeV$^2$, and conclude that they
agree between the ALEPH and OPAL fits within (sometimes fairly large) errors.
However, comparing
Tab.~\ref{VAwtaupaper} above with Tab.~5 of Ref.~\cite{alphas2},
one observes that the OPE and DV parameters for the axial channel
agree less well between the ALEPH and OPAL fits. This may be an
indication that it is safer to restrict our fits to the vector channel.
Results for $\a_s(m_\tau^2)$ are, nevertheless,
found to be consistent between pure-$V$
and combined $V$ and $A$ fits, both in this article
and in Ref.~\cite{alphas2}.

\subsection{\label{final} Final results for the strong coupling from ALEPH and OPAL data}

To conclude this section, we present our best values for the strong coupling
at the $\tau$ mass extracted from the ALEPH and OPAL data for hadronic $\tau$ decays,
and based on the assumptions that underlie our analysis.   The FOPT and CIPT
averages, weighted according to the errors in Eqs.~(\ref{alphasfinal}) and~(\ref{alphasOPAL}),
are
\begin{eqnarray}
\label{alphasALEPHOPAL}
\a_s(m_\tau^2)&=&0.303\pm 0.009\ ,\qquad(\overline{\mbox{MS}},\ n_f=3,
\ \mbox{FOPT,\ ALEPH\ \&\ OPAL})\ ,\\
&=&0.319\pm 0.012\ ,\qquad(\overline{\mbox{MS}},\ n_f=3,
\ \mbox{CIPT,\ ALEPH\ \&\ OPAL})\ .\nonumber
\end{eqnarray}
These convert to the values
\begin{eqnarray}
\label{alphasALEPHOPALZ}
\a_s(m_Z^2)&=&0.1165\pm 0.0012\ ,\qquad(\overline{\mbox{MS}},\ n_f=5,
\ \mbox{FOPT,\ ALEPH\ \&\ OPAL})\ ,\\
&=&0.1185\pm 0.0015\ ,\qquad(\overline{\mbox{MS}},\ n_f=5,
\ \mbox{CIPT,\ ALEPH\ \&\ OPAL})\ .\nonumber
\end{eqnarray}

\section{\label{ALEPH} The analysis of Ref.~\cite{ALEPH13}}
We now turn to a discussion of what we have referred to as the
standard analysis, which was used in Refs.~\cite{ALEPH13,ALEPH,ALEPH08,OPAL},
and is based on Ref.~\cite{DP1992}. We begin with a brief overview of what
is done in this approach.
One considers spectral moments with the weights
\begin{eqnarray}
\label{stweights}
w_{k\ell}(x)&=&(1-x)^2(1+2x)(1-x)^k x^\ell\ ,\\
x&=&s/s_0\ ,\nonumber
\end{eqnarray}
choosing $(k,\ell)\in\{(0,0),\,(1,0),\,(1,1),\,(1,2),\,(1,3)\}$,
and evaluating these moments at $s_0=m_\tau^2$ only.
Ignoring logarithms,\footnote{Which appear in subleading terms in $\a_s$
at each order in the OPE.} terms in the OPE contribute to these weights
up to $D=16$. The five $s_0=m_\tau^2$ moment values
are, of course, insufficient to determine the eight OPE parameters
$\alpha_s(m_\tau^2)$, $\langle {\frac{\a_s}{\p}} GG\rangle$, $C_6$, $C_8$,
$C_{10}$, $C_{12}$, $C_{14}$ and $C_{16}$, so some truncation is
necessary. The standard analysis approach to this problem is
to assume the OPE coefficients $C_{D=2k}$ for $D>8$ are
small enough that they may all be safely neglected in all of the
FESRs under consideration, despite numerical enhancements of their
contributions via larger coefficients in some of the higher degree
weights. Duality violations are, similarly,
assumed to be small enough that $\D(s)$ in
Eq.~(\ref{DVdef}) can be ignored as well, at least for $s_0$
close to $m_\tau^2$. With these assumptions, the remaining
OPE parameters $\a_s(m_\tau^2)$, $\langle {\frac{\a_s}{\p}} GG\rangle$,
$C_6$ and $C_8$ are fitted using the $s_0=m_\tau^2$ values of
the five $w_{k\ell}$ spectral moments noted above, for
each of the channels $V$, $A$, and $V+A$.
The central values and errors for $\a_s(m_\tau^2)$
are taken from the fits (FOPT and CIPT) to the $V+A$ channel,
based on the VSA-motivated expectation of
significant $D=6$ cancellation and the hope of similar strong
DV cancellations in the $V+A$ sum. However, as 
we have seen in Eq. (\ref{rho1rho5}), VSA is a rather poor approximation. 
Furthermore, the fact that the spectral function for the $V+A$ combination 
is flatter in the region between $2$ and $3$ GeV$^2$ than is the case for 
the $V$ or $A$ channels separately may 
mislead one into believing 
that DVs are already negligible at these scales for the $V+A$
combination. In actual fact, however, though somewhat reduced in the $V+A$ sum,
DV oscillations are still evident in the ALEPH $V+A$
distribution. In addition, since we have a good representation of the 
individual $V$ and $A$ channels, we also have a good representation of 
their sum. The fact that our fits yield results for $\gamma_A$ 
significantly larger than those for $\gamma_V$ implies that the 
level of reduction of DV contributions in going from the separate 
$V$ and $A$ channels to the $V+A$ sum is accidental in the window between
$2$ and $3$ GeV$^2$, and does not persist to
higher $s$, where the stronger exponential damping in the $A$ channel
would drive the result for the $V+A$ sum towards 
that for the $V$ channel alone.

These assumptions should be compared with those
that have to be made
in order to carry out the analysis presented in this article
(as well as in the OPAL-based analyses
of Refs.~\cite{alphas1,alphas2}). DVs are unambiguously
present in the spectral functions, as can be seen, for
example, in the relevant panels of
Figs.~\ref{CIFOw0fit}, \ref{CIFOw023fit}, \ref{CIFOw0VAfit} and
\ref{CIFOw023VAfit}. In the standard analysis, the hope is that the
double or triple pinching of the weights in
Eq.~(\ref{stweights}) is sufficient to allow DVs to be ignored
altogether, and indeed, for example Fig.~\ref{CIFOw023fit}, shows that
pinching significantly reduces the role of DV contributions,
especially near $s_0=m_\tau^2$. However, if, as in
the standard analysis, one restricts one's attention to $s_0=m_\tau^2$,
and wishes to employ only weights which are at least doubly pinched, the
number of OPE parameters to be fit will necessarily exceed the number
of weights employed, making additional assumptions, such as the
truncation in dimension of the OPE described above,
unavoidable.%
\footnote{For a detailed discussion of this
point, see Ref.~\cite{alphas1}.} With the standard-analysis
choice of the set
of weights of Eq.~(\ref{stweights}),
one finds that the OPE must be truncated at
dimension $D=8$ in order to leave at least one residual degree of
freedom in the fits.
In our analysis, in contrast, we choose not to ignore
DVs {\it a priori}. This requires us to model their contribution to the
spectral functions (as we did through Eq.~(\ref{ansatz})), and to use
not just the single value $s_0=m_\tau^2$, but rather a range of $s_0$
extending down from $m_\tau^2$. The one assumption we {\it do} have
to make is that
the \ansatz~(\ref{ansatz}) provides a sufficiently accurate description of DVs
for values of $s_0$ between approximately $1.5$~GeV$^2$ and $m_\tau^2$.

Clearly, whatever choice is made, it needs to be tested.
For our analysis framework, we have presented detailed tests already
above. In this section we consider primarily the standard
analysis, most recently used in Ref.~\cite{ALEPH13}. Our conclusion,
from what follows below, is that the assumptions made in this framework
do not hold up to
quantitive scrutiny,
and hence that the standard analysis approach
should no longer be employed in future analyses.%
\footnote{We point out that the inadequacy of the
standard analysis framework
was already demonstrated in Refs.~\cite{MY08,CGP,alphas1,alphas2}, but it
appears important to re-emphasize this point
in view of the continued use of this framework in the literature, in
particular in the updated analysis of Ref.~\cite{ALEPH13}.}

The results presented in Tab.~4 of Ref.~\cite{ALEPH13} already indicate that
there are problems with the standard analysis.
Let us consider the values obtained for the gluon
condensate, $\langle{\frac{\a_s}{\p}} GG\rangle$, in the different
channels, together with the $\chi^2$ value for each fit (recall that for
each of these fits there is only one degree of freedom):
\begin{eqnarray}
\label{gluon}
\langle\frac{\a_s}{\p}GG\rangle&=&(-0.5\pm 0.3)\times
10^{-2}~\mbox{GeV}^4\ ,\qquad \chi^2=0.43\qquad V\ ,\\
&&(-3.4\pm 0.4)\times 10^{-2}~\mbox{GeV}^4\ ,\qquad
\chi^2=3.4~\qquad A\ ,\nonumber\\
&&(-2.0\pm 0.3)\times 10^{-2}~\mbox{GeV}^4\ ,\qquad
\chi^2=1.1~\qquad V+A\ .\nonumber
\end{eqnarray}
The $\chi^2$ values correspond to $p$-values of 51\%, 7\%, and 29\%,
respectively, indicating that all fits are
acceptable. For these fits to be taken as meaningful,
however, their results should satisfy known physical constraints.
One such constraint is that there is only one effective gluon condensate,
whose values should therefore come out the same in all of the
$V$, $A$ and $V+A$ channels. This is rather far from the case
for the results quoted in Eq.~(\ref{gluon}), where, for example,
the $V$ and $V+A$ channel fit values differ very
significantly.
It is, moreover, problematic to accept the
$V+A$ channel value and ignore the $V$ channel one when the
$p$-value of the $V$-channel fit is, in fact, larger than that
of the $V+A$ channel.

There can be several reasons for the inconsistencies in
the results of Ref.~\cite{ALEPH13}.
One possibility is that some of the weights~(\ref{stweights})
have theoretical problems already in perturbation
theory, as argued in Ref.~\cite{BBJ12}. Another possibility is that
the assumptions underlying the standard analysis do not hold.
Whatever the reason, the discrepant gluon
condensate values point to a serious problem with the
standard analysis framework.\footnote{This problem
already existed in earlier ALEPH
analyses \cite{ALEPH,ALEPH08}, but in principle it might have
been due to the problem with the data itself. Note that OPAL
enforced equality
of the gluon condensate between various channels, and were
able to obtain reasonable fits as judged by the $\chi^2$,
possibly because of the larger data errors.}

We now turn to quantitative tests of the OPE fit results
reported in Ref.~\cite{ALEPH13}. We focus on the $V+A$ channel,
where DVs and $D>4$ OPE contributions were expected to
play a reduced role, and on the CIPT $D=0$ treatment, since
this is the only case for which the OPE fit parameter values
are quoted in Ref.~\cite{ALEPH13}. The tests consist of comparing
the weighted OPE and spectral integrals for the weights
$w_{k\ell}$ employed in the analysis of Ref.~\cite{ALEPH13}, not
just at $s_0=m_\tau^2$, but over an interval of $s_0$ extending
below $m_\tau^2$. If the assumptions made about $D>8$ OPE and
DV contributions being negligible are valid at $s_0=m_\tau^2$
they should also be valid in some interval below this point.
A good match between the weighted spectral integrals and the
corresponding OPE integrals, evaluated using the results
for the OPE parameters quoted in Ref.~\cite{ALEPH13}, should thus
be found over an interval of $s_0$. If, on the other hand,
these assumptions are not valid, then the fit parameter values
will contain contaminations from DV contributions and/or contributions
with higher $D$, both of which scale differently with $s_0$
than do the $D=0$, $4$, $6$ and $8$ contributions appearing in
the truncated OPE form. Such contamination will show up as
a disagreement between the $s_0$ dependence of the fitted
OPE representations and the experimental spectral integrals.

It is worth expanding somewhat on this latter point since 
the agreement of the OPE and spectral integrals at $s_0=m_\tau^2$ for 
the weights $w_{k\ell}$ employed in the standard analysis is sometimes 
mistakenly interpreted as suggesting the validity of the assumptions 
underlying the standard analysis at $s_0=m_\tau^2$. However, while the 
agreement is certainly a necessary condition for the validity of these 
assumptions, it is not in general, a sufficient one. This caution is 
particularly relevant since four parameters are being fit using 
only five data points, making it relatively easy for the effects of 
neglected, but in fact non-negligible, higher-$D$ and/or DV contributions 
to be absorbed, {\it at a fixed} $s_0$, into the values of the four fitted 
lower-$D$ parameters. That this is a realistic possibility is demonstrated 
by the alternate set of OPE fit parameters obtained in the analysis of 
Ref.~\cite{MY08}, which neglected DV contributions, but not OPE 
contributions with $D>8$. The results of this fit, including non-zero 
$C_D$ with $D>8$ and an $\alpha_s(m_\tau^2)$ significantly different 
from that obtained via the standard analysis of the same data~\cite{ALEPH08}, 
produced equally good agreement between the $s_0=m_\tau^2$ 
OPE and spectral integral results for all the $w_{k\ell}$ employed in
the standard analysis fit of Ref.~\cite{ALEPH08}, conclusively demonstrating 
that this agreement does not establish the validity of the standard analysis 
assumptions. So long as one works at fixed $s_0=m_\tau^2$, there is no way to 
determine whether the results of the standard analysis are, in fact, 
contaminated by neglected higher-$D$ OPE and/or DV effects or not. 
One may, however, take advantage of the fact
that different contributions to the theory sides of the various FESR 
scale differently with $s_0$, with integrated DV contributions 
oscillatory in $s_0$ and integrated $D=2k$ OPE contributions scaling 
as $1/s_0^k$. If the $D=0$, $4$, $6$ and $8$ parameters obtained
from the fixed-$s_0=m_\tau^2$ standard analysis fit have, in fact, absorbed
the effects of $D>8$ and/or DV contributions, the fact that the nominal 
lower-$D$ $s_0$-scaling does not properly match that of the higher-$D$ and/or DV 
contaminations will be exposed when one considers the same FESR, with the 
same standard analysis OPE fit parameter values,
at lower $s_0$. A breakdown of the standard analysis assumptions
will thus be demonstrated by a failure of the agreement of the OPE 
and spectral integrals observed at $s_0=m_\tau^2$ to persist over a 
range of $s_0$ below $m_\tau^2$. Such $s_0$-dependence tests represent 
important self-consistency checks for all FESR analyses.

Before carrying out these self-consistency
tests on the results of the standard analysis, it is useful to make 
explicit the relative roles of the various different $D$ contributions 
entering the $s_0=m_\tau^2$ results for the $w_{k\ell}$-weighted OPE 
integrals employed in the $V+A$ CIPT fit of Ref.~\cite{ALEPH13}.
For the $D=0$ contributions, it is important to remember that
the leading one-loop contribution is independent of both $s_0$ and $\a_s$.
It is thus the difference of the full $D=0$ contribution and
this leading term which determines the $\alpha_s$ dependence of
the $D=0$ contributions, and which is relevant to the determination
of $\alpha_s(m_\tau^2)$. Tab.~\ref{aleph13ope} shows the
$s_0=m_\tau^2$ results for (i) the $\alpha_s$-dependent $D=0$ contributions
and (ii) the $D=4$, $6$, and $8$ contributions corresponding to the CIPT
fit results of Tab.~4 of Ref.~\cite{ALEPH13}, for each of the
$w_{k\ell}$ employed in that analysis.
The sum of the
$D=6$ and $8$ contributions, which is $\sim 1-2\%$ of the
$\a_s$-dependent $D=0$ contribution for $w_{00}$ and
$w_{10}$, is, in contrast, $\sim 10-25\%$ of the corresponding $D=0$
contributions for the $w_{11}$, $w_{12}$ and $w_{13}$ cases.
Furthermore, for
$w_{11}$, the $D=4$ contribution
is essentially the same size as the $\alpha_s$-dependent $D=0$
one.

\begin{table}[h]
\begin{center}
\vspace{2ex}
\begin{tabular}{|c|c|c|c|c|}
\hline
$(k,\ell )$&$\alpha_s$-dependent $D=0$&$D=4$&$D=6$&$D=8$\\
\hline
$(0,0)$&0.005173&-0.000008&-0.000117&\ 0.000033\\
$(1,0)$&0.004399&-0.000361&-0.000117&\ 0.000082\\
$(1,1)$&0.000365&\ 0.000350&-0.000039&-0.000049\\
$(1,2)$&0.000208&\ 0.000002&\ 0.000039&-0.000016\\
$(1,3)$&0.000081&\ 0.000000&\ 0.000000&\ 0.000016\\
\hline
\end{tabular}
\end{center}
\begin{quotation}
\floatcaption{aleph13ope}{{\it The $D=4$,
$6$ and $8$ and $\a_s$-dependent $D=0$
contributions to the $s_0=m_\tau^2$, $V+A$, $w_{k\ell}$
moments corresponding to the $V+A$ OPE fit parameter results of Tab.~4
of Ref.~\cite{ALEPH13}.}}
\end{quotation}
\vspace*{-4ex}
\end{table}%

It is clear from these observations that it is the $w_{11}$, $w_{12}$
and $w_{13}$ moments which dominate the determinations of
the $D=4,\, 6$ and $8$ OPE parameters in the analysis of Ref.~\cite{ALEPH13}.
Bearing in mind the very slow variation with $s_0$ of the $D=0$ 
contributions to the dimensionless OPE integrals and the $1/s_0^k$
scaling of the $D=2k$ contributions, it is, moreover, 
clear that the relative roles of the non-perturbative contributions will 
grow significantly relative to the $\alpha_s$-dependent $D=0$ ones as $s_0$ 
is decreased. Studying the $s_0$ dependence of the match of the OPE to the
corresponding spectral integrals for the $w_{11}$, $w_{12}$ and $w_{13}$
spectral weights thus provides a particularly
powerful test of the reliability of the values for the $D=4,\, 6$ and $8$ 
parameters obtained in the fits of Ref.~\cite{ALEPH13}.

The results of these tests are shown in Fig.~\ref{aleph13fitprobs}.
It is clear that the $s_0$-dependence of the experimental
spectral integrals and fitted OPE integrals is very different,
demonstrating conclusively the unreliability of the $D=4,\, 6$ and $8$
fit parameter values obtained in Ref.~\cite{ALEPH13}. 
Changes in the values of the $D=6$ and $8$ parameters, which
enter the $w_{00}$ FESR, would of course also force a change
in the $\alpha_s{(m_\tau^2)}$ required to produce a match
between the $s_0=m_\tau^2$ $w_{00}$-weighted OPE and spectral
integrals.

\begin{figure}[t]
\begin{center}
\includegraphics*[width=7cm]{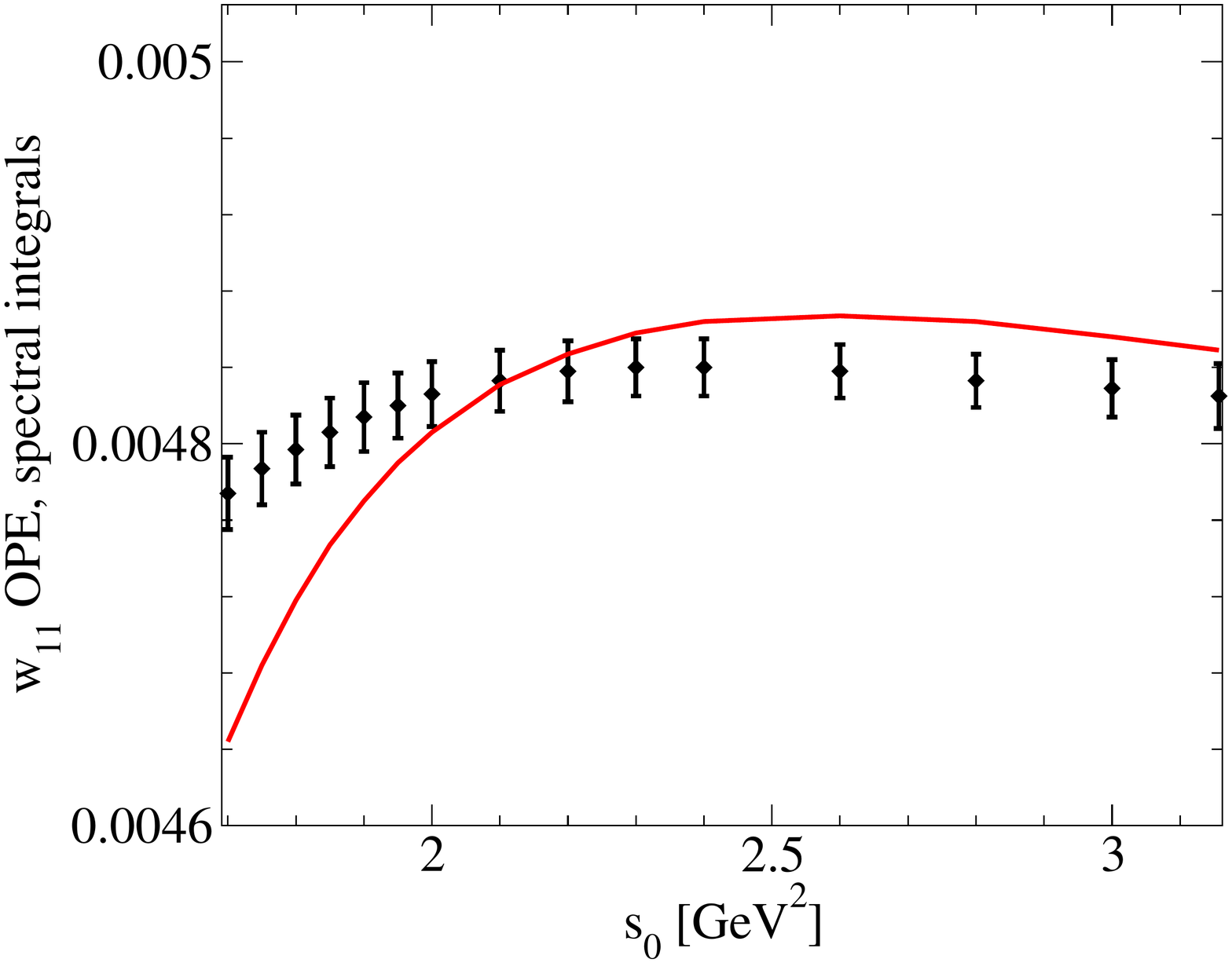}
\hspace{0.5cm}
\includegraphics*[width=7cm]{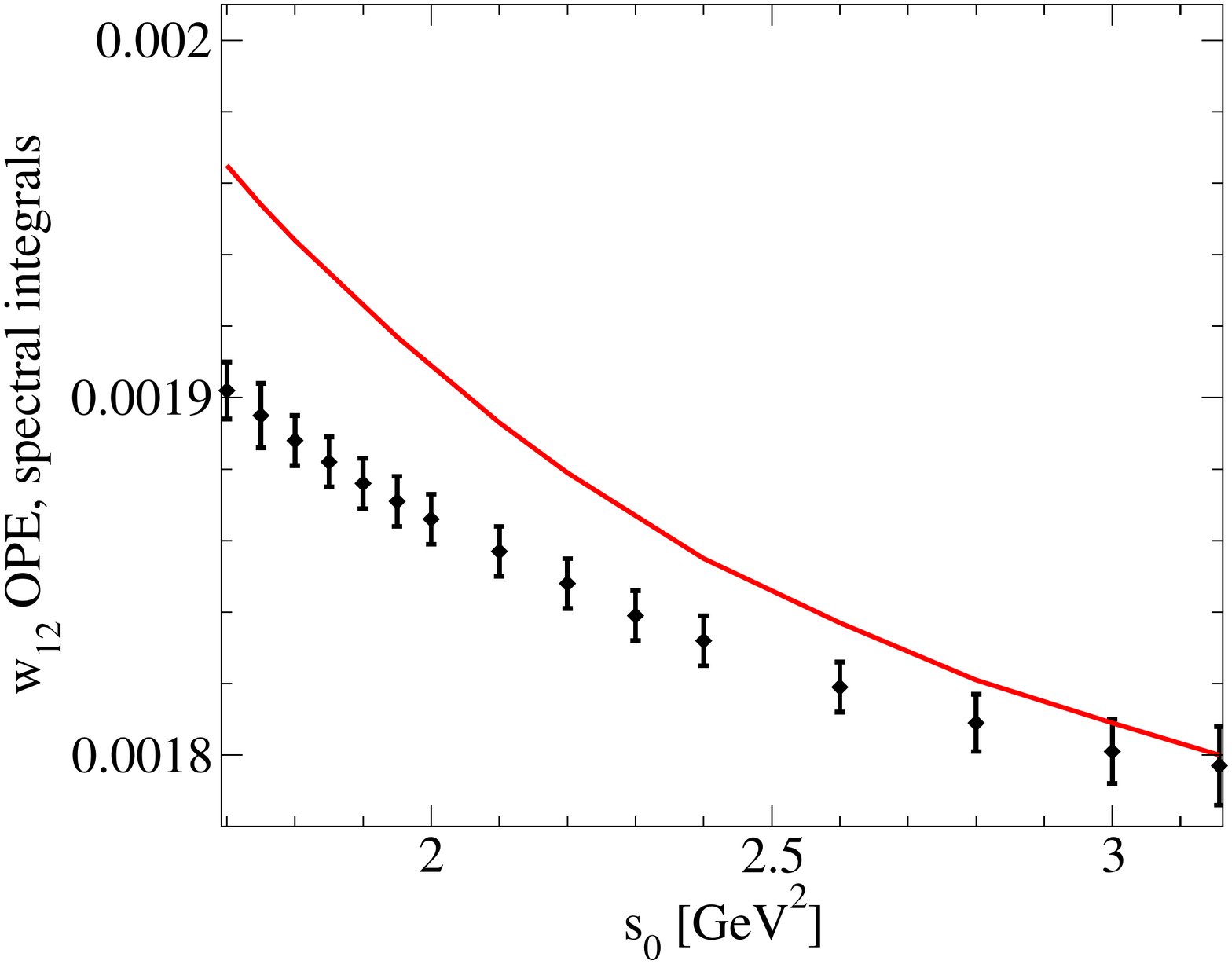}
\vspace{0.5cm}
\includegraphics*[width=7cm]{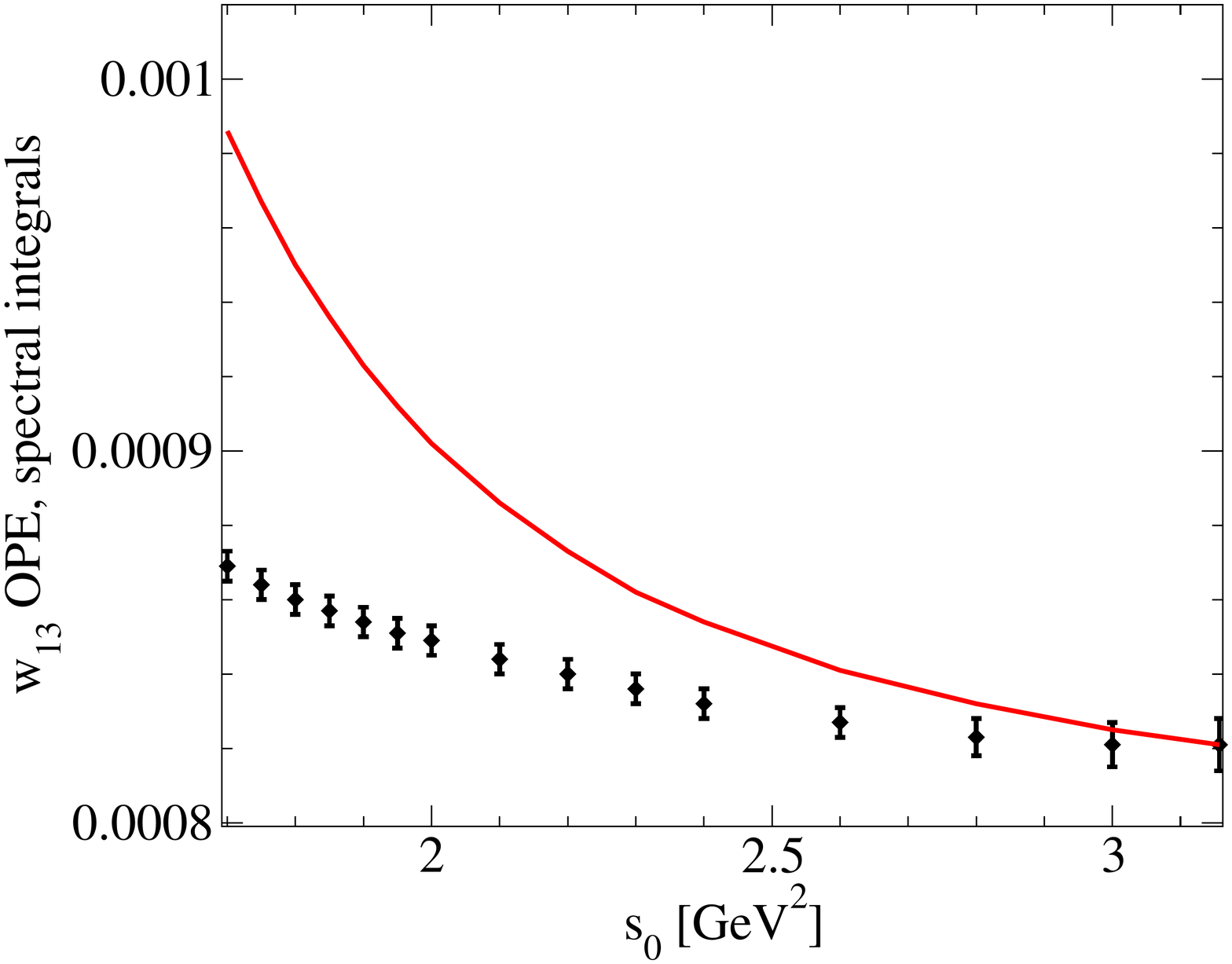}
\end{center}
\vspace*{-6ex}
\begin{quotation}
\floatcaption{aleph13fitprobs}{\it Comparisons of the $s_0$ dependence of the
$w_{k\ell}=w_{11}$, $w_{12}$ and $w_{13}$ $V+A$ spectral integrals to that
of the corresponding OPE integrals evaluated employing as input the results
of the CIPT fit for the OPE parameters from Tab.~4 of Ref.~{\rm \cite{ALEPH13}}.
}
\end{quotation}
\vspace*{-4ex}
\end{figure}

It is worth expanding somewhat on these observations
for the $w_{13}$ case, where the source of the problem with the
fit of Ref.~\cite{ALEPH13} becomes particularly obvious. Because of
the $x^3$ factor present in $w_{13}(x)$, the $D=2$ and $4$ contributions
to the OPE part are completely negligible numerically, leaving the
standard analysis version of the $w_{13}$-weighted OPE integral
entirely determined by the parameters $\alpha_s(m_\tau^2)$ and
$C_{8,V+A}$. With the results and errors for these quantities
from Tabs.~4 and 5 of Ref.~\cite{ALEPH13}, one finds that,
as $s_0$ is decreased from $m_\tau^2$ to \eg\ $2$ GeV$^2$,
the $\alpha_s$-dependent
$D=0$ contribution {\it decreases} by $0.000001(0)$, while the
$D=8$ contribution {\it increases} by $0.000086(20)$. This is to be
compared to the {\it increase} in the corresponding spectral
integral, which is $0.000028(8)$. 
Evidently the disagreement between the $w_{13}$-weighted OPE and spectral 
integral results seen in Fig.~\ref{aleph13fitprobs} results from a problem
with the fit value for $C_{8,V+A}$. Trying to 
fix the problem with the $w_{13}$ FESR through a change in $C_{8,V+A}$ alone
turns out to exacerbate the problem with the $w_{12}$ FESR. Working
backward, one finds that attempting to change $C_{4,V+A}$, $C_{6,V+A}$ and 
$C_{8,V+A}$ so as to improve the match between the $s_0$ dependences
of the OPE and spectral integrals for the $w_{11}$, $w_{12}$ and $w_{13}$
FESRs without any change in $\alpha_s(m_\tau^2)$ produces
changes in the $D\ge 4$ contributions to the $w_{10}$ and $w_{00}$ FESRs that
can only be compensated for by a decrease in $\alpha_s(m_\tau^2)$.
The problem of the discrepancies between the $s_0$-dependences
of the OPE and spectral integrals in the $w_{11}$, $w_{12}$ and $w_{13}$ 
FESR parts of the standard analysis can thus not be resolved simply
through shifts in $C_{4,V+A}$, $C_{6,V+A}$ and $C_{6,V+A}$ which leave the 
target of the analysis, namely the output $\alpha_s(m_\tau^2)$
value, unchanged.

A natural question, given the discussion above, is whether our
approach produces a better match between experiment and theory
for the higher spectral weights.  The answer, as we will see below, is yes. Before
embarking on this investigation, however, it is important to emphasize the
non-optimal nature of the  FESRs with weights $w_{10}$, $w_{11}$,
$w_{12}$, and $w_{13}$. First, all of these weights contain a term linear in
the variable $x$, a fact which, according to the arguments of Ref.~\cite{BBJ12},
should make standard methods of estimating the
uncertainty associated with truncating the integrated perturbative
series for these weights much less reliable than is the case for the weights
employed in our analysis.
Second, the values of the $C_D$ with $D>8$  obtained
from the fits reported in Ref.~\cite{MY08} were found to produce very strong
cancellations amongst higher-$D$ OPE contributions when employed in the
higher $(k,\ell )$ $w_{k\ell}$ FESRs, making these FESRs particularly
sensitive to any shortcomings in the treatment of higher-$D$ OPE
contributions, as well as a poor choice for use in attempting to fit
the values of $C_D$ with $D>8$. The strong cancellation
amongst higher-$D$ OPE contributions for the higher-$(k,\ell )$
$w_{k\ell}$ moments turns out to be also a feature of the results
of our extended analysis below, and hence not attributable
 to the neglect of DV contributions in Ref.~\cite{MY08}.
Because of these strong cancellations, the use of the higher-$(k,\ell )$
$w_{k\ell}$ should be avoided in future analyses, and
we consider them below only for the sake of comparison with the
results of the analysis of Ref.~\cite{ALEPH13}. In making this
comparison, we will focus on the CIPT resummation of perturbation theory,
with the CIPT version of the standard analysis being the only one for
which quantitative fit results are reported in Ref.~\cite{ALEPH13}.

To evaluate the OPE contributions to the $w_{10}$, $w_{11}$, $w_{12}$
and $w_{13}$ FESRs requires knowledge of five new quantities,
$C_{4,V+A}$, $C_{10,V+A}$, $C_{12,V+A}$, $C_{14,V+A}$ and $C_{16,V+A}$, in
addition to the OPE and DV parameters already obtained in our analysis.
We estimate these using the $w(s)=(s/s_0)^{k-1}$ versions
of the FESR Eq.~(\ref{sumrule}), neglecting, as before, sub-leading contributions
at each order $D>2$ in the OPE. This yields, for $D=2k>2$,
\begin{eqnarray}
\label{OPEnFESR}
(-1)^{k+1} C_{2k,V+A}
&=&2f_\p^2 m_\p^{2(k-1)}+\int_0^{s_0}ds\,s^{k-1}\,\r^{(1)}_{V+A}(s)\\
&&+\int_{s_0}^\infty ds\,s^{k-1}\,\r_{V+A}^{\rm DV}(s)
+\frac{1}{2\p i}\oint_{|z|=s_0}dz\,z^{k-1}\,\P_{V+A}^{PT}(z)\ ,\nonumber
\end{eqnarray}
where $\P^{PT}$ is the perturbative contribution to $\P(z)$,
corresponding to the $D=0$ term in Eq.~(\ref{OPE}). The choices
$k=2,\, \cdots ,\, 8$ yield $C_4,\, \cdots ,\, C_{16}$, respectively.
With $\alpha_s(m_\tau^2)$ and the $V$ and $A$ channel DV parameters
from the $s_{\rm min}=1.55$ GeV$^2$ combined $V$ and $A$ CIPT fit
of Tab.~\ref{VAwtaupaper}, we find, for the central values,
\begin{eqnarray}
\label{OPEn}
C_{4,V+A}&= &0.00268\ {\rm GeV}^4\ ,\\
C_{6,V+A}&= &-0.0125\ {\rm GeV}^6\ , \nonumber\\
C_{8,V+A}&= &0.0349\ {\rm GeV}^8\ , \nonumber\\
C_{10,V+A}&= &-0.0832\ {\rm GeV}^{10}\ , \nonumber\\
C_{12,V+A}&= &0.161\ {\rm GeV}^{12}\ ,\nonumber\\
C_{14,V+A}&= &-0.191\ {\rm GeV}^{14}\ ,\nonumber\\
C_{16,V+A}&= &-0.233\ {\rm GeV}^{16}\ .\nonumber
\end{eqnarray}
For $C_{6,V+A}$ and $C_{8,V+A}$ the agreement with the values
in Tab.~\ref{VAwtaupaper} is excellent. 
With such values of the $C_D$, $D>8$ contributions
are far from negligible compared to the $D=6$ and $8$ ones for the
$w_{k\ell}$ spectral weights with degree higher than three; the maximum
scale, $m_\tau^2$, accessible in hadronic $\tau$ decays is not,
it turns out, high enough to ensure that the OPE series is rapidly
converging in dimension.

The theory parts $I_{\rm th}^{(w_{k\ell})}(s_0)$ of the $w_{10}$, $w_{11}$,
$w_{12}$ and $w_{13}$ FESRs
produced by the results of Eq.~(\ref{OPEn}) and Tab.~\ref{VAwtaupaper}
are compared to the corresponding spectral integrals in
Fig.~\ref{w10w11theoryspecintcomp} as a function of $s_0$.
The agreement is
obviously excellent, and far superior to that obtained
from the standard analysis of Ref.~\cite{ALEPH13}.   
This excellent agreement, over the
whole range of $s_0$ shown, is completely destroyed if one removes
the $D>8$ contributions from the theory sides of the $w_{10}$,
$w_{11}$, $w_{12}$ and $w_{13}$ FESRs.
We emphasize again that the
aim here is not a reliable determination of the OPE coefficients $C_{4-16}$, but
a proof of existence of a set of values which, combined with our values for
$\a_s(m_\tau^2)$ and the DV parameters, give an excellent representation of the
$s_0$ dependence of the moments with the weights $w_{10}$, $w_{11}$,
$w_{12}$ and $w_{13}$ (in addition, of course, to the weights included in
our fits, in particular $w_{00}=\hw_3$).

\begin{figure}[t!]
\begin{center}
\includegraphics*[width=7cm]{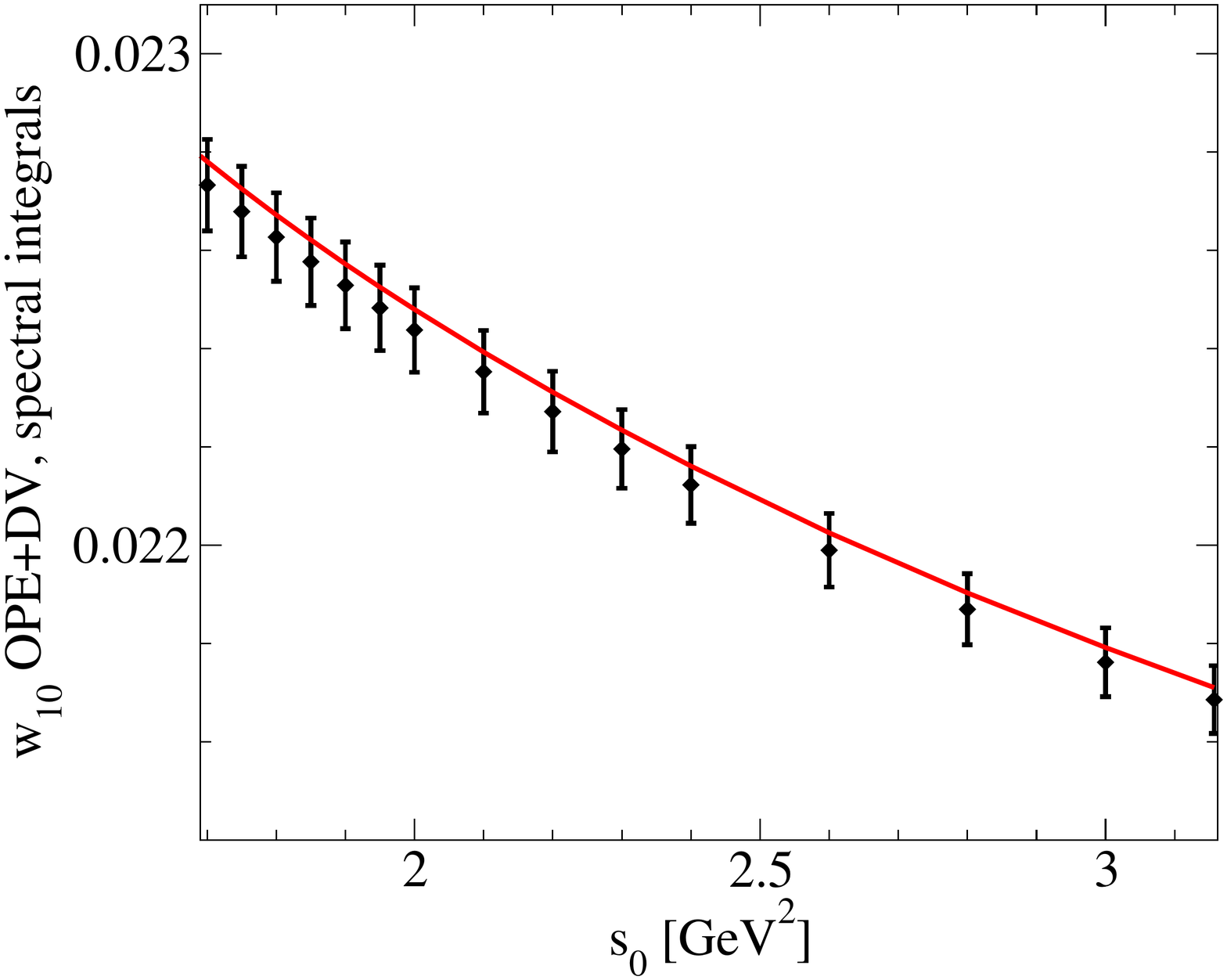}
\hspace{0.5cm}
\includegraphics*[width=7cm]{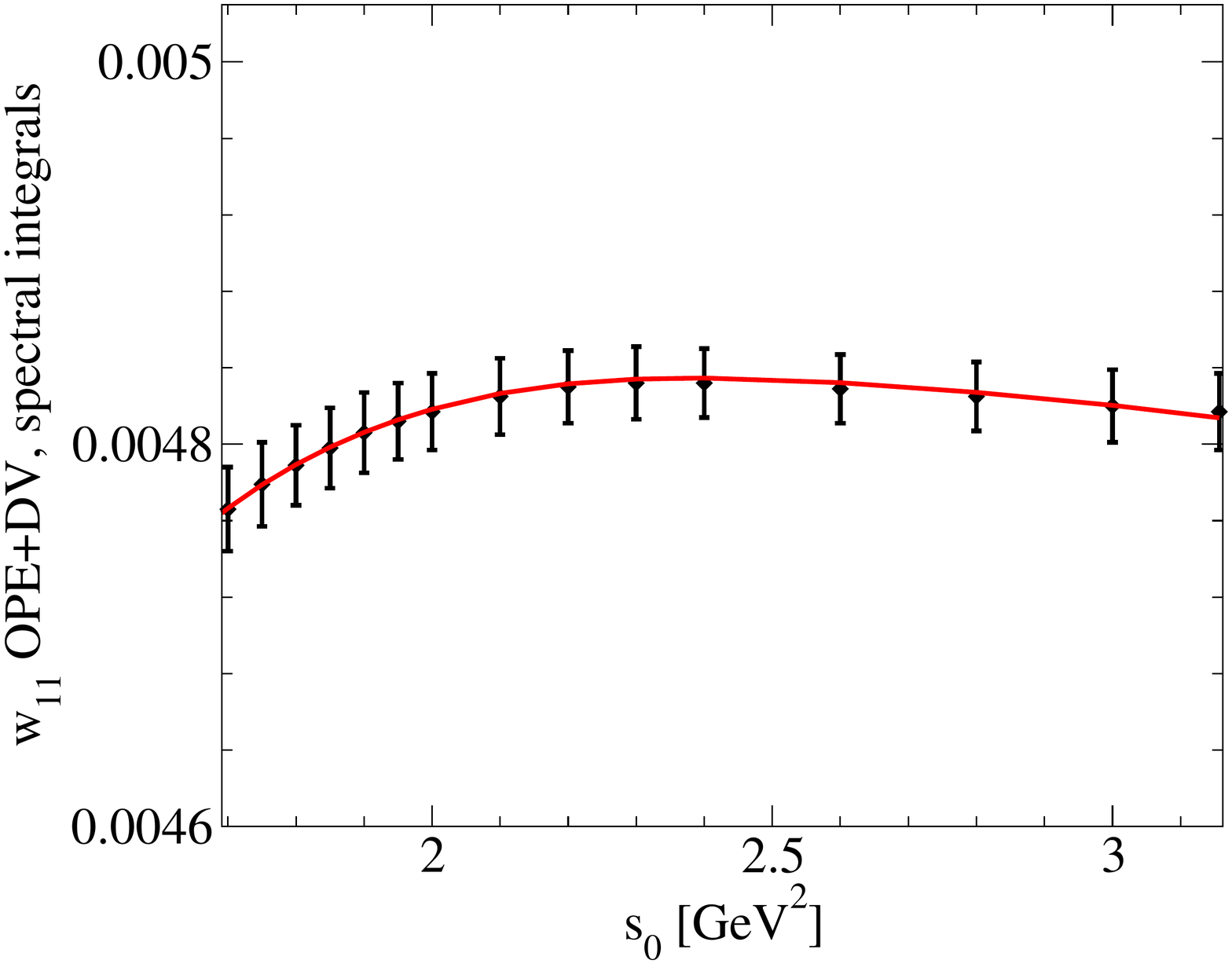}
\vspace{0.5cm}
\includegraphics*[width=7cm]{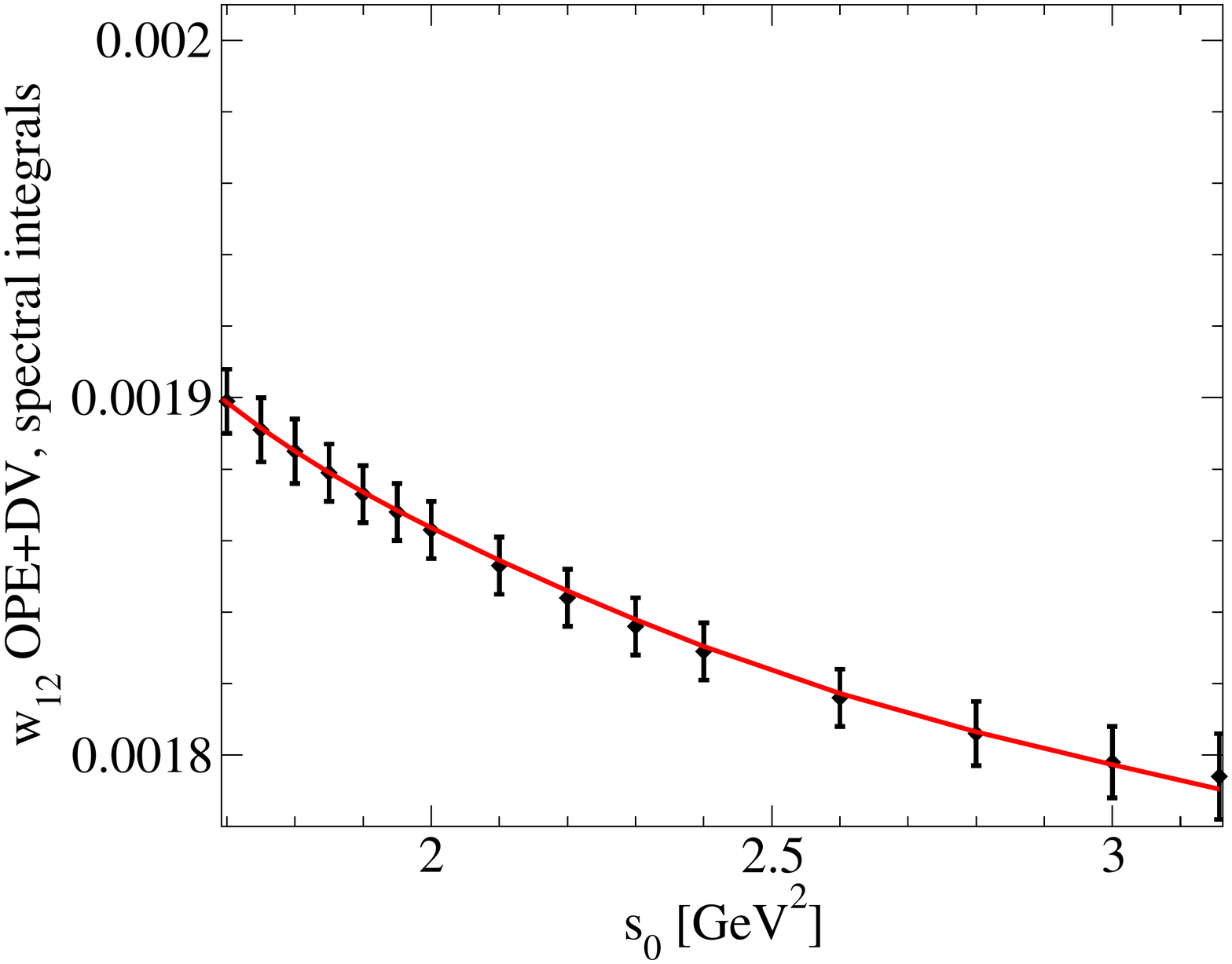}
\hspace{0.5cm}
\includegraphics*[width=7cm]{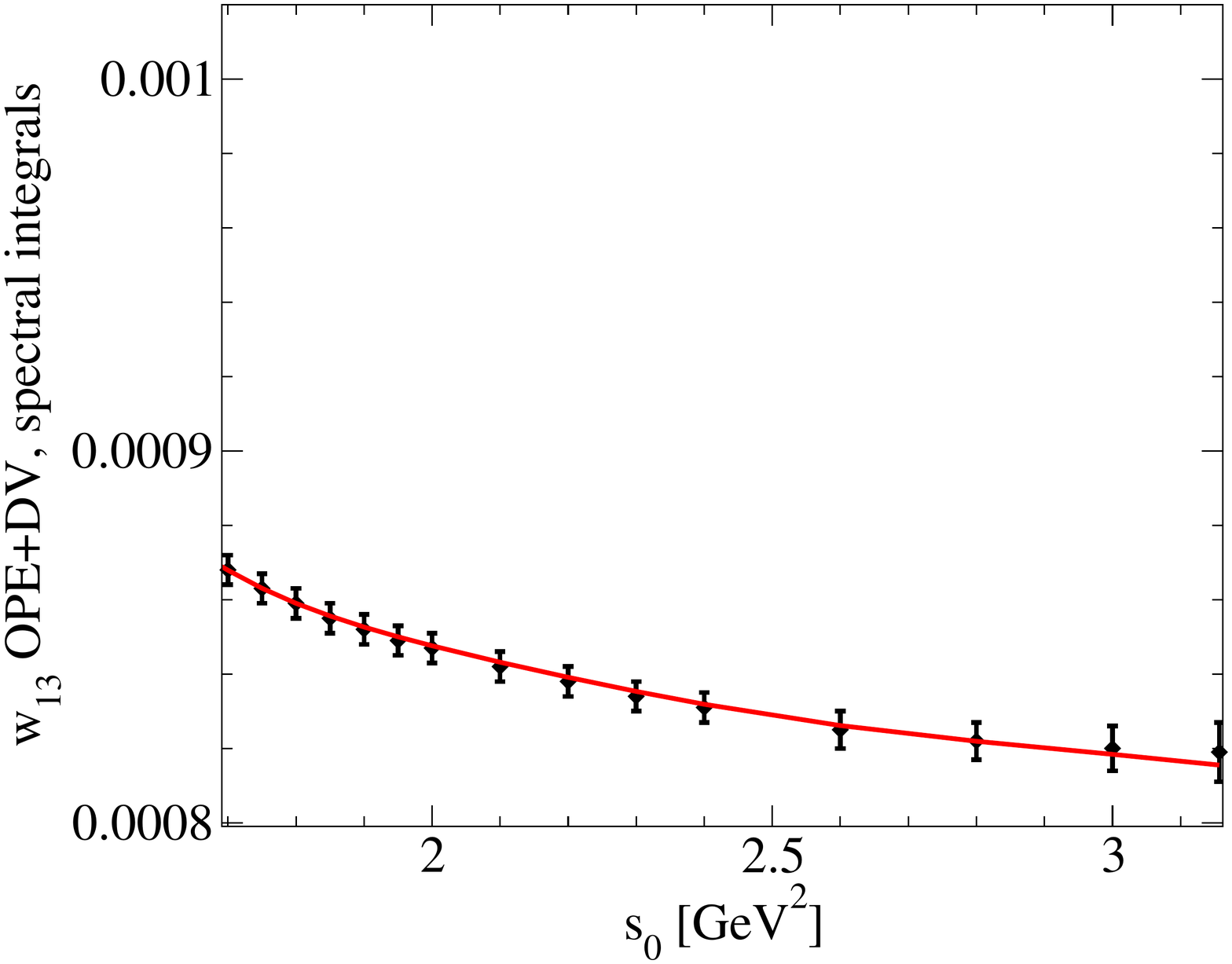}
\end{center}
\begin{quotation}
\floatcaption{w10w11theoryspecintcomp}{\it Comparison of
$I_{\rm th}^{(w_{k\ell})}(s_0)$ (\seef\ Eq.~(\ref{sumrule}) for $w_{k\ell}=w_{10}$,
$w_{11}$, $w_{12}$ and $w_{13}$ with $I_{\rm ex}^{(w_{k\ell})}(s_0)$
for the $V+A$ channel using
the results of the $s_{\rm min}=1.55$ GeV$^2$, combined $V$ and $A$
CIPT fit of Tab.~\ref{VAwtaupaper} and Eq.~(\ref{OPEn}).
Top left panel:
the $w_{10}$ case; top right panel: the $w_{11}$ case,
bottom left panel: the $w_{12}$ case; bottom right panel:
the $w_{13}$ case.
}
\end{quotation}
\vspace*{-4ex}
\end{figure}

The problems demonstrated above with the
standard analysis results of Ref.~\cite{ALEPH13} could be a consequence
of the neglect of non-negligible DVs, the breakdown of
the assumption that $D>8$ OPE contributions are
negligible for all of the $w_{k\ell}$ employed, or both.
In an attempt to clarify the situation, it is useful to consider a fit
in which the potentially dangerous assumption about $D>8$ OPE
contributions is avoided. As an example, we consider a fit
to the doubly pinched $\hat{w}_3=w_{00}$ FESR in the $V+A$ channel
ignoring DV contributions. Since the weight is doubly pinched,
one expects DV contributions to be significantly suppressed,
though the actual amount of suppression is not clear {\it a priori}.
Since the OPE integrals
still depend on three parameters, $\alpha_s(m_\tau^2)$, $C_{6,V+A}$
and $C_{8,V+A}$, it is, of course, necessary to consider the
fit over a range of $s_0$. To be specific, we focus on fits
employing the FOPT resummation of perturbation theory.
This exercise results in apparently perfectly acceptable fits,
with $p$-values $10\%$ and higher
for $s_{\rm min}\ge 1.95$~GeV$^2$. The fit quality drops dramatically as
$s_0$ is lowered beyond this point, with $p$-values already at
the $0.2\%$ level for $s_{\rm min}=1.90$~GeV$^2$. The highest $p$-value,
$57\%$, occurs for $s_{\rm min}=2.2$~GeV$^2$, and corresponds~to
\begin{eqnarray}
\label{w00noDVsfitvals}
\alpha_s(m_\tau^2)&=&0.330\pm 0.006\ ,\\
C_{6,V+A}&=&0.0070\pm 0.0022\ {\rm GeV}^6\ ,\nonumber\\
C_{8,V+A}&=&-0.0088\pm 0.0042\ {\rm GeV}^8\ .\nonumber
\end{eqnarray}

\begin{figure}[t]
\begin{center}
\includegraphics*[width=7cm]{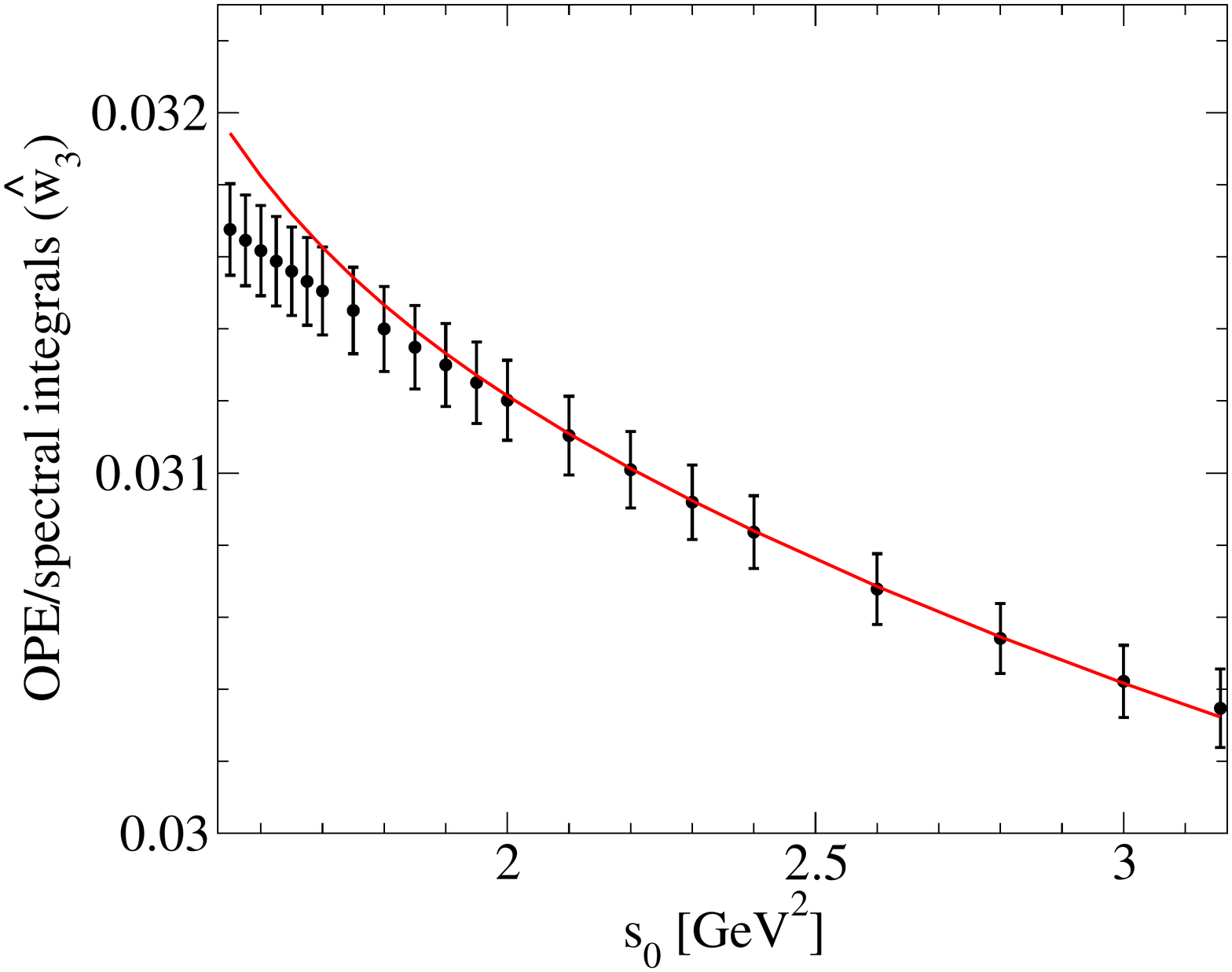}
\hspace{0.0cm}
\includegraphics*[width=7cm]{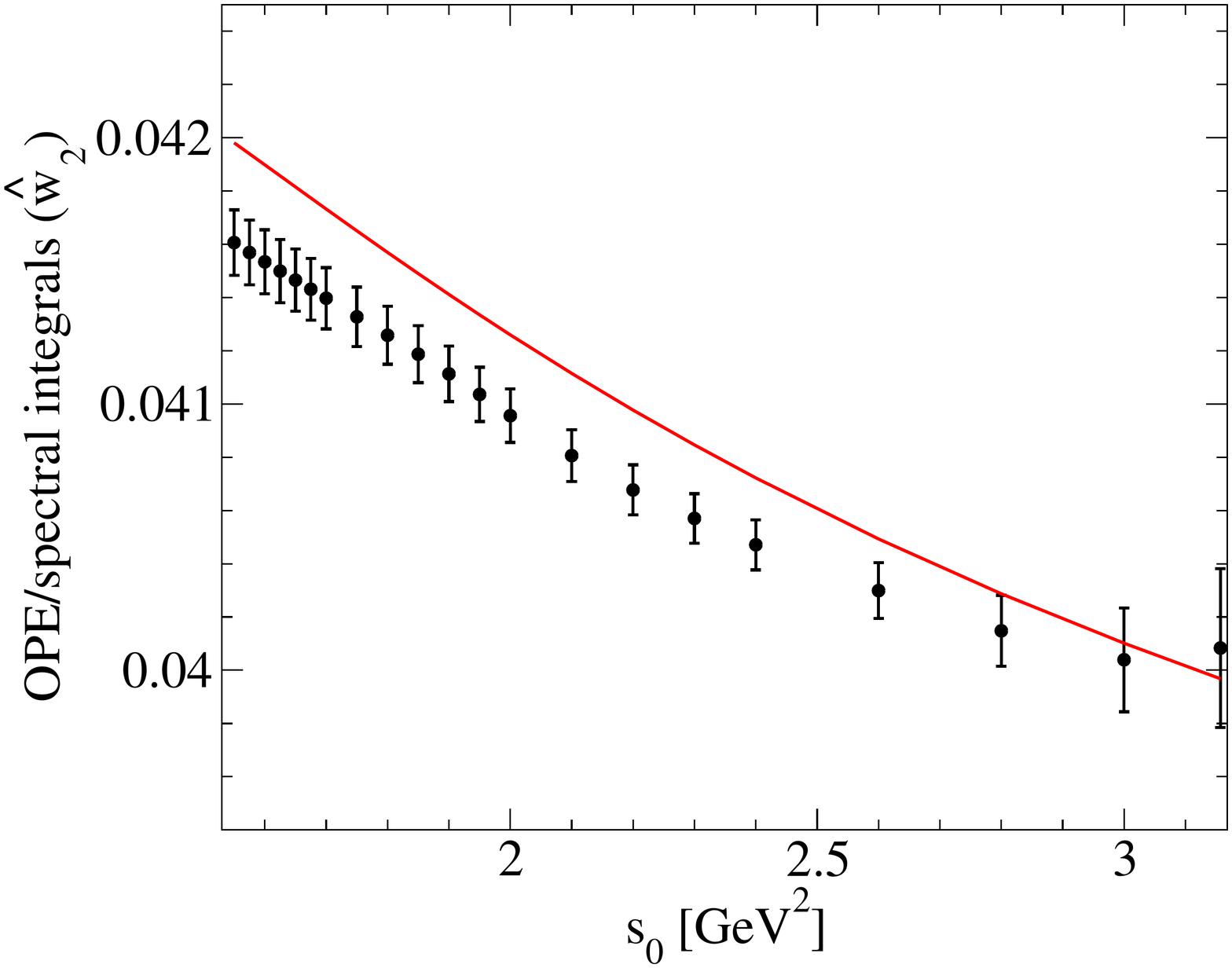}
\end{center}
\begin{quotation}
\floatcaption{w00noDVscomp}{\it Comparison of the
$\hw_3=w_{00}$-weighted spectral integrals (left panel) and
$\hw_2$-weighted spectral integrals (right panel) with the corresponding
OPE integrals evaluated using the results of the no-DV fit
given in Eq.~(\ref{w00noDVsfitvals}).}
\end{quotation}
\vspace*{-4ex}
\end{figure}

The quality of the resulting match between
the fitted OPE and spectral integrals for $s_{\rm min}=2.2$ GeV$^2$, shown
in the left panel of Fig.~\ref{w00noDVscomp}, is  excellent. Despite this good
quality match, the results of Eq.~(\ref{w00noDVsfitvals}) are
incomplete, in the sense that, in addition to the fit error induced by
the covariances of the $V+A$ spectral data, there is an unspecified
(and hence unquantified)
systematic error associated with the neglect of DV
contributions in the fit. Since the DV contribution to the
FESR~(\ref{sumrule}) involves the weighted integral of the DV
component of the spectral function in the interval
$s\ge s_0$, neglecting this systematic error would
be reasonable if the $V+A$ spectral distribution showed
no signs of DVs in the region $s>2.2$~GeV$^2$. This is, however,
rather far from being the case, making the absence of an
estimate for the residual systematic error associated with
neglecting DV contributions problematic. One internally
consistent way to test whether DV contributions are sufficiently
small to be neglected for the $\hat w_3$ FESR is to demonstrate
that they are already small
for the singly pinched $\hat{w}_2$ FESR.
Whether or not this is the case can be
investigated by
comparing the $\hat w_2$-weighted OPE and spectral integrals, in the same
$s_0$ range, using parameters obtained
from the no-DV fit to $\hat w_3$, Eq.~(\ref{w00noDVsfitvals}).
The results of this test are shown
in Fig.~\ref{w00noDVscomp} (right panel). The agreement
between the OPE and spectral integrals is clearly not
good, indicating the presence of significant DV contributions
in the $\hat{w}_2$ FESR.  This, together with the rapid deterioration
of the
$\hat w_3$ no-DV fit quality for $s_{\rm min}\le 1.95$~GeV$^2$, suggests
that neglecting DV contributions to the $\hat w_3$
FESR is also dangerous.

The hope underlying existing FESR analyses which ignore
DV effects is that the double pinching of the weight $w_{00}=\hw_3$
is sufficient to make the residual DV contributions very small.
While the arguments above make this possibility
unlikely, it is still logically possible that, although DVs cannot be ignored
in the singly-pinched $\hw_2$ FESR, they can be ignored in the
doubly-pinched $\hw_3$ FESR.
Let us therefore consider
again the FOPT version of the $\hat{w}_3$ FESR in the $V+A$ channel,
but now, rather than ignoring DVs,
taking as external input the results for the DV parameters from
the $s_{\rm min}=1.55$~GeV$^2$ FOPT fit of Tab.~\ref{VAwtaupaper}
and fitting the remaining OPE parameters
$\a_s(m_\tau^2)$, $C_{6,V+A}$, and $C_{8,V+A}$, to the $\hw_3$
weighted spectral integral in the $V+A$ channel in the presence of
this estimate of the DV contributions.
The results of this exercise, which are to be
compared with Eq.~(\ref{w00noDVsfitvals}),~are
\begin{eqnarray}
\label{w00externalDVinputfitvals}
\alpha_s(m_\tau^2)&=&0.301\pm 0.006 \pm 0.009\ ,\\
C_{6,V+A}&=&-0.0127 \pm 0.0020 \pm 0.0066  \ {\rm GeV}^6\ ,\nonumber\\
C_{8,V+A}&=&0.0399 \pm 0.0040 \pm 0.021\ {\rm GeV}^8\ ,\nonumber
\end{eqnarray}
where the first error is statistical and
the second is that induced by the correlated uncertainties of
the external input DV parameters.
The inclusion of the DV contributions
induces  a significant decrease in the value of $\a_s(m_\tau^2)$
and  significant changes in the results for $C_{6,V+A}$ and
$C_{8,V+A}$ (including changes in sign for both) as compared to
the no-DV fit results of Eq.~(\ref{w00noDVsfitvals}). The fit parameters
are all changed in the direction of the results of the
more detailed combined $V$ and $A$ fits discussed in Sec.~\ref{fits}. 
This exercise clearly
demonstrates that the effects of DVs on the parameters
obtained from the $V+A$ $\hat{w}_3$ FESR analysis
are much
larger than the nominal errors obtained on those parameters
from the no-DV fit.  This provides a further indication of
the necessity of modeling DV effects
in analyses attempting to extract $\a_s(m_\tau^2)$ from
hadronic $\tau$-decay data.

\section{\label{conclusion} Conclusion}
In this article, we reanalyzed the recently revised ALEPH data \cite{ALEPH13}
for non-strange hadronic $\tau$ decays, with as primary goal the extraction
of the strong coupling $\a_s$ at the scale $m_\tau$. The rather low value
of $m_\tau$ raises the question of to what extent
the determination of a perturbative quantity like $\a_s$ in such an
analysis might be ``contaminated'' by non-perturbative effects.
Our specific aim was to take all known non-perturbative effects into
account and arrive at a realistic estimate of the systematic error
on the value of $\a_s$ extracted using hadronic $\tau$ data.
This is important for three reasons. First, the value of $\a_s$
from $\tau$ decays, evolved to the $Z$ mass, has long been claimed
to be one of the most precise values available. Second, because the
$\tau$ mass is so much smaller than other scales at which the strong
coupling has been determined, $\a_s(m_\tau^2)$ provides a powerful test of
the QCD running of the strong coupling, with the corresponding $\b$ function
known to four-loop order.
Finally, there continues to be some tension
between the values of the $n_f=5$ coupling $\alpha_s(M_Z^2)$
obtained from different sources. While lattice determinations
involving analyses of
small-size Wilson loops~\cite{hpqcdalphas,adelaidehpqcd},
$c\bar{c}$ pseudoscalar correlators~\cite{hpqcdcharmpsalphas},
the relevant combination of ghost and gluon two-point
functions~\cite{sternbeckggalphas,frenchgp}, and employing
the Schr\"odinger functional scheme~\cite{pacscsalphas} yield values,
$0.1183(8)$~\cite{hpqcdalphas}, $0.1192(11)$~\cite{adelaidehpqcd},
$0.1186(5)$~\cite{hpqcdcharmpsalphas}, $0.1196(11)$~\cite{frenchgp},
and $0.1205(20)$~\cite{pacscsalphas}, compatible both amongst one another and
with the central value of the global electroweak fit result,
$\alpha_s(M_Z^2)=0.1196(30)$~\cite{globalewalphas},
lower values have been obtained in a number of other
analyses, \eg, $0.1174(12)$ from lattice analyses of
$f_\pi /\Lambda_{QCD}$~\cite{fpilattalphas}, $0.1166(12)$ from an
analysis of the static quark energy~\cite{staticValphas},
$0.1118(17)$ from the recently revised JLQCD lattice
determination from current-current two-point functions~\cite{jlqcdalphas}, and
values in the range $0.1130-0.1160$ from analyses of DIS data
and shape observables in $e^+e^-$~\cite{disjetsthrustetc}.

We have employed our analysis method previously
\cite{alphas1,alphas2}, using the OPAL data \cite{OPAL}, but the
revised ALEPH data have significantly smaller errors, and thus provide
a more stringent test of our analysis method.

The fact that at such low scales non-perturbative effects are not
negligible has of course been long known, and has been taken into
account in the analysis of hadronic $\tau$ decays through the inclusion
of higher-dimension condensate terms in the OPE.  However, the
experimental data are provided in the form of spectral functions, \ie,
as functions of $s=q^2$ with $q$ denoting momentum in Minkowski space.
Such values of $q^2$, viewed as a complex variable, are outside the
domain of validity of the OPE. While this is well known, it can also
easily be inferred from the form of the vector spectral function in
Fig.~\ref{ALEPH-OPAL}, which clearly shows oscillations that cannot be
represented by the OPE.  These oscillations
lead unavoidably to the conclusion that violations of quark-hadron
duality are, in general, significant at the scales accessible through
experimental hadronic $\tau$ decay data.

It follows that in order to investigate the effect of duality
violations on the extraction of $\a_s$ from $\tau$ decay data, they need to
be taken into account. Unfortunately, a model is needed in order to
parametrize the oscillations in the spectral functions, and this
modeling necessitates making some assumptions on which to base the
analysis.  This is, however, true for any such analysis: the
assumption that duality violations can be ignored in a given analysis
 amounts to assuming a model as
well; in terms of the \ansatz~(\ref{ansatz}) it corresponds to taking
the parameters $\d_{V,A}$ to $\infty$. 
 We have, instead, assumed that
this \ansatz\ (with finite $\d$) provides a reasonable model of the
resonance features present in the spectral functions for values of $s$
in some region below $m_\tau^2$ in which perturbation theory is still
meaningful.\footnote{Up to the order considered \cite{PT}.}  As much
as our aim is to find the most accurate value of $\a_s(m_\tau^2)$
possible given the data, an equally important goal was to test the
validity of our approach, with the increased precision of the ALEPH
data as compared to the OPAL data being particularly
useful in this regard.
This increased precision is, moreover, found to
produce unique fit minima in the {\tt hrothgar} studies of the multi-dimensional
fit parameter space, improving the situation found for the corresponding fits to
the OPAL data, and confirming that the precision of the ALEPH data is
more than good enough to support fits incorporating an explicit representation
of DV contributions.

Despite the recent resurgence of interest in this problem, triggered
by the completion of the five-loop calculation of the Adler function
in Ref.~\cite{PT}, very few investigations have carried out a complete
analysis starting from the data.  In essence, only two methods have
been proposed through which to investigate non-perturbative effects,
with the first being the method based on Refs.~\cite{BNP,DP1992}, which
was employed by Refs.~\cite{ALEPH13,ALEPH,ALEPH08,OPAL}, and the second
being the method we employed in this article, applying and extending
ideas proposed in earlier work \cite{MY08,alphas1,alphas2,BBJ12}.  In
the absence of a detailed theoretical understanding of duality
violations, it is important to test for the self-consistency of either
analysis method using the data employed in the analysis.

In Sec.~\ref{ALEPH} we demonstrated that the first method, used in
Ref.~\cite{ALEPH13}, does not pass such tests.  Indications supporting
this conclusion have been published in earlier work, but now that the
revised data are available, and in view of our critique in
Sec.~\ref{ALEPH}, we conclude that this method suffers from numerically
significant systematic uncertainties not quantifiable within the
analysis framework employed in Ref.~\cite{ALEPH13}, and hence must be
discarded. The second method, employed in this article, does a much
better job in fully describing the data, as we have shown in great
detail in Secs.~\ref{fits}, \ref{results} and \ref{ALEPH} above.
However, there are some signs that also the limits of this method
maybe in view. Fit qualities are typically larger than in the case of
our analysis of the OPAL data \cite{alphas2}, and a comparison of
results based on ALEPH and OPAL data also shows some tension, even
though errors are too large to say anything more conclusive.  While
these tensions may be caused by imperfections in the data (for
instance slight discrepancies in the spectral function data visible in
Fig.~\ref{ALEPH-OPAL}), it is by no means excluded that they point to
shortcomings of the theory description as well.

We briefly reviewed, in Sec.~\ref{theory},
why we consider the
DV parametrization in Eq. (\ref{ansatz}) 
a physically sensible one. 
However, it remains relevant to test this
form more quantitatively using experimental data. In
this regard, we would like to stress
that the exercise involving the $x^N$ FESRs leading 
to the results of Eq.~(\ref{OPEn})
represents a highly non-trivial test of
this type.
This follows from the
fact that DV contributions to the $x^N$ FESRs
are generally not small, and oscillate with $s_0$.
The $D=0$ OPE and DV contributions to the theory side of the $x^N$ FESR
for each $N$
are, in this exercise, fixed by the results of the earlier fits involving
the \ansatz~(\ref{ansatz}), leaving only a $D=2N+2$ OPE contribution
controlled by $C_{2N+2}$ to complete the theory side of the FESR.
The different $x^N$ considered 
provide very different weightings on the interval from $s_0$ to $\infty$,
and the different $s_0$ considered represent
integration over different portions of the oscillations in the
experimentally accessible region.  Therefore, a problem with the DV \ansatz\ would be
expected to show up as an inability to successfully fit, with the single
parameter $C_{2N+2}$, the $s_0$-dependent difference between the
experimental spectral integrals and the sum of the previously fixed
$D=0$ OPE and DV theory integral contributions. In fact, as we have seen,
a set of $C_{2N+2}$ exist which produce excellent matches to the
experimental spectral interals over a sizeable range of $s_0$ for all
$N$ ($N=1, \cdots , 7$) required to generate the results, shown in
Fig.~\ref{w10w11theoryspecintcomp}, for the weights $w_{k\ell}$ employed in
Ref.~\cite{ALEPH13}. The fact that the form~(\ref{ansatz}) conforms
to the qualitative features expected of the contribution representing
the residual error of an asymptotic series, and the success of the
detailed self-consistency tests just described, confirms
that the \ansatz~(\ref{ansatz}) provides a good representation of
DV effects in the channels of interest. Possible residual inaccuracies
in this representation should, in any case, not be turned into an argument
to not include DVs at all, since that strategy
would lead to the presence of unquantifiable systematic errors
which use of our \ansatz\ strongly indicates are unlikely
to be small.

It is interesting to compare the values of $\a_s(m_\tau^2)$ from the
various analyses. First, the half differences between our ALEPH- and
OPAL-based values are 0.015 (FOPT)
and 0.019 (CIPT),
while the  average (between FOPT and CIPT) fit
errors is about 0.012
for fits to ALEPH data (\seef\ Eq.~(\ref{alphasfinal})), and about double
that for fits to OPAL data. Finally, the difference between the FOPT and
CIPT values is 0.014 for the ALEPH-based values, and 0.022
for the OPAL-based values. These differences and errors are all comparable in size, and it
appears reasonable to conclude that they reflect both the data and theory
limitations on the accuracy with which $\a_s(m_\tau^2)$ can be obtained from
analyses of hadronic $\tau$ decay, at least at present.
We do not believe that it is meaningful to condense these
results in the form of one central value and one aggregate
error for $\a_s(m_\tau^2)$. Clearly, our ALEPH-based values are
not in agreement with the value obtained in 
Ref.~\cite{ALEPH13}, despite using the same data.
Averaging the values of Eq.~(\ref{alphasfinal}) and adding half the difference
between the two values as an error estimate for the CIPT/FOPT perturbative
uncertainty, we would find a value $\a_s(m_\tau^2)=0.303\pm 0.014$, to be
compared with the value $0.332\pm 0.012$ quoted in Ref.~\cite{ALEPH13}.
It should be emphasized again that the error in the latter value does
not include a component accounting for the systematic
problems identified in Sec.~\ref{ALEPH}.

One may ask whether one can do better. First, it would be interesting
to apply our analysis method to data with better statistics,
and such data are in principle available from the BaBar and
Belle experiments. Such data would allow us to scrutinize
our theoretical understanding in more detail and would, as can be seen
from Fig.~\ref{ALEPH-OPAL}, be especially useful in the upper part 
of the spectrum. However, to date the analyses required to produce
inclusive hadronic spectral functions from these data
are not complete, and thus such an investigation must be postponed
until they become available. Second, it would be nice to develop a
deeper insight into the theory itself, or, lacking that, to develop
new tools for testing any given model for duality violations.
A recent idea in this direction based on functional analysis can be found in
Ref.~\cite{CGP14}. Finally, we note that the
difference between the results for $\alpha_s(m_\tau^2)$
obtained using the FOPT and CIPT resummation schemes represents,
at present, an important limitation on the accuracy with which
$\alpha_s$ can be obtained at a scale as low as $m_\tau^2$; further
progress will require an improved understanding of this issue. 

\vspace{3ex}
\noindent {\bf Acknowledgments}
\vspace{3ex}

We would like to thank Matthias Jamin for useful discussions, and
Andy Mahdavi for generous help with {\tt hrothgar}. MG thanks IFAE and
the Department of Physics at the UAB, and  KM and SP thank the
Department of Physics and Astronomy at SFSU for hospitality.
The work of DB was supported by the Gottfried Wilhelm Leibniz programme
of the Deutsche Forschungsgemeinschaft (DFG) and the Alexander von
Humbodlt Foundation.
MG is supported in part by the US Department of Energy under contract
DE-FG02-92ER40711, and
JO is supported by the US Department of Energy under contract DE-FG02-95ER40896.
SP is supported by CICYTFEDER-FPA2011-25948, 2014~SGR~1450, and
the Spanish Consolider-Ingenio 2010 Program CPAN (CSD2007-00042).
KM is supported by a grant from the Natural Sciences and
Engineering Research Council of Canada.

\end{document}